\documentclass[traditabstract]{aa} % for the abstract without structuration 
\usepackage{graphicx}
\usepackage{txfonts}
\usepackage{wasysym}
\newcommand{\keplere}{{\em KeplerE}}
\newcommand{\variants}{{\em Variants}}
\newcommand{\kepler}{{\em Kepler}}
\newcommand{\meuscat}{{\em Meus Cat}}
\newcommand{\progym}{{\em Progym}}
\newcommand{\dreyer}{{\em Dreyer}}
\newcommand{\manuscript}{{\em Manuscript}}

\begin{document}
  \title{Three editions of the Star Catalogue of Tycho Brahe}
  \subtitle{Machine-readable versions and comparison with the modern Hipparcos Catalogue}

  \author{Frank Verbunt\inst{1} \and Robert H. van Gent\inst{2,3}}

  \institute{Astronomical Institute, Utrecht University, PO Box 80\,000,
    3508 TA Utrecht, The Netherlands; \email{f.w.m.verbunt@uu.nl}
  \and URU-Explokart, Faculty of Geosciences, Utrecht University, PO Box 80\,115,
    3508 TC Utrecht, The Netherlands
  \and Institute for the History and Foundations of Science,  PO Box 80\,000,
    3508 TA Utrecht, The Netherlands }

  \date{Received \today / Accepted}

  \abstract{Tycho Brahe completed his catalogue with the positions and
    magnitudes of 1004 fixed stars in 1598. This catalogue circulated
    in manuscript form. Brahe edited a shorter version with 777 stars,
    printed in 1602, and Kepler edited the full catalogue of 1004
    stars, printed in 1627. We provide machine-readable 
    versions\thanks{The full Tables \keplere\ and \variants\ (see 
    Table\,\ref{t:machine}) and the Table with the latin descriptions of the
    stars are available in electronic from only at the CDS via
    anonymous ftp to cdsarc.u-strasbg.fr (130.79.128.5)
    or via http://cdsweb.u-strasbg.fr/cgi-bin/qcat?J/A+A/} of
    the three versions of the catalogue, describe the differences
    between them and briefly discuss their accuracy on the basis of
    comparison with modern data from the Hipparcos Catalogue. We also
    compare our results with earlier analyses by Dreyer (1916) and
    Rawlins (1993), finding good overall agreement. The magnitudes
    given by Brahe correlate well with modern values, his longitudes
    and latitudes have error distributions with widths of 2\arcmin,
    with excess numbers of stars with larger errors (as compared to
    Gaussian distributions), in particular for the faintest stars.
    Errors in positions larger than $\simeq$10\arcmin, which comprise
    about 15\%\ of the entries, are likely due
    to computing or copying errors.  
   }

  \maketitle

  \section{Introduction}

The astronomical observations of Tycho Brahe improved by an order of
magnitude on the positional accuracy achieved by his predecessors.
His measurements of the positions of stars on the celestial sphere
resulted in a manuscript catalogue in 1598 (Brahe
1598).\nocite{brahe98} Astronomers and mapmakers throughout Europe
used handwritten copies of this catalogue. Brahe edited a shorter
version, with 777 stars, which was printed in 1602 as part of
\textit{Astronomiae Instauratae Progymnasmata} (Brahe
1602).\nocite{brahe02} The full list of 1004 entries with some
modifications was published by Johannes Kepler in 1627 as part of the
\textit{Tabulae Rudolphinae} (Kepler 1627).\nocite{kepler}

These catalogues are a monument in the history of astronomy, and as
such have been studied repeatedly (e.g. Baily 1843, Dreyer 1890,
Rawlins 1993).
\nocite{baily}\nocite{dreyer}

In this paper we describe machine-readable versions of the
catalogues. In addition to the numbers given by Brahe (and Kepler) the
machine-readable tables provide cross-references between the
catalogues, identifications with stars from the (modern!)
\textit{Hipparcos Catalogue} (ESA 1997) and on the basis of these the
accuracy of the positions and magnitudes tabulated in the old
catalogues.
\nocite{esa}

The accuracy of the measurements of Brahe is best studied by reference
to his observational logs rather than by reference to his reduced
data, and Wesley (1978) has shown that the measurement accuracy varies
between the instruments used by Brahe.  For the mural quadrant, for
example, the average error is 34\farcs6.  In his comprehensive study 
of Brahe's star catalogue, Rawlins (1993) also refers to the
observational logs to correct errors that Brahe made in producing his
star catalogue. An important conclusion drawn by Rawlins is that
errors exceeding 6\arcmin\ are usually the consequence of errors in the
reduction of the measurements, rather than in the measurements
themselves. Such errors include repetition of stars in several
entries, mixing data for different stars for one entry, and spurious
entries. Rawlins (1993) produces an improved version of the star
catalogue of Brahe, the best version that Brahe could have published
on the basis of his measurements.  The goal of {\em our} edition is to
present the star catalogue of Brahe as it was available for use to a
17th century astronomer or map maker, i.e.\ the versions as given by
Brahe and Kepler. 

Our analysis improves on earlier ones in three ways. First and
foremost, our analysis is based on the \textit{Hipparcos Catalogue}
(ESA 1997), which is more accurate, complete and homogeneous than 
the stellar catalogues used in earlier analyses. Second, we 
grade each identification, discriminating between secure, probable
and merely possible. Third, we provide images of each constellation
comparing the positions and magnitudes from the Brahe catalogue with 
the posistion and magnitudes of {\em all stars in the field}, thus
illustrating which stars were selected by Brahe for measurements
and which ones not.

In describing the different versions of the Catalogues we use the 
following notation: \manuscript\ refers to the Manuscript version
(Brahe 1598), \progym\ refers to the version edited and printed by Brahe (1602),
\kepler\ refers to the version edited by Kepler (1627), and \keplere\ to
our emended version of the latter catalogue. Individual entries are
numbered according to the order in which they appear in the different
versions. For example, the third entry in \kepler\ corresponds to
% entry 309 in \progym\ and to 
entry 338 in \manuscript, which we denote as K\,3 and M\,338,
respectively. (Thus, our M numbers correspond to D numbers of Rawlins
1993.) The sequence number within a constellation is indicated
by a number following the abbreviated constellation name: Oph\,14 is
the 14th star of Ophiuchus. 

  \section{Description of the Catalogues \label{s:description}}

The Star Catalogue of Brahe is organized by
constellation. \manuscript\ starts with the zodiacal constellations,
followed first by the Northern and then by the Southern
constellations.  Brahe added two constellations with respect to those
given by Ptolemaios: Coma Berenices (between Bootes and Corona
Borealis), and Antinous (between Aquila and Delphinus) and omitted the
southern constellations that follow Centaurus in the Catalogue of
Ptolemaios, viz.\ Lupus, Ara, Corona Australis, and Piscis
Austrinus. For many constellations the list of stars belonging to it
is followed by a list of additional stars near to, but not considered
to be part of, the constellation. In the case of Ophiuchus and Serpens
the additional stars precede the stars of the constellations.  The
total number of entries in \manuscript\ is 1004, but as we will see
below up to fourteen stars occur more than once, so that the actual
number of stars present in the catalogue is closer to 990.  The
version we use is the one given in Volume 3 of the complete works by
Tycho Brahe as edited by Dreyer (1916). Dreyer based his edition
mainly on Codex 306 in the Copenhagen Royal Library, but also
consulted various other manuscripts (Dreyer 1916:
p.332).\nocite{dreyer16}

In \progym\ the additional stars and occasionally some more
stars at the end of the constellation have been omitted, and the
constellation Centaurus is absent, reducing the total number of entries
to 777. Coma Berenices is listed here between Ericthonius and
Ophiuchus. No star occurs twice in this shorter catalogue, and the
number of stars in it equals the number of entries. The version we use
is the reprint in Volume 2 of the complete works by Tycho Brahe
edited by Dreyer (1915).\nocite{dreyer15}

\kepler\ restores the order of constellations as used by Ptolemaios;
it also restores most stars omitted in \progym.  For some stars Kepler
adds an alternative position, indicated with {\em meus catalogus} (`my
catalogue') or with {\em Piferus} or abbreviated versions of these
indications.  For this catalogue we use the orginal edition (Kepler
1627).  \nocite{kepler} Piferus refers to Francesco Pifferi whose
translation with commentary of Sacrobosco's {\em Tractatus de Sphaera}
was printed in Siena in 1604.  In comments with some entries Kepler
(1627) refers to {\em Gr\"unbergerus}, i.e.\ Christoph Gr\"unberger,
who in 1612 brought out an edition of Brahe's star list.

\begin{table}
\caption[]{Constellations and numbers of stars in them.
  \label{t:numbers}}
\begin{tabular}{rrcrrrl}
$C_B$ & C &  & $N$\, & $N_a$ & K & constellation \\
13 &  1 & UMi &  7 & 13 & 1  & Ursa Minor, Cynosura \\
14 &  2 & UMa & 29 & 27 & 21 & Ursa Maior, Helice \\
15 &  3 & Dra & 32 &  0 & 77 & Draco\\
16 &  4 & Cep &  4 &  7 & 109 & Cepheus\\
17 &  5 & Boo & 18 & 10 & 120 & Bootes, Arctophylax \\
18 &  6 & CrB &  8 &  0 & 148 & Corona Borea \\
19 &  7 & Her & 28 &  0 & 156 &  Engonasi, Hercules \\
20 &  8 & Lyr & 11 &  0 & 184 & Lyra, Vultur Cadens \\
21 &  9 & Cyg & 18 &  9 & 195 & Olor, Cygnus \\
22 & 10 & Cas & 26 & 20 & 222 & Cassiopeia \\
23 & 11 & Per & 29 &  4 & 268 & Perseus \\
24 & 12 & Aur &  9 & 18 & 301 & Auriga, Heniochus\\
%, Ericthonius \\
26 & 13 & Oph & 15 & 22 & 328 & Ophiuchus, Serpentarius \\
27 & 14 & Ser & 13 &  0 & 365 & Serpens Ophiuchi \\
28 & 15 & Sge &  5 &  3 & 378 & Sagitta sive Telum \\
29 & 16 & Aql & 12 &  0 & 386 & Aquila seu Vultur Volans \\
30 & 17 & Atn &  3 &  4 & 398 & Antinous \\
31 & 18 & Del & 10 &  0 & 405 & Delphinus \\
32 & 19 & Equ &  4 &  0 & 415 & Equuleus, Equi Sectio \\
33 & 20 & Peg & 19 &  4 & 419 & Pegasus, Equus Alatus \\
34 & 21 & And & 23 &  0 & 442 & Andromeda \\
35 & 22 & Tri &  4 &  0 & 465 & Triangulus, Deltoton \\
25 & 23 & Com & 14 &  1 & 469 & Coma Berenices \\
 1 & 24 & Ari & 21 &  0 & 484 & Aries \\
 2 & 25 & Tau & 43 &  0 & 505 & Taurus \\
 3 & 26 & Gem & 25 &  4 & 548 & Gemini \\
 4 & 27 & Cnc & 15 &  0 & 577 & Cancer \\
 5 & 28 & Leo & 30 & 10 & 592 & Leo \\
 6 & 29 & Vir & 33 &  6 & 632 & Virgo \\
 7 & 30 & Lib & 10 &  8 & 671 & Libra \\
 8 & 31 & Sco & 10 &  0 & 689 & Scorpius \\
 9 & 32 & Sgr & 14 &  0 & 699 & Sagittarius \\
10 & 33 & Cap & 28 &  0 & 713 & Capricornus \\
11 & 34 & Aqr & 41 &  0 & 741 & Aquarius \\
12 & 35 & Psc & 36 &  0 & 782 & Pisces \\
36 & 36 & Cet & 21 &  0 & 818 & Cetus \\
37 & 37 & Ori & 42 & 20 & 839 & Orion \\
38 & 38 & Eri & 10 &  9 & 901 & Eridanus \\
39 & 39 & Lep & 13 &  0 & 920 & Lepus \\
40 & 40 & CMa & 13 &  0 & 933 & Canis Maior \\
41 & 41 & CMi &  2 &  3 & 946 & Canis Minor, Procyon \\
42 & 42 & Arg &  3 &  8 & 951 & Argo Navis \\
43 & 43 & Hya & 19 &  5 & 962 & Hydra \\
44 & 44 & Crt &  3 &  5 & 986 & Crater \\
45 & 45 & Crv &  4 &  3 & 994 & Corvus \\
-- & 46 & Cen &  0 &  4 &1001 & Centaurus \\
   &    &     &777 & 227 \\
\end{tabular}
\tablefoot{ For each constellation the columns
  give the sequence number in \progym\ ($C_B$) and \kepler\ ($C$), the
  abbreviation we use, the number $N$ of stars in \progym, the number
  $N_a$ {\em added} in \kepler, the K-number of the first star in it
  (for quick reference of a K-number to its constellation), and the
  names as given in \kepler.  Note that the modern constellations do
  not exactly match those used by Brahe; in particular the
  constellation Antinous has been subsumed in Aquila, and Argo is split
  since the 18th century in Carina, Pyxis, Puppis and Vela.}
\end{table}

The ordering of the constellations and the numbers of stars in each
constellation is given for the two printed editions (1602 and 1627) in
Table\,\ref{t:numbers}. Generally, \manuscript\ has the same
stars in the same order within each constellation as \kepler.
Exceptions are
\begin{itemize}
\item  \kepler\ ads new stars in Cygnus (Nova 1600 = P\,Cygni) and 
Cassiopeia (Nova 1572) at the end of these constellations. (Remarakbly,
the supernova of 1604, now named after Kepler, is not in the 
catalogue.\footnote{It is however in the {\em Secunda Classis}, a
list which immediately follows the star catalogue of Brahe in
Kepler's (1627) edition, and which contains stars not in the Brahe catalogue})
\item \manuscript\ lists thirteen stars denoted as {\em
Sequentes pertinent ad Ophiuchum et eius Serpentem} (the following
belong to Ophiuchus and his Snake) before the stars of Ophiuchus.
\manuscript\ also erroneously gives the position of the Oph\,14 as
identical to that of Oph\,12. \kepler\ places these thirteen
stars in his catalogue as follows: the first one {\em In dextra tibia
Ophiuchi} (in the right shinbone of Ophiuchus) is identified with 
Oph\,14 {\em In dextra tibia} and gives
it its correct position; the remaining twelve are listed at the end of
Ophiuchus, as Oph\,26 to Oph\,37.
\item \kepler\ omits Gem\,30 of \manuscript. (As one
of the stars additional to Gemini in \manuscript, this star is
also absent from \progym.)
\item In \kepler\ Com\,15 is identical to Com\,2. Com\,15 is not in
\manuscript\ and Com\,2 is not in \progym.
\item \manuscript\ contains a star which would have been between
Eri\,13 and Eri\,14 in \kepler, but Kepler omits it. Its position
is almost the same as that of Eri\,8, and it certainly refers to the same
star. \progym\ lists only 10 stars for Eridanus.
\end{itemize}
If we do the counting, we see that Kepler adds three stars (the Novae and
Com\,15) and omits three stars (one each in Ophiuchus, Gemini, Eridanus)
with respect to \manuscript, and thus ends up with the same
total number of 1004 entries.
With these exceptions, the order of the stars within each
constellation is preserved between the various versions of the
catalogue. Thus, UMa\,13 is the same star in all three versions, and
UMa\,33 in \manuscript\ and \kepler\ is absent from \progym.

For each star a short description is followed by the ecliptic longitude
and latitude, and the magnitude.  The longitude is given in degrees ($G$:
integer), minutes ($M$: integer or integer plus 0.5), and zodiacal
sign. The zodiacal sign is indicated with its symbol in the original
Catalogues, which we replace with an integer number $Z$ from 1 to 12
as shown in Table\,\ref{t:signs}.  The longitude in decimal degrees
follows as
$$\lambda = (Z-1)*30 + G + M/60 \nonumber$$
The latitude is given in degrees  ($G$: integer) and minutes ($M$: integer
or integer plus 0.5), plus a B or A indicating north (Borealis) or south
(Australis). In the 1627 edition the B or A is often omitted, assumed
implicitly to be the same as for the previous star.
The latitude in decimal degrees follows as
$$\beta = \pm (G + M/60); \qquad +/- \,\mathrm{for\, B/A}$$

The equinox of the coordinates is given in all three versions of the
catalogue as 1600 (MDC) \textit{annum completum}, or in modern
terms AD 1601.00.

The last number given for each star is the magnitude. This is given as an
integer, occasionally followed in \progym\ by a colon (:) or 
decimal point (.). According to Dreyer (1890, p.354) these qualifications
indicate somewhat brighter (:) and fainter (.), respectively.
\nocite{dreyer}
\manuscript\ and \kepler\ omit the magnitude qualifications.
In lieu of a magnitude the original catalogues occasionally refer
to a star as {\em ne}, for nebulous.

\begin{table}
\caption{Numbers $Z$ replacing the zodiacal signs in the machine-readable
longitudes. \label{t:signs}}
\begin{tabular}{rc|rc|rc|rc|rc|rc}
1 & \aries  &  3 & \gemini & 5 & \leo   & 7 & \libra & 9 & \sagittarius & 11 
 & \aquarius \\
\hline
2 & \taurus &  4 & \cancer & 6 & \virgo & 8 & \scorpio & 10 & \capricornus
 & 12 & \pisces
\end{tabular}
\end{table}

\section{Identification procedure \label{s:identification}}

To identify each star from the Star Catalogue of Brahe with a modern
counterpart, we start with the selection from the \textit{Hipparcos
Catalogue} (ESA 1997) of all stars brighter than Johnson visual
magnitude 6.0.\nocite{esa} We correct the equatorial
\textit{Hipparcos} star positions for proper motion, then precess the
coordinates to equinox 1601.00 (old style, JD\,2305824), and finally
convert equatorial to ecliptic coordinates.  For each entry in the old
catalogue, we find the \textit{Hipparcos} entry at the smallest
angular distance.

\subsection{Conversion of the coordinates \label{s:conversion}}

The \textit{Hipparcos} coordinates are equatorial, for epoch 1991.25
(JD\,2448349.0625)
and equinox 2000.0.  The proper motion thus has to be corrected for
$-$390.25 years.  We apply linear corrections separately to right
ascension and declination, i.e.\ we ignore curvature in the proper
motion.  A major reason for using the \textit{Hipparcos Catalogue} is
that it provides errors on the proper motion, and that these errors
are small.  The typical accuracy of the proper motions given in the
\textit{Hipparcos Catalogue} is a few mas/yr, corresponding to less
than an arcsec over the 390.25 period.  The best measurements by Brahe
have an accuracy of the order of an arcminute.  The error in the
Hipparcos position computed for 1601 contributes less than one percent
to the error in the difference between the Brahe and Hipparcos
positions if the error in the proper motion is less than 0\farcm1 in
390.25 yrs, i.e.\ less than 0\farcs015/yr. This is the case for all
but eight of the stars with $V<6.0$.  The star with the largest error
in proper motion among these eight is $\alpha^2$\,Cen; at
$\sigma_\mu=0\farcs033$/yr, this error is less than 1 percent of the
proper motion itself, and leads to a positional accuracy of 13\arcsec\
for epoch 1601.  Note that 3869 and 645 stars with $V<6.0$ have proper
motions leading to displacements larger than 0\farcm1 and 1\arcmin\
over 390.25 yr, respectively.  We conclude that correction for proper
motion is important, but that the errors in this correction may be
neglected.

For precession of the equatorial coordinates from 2000.0 to 1601.0 we
use equations Eq.3.211 and 3.212-2 of Seidelman (1992).\nocite{seidelman}

Finally, for conversion of equatorial to ecliptic coordinates we use
the value for the obliquity in 1601, computed with Eq.3.222-1 of
Seidelman (1992).

\subsection{Finding the nearest counterpart}

The angle $d$ between two stars with position $\lambda_1,\beta_1$ and
$\lambda_2,\beta_2$ is computed from
\begin{equation}
\cos d = \sin\beta_1\sin\beta_2+\cos\beta_1\cos\beta_2\cos(\lambda_1-\lambda_2)
\label{e:angle}\end{equation}
For small angles $d$, computations based on this equation suffer from
round-off errors.  Therefore, if Eq.\,\ref{e:angle} results in
$d<10\arcmin$, we compute $d$ from
\begin{equation}
d^2 = \cos\beta_1\cos\beta_2(\lambda_1-\lambda_2)^2 +(\beta_1-\beta_2)^2
\label{e:apprangle}\end{equation}

\subsection{Deciding on the identification}

In many cases, the nearest {\em Hipparcos} match to the star in the
Brahe Catalogue, as found with Eq.\,\ref{e:angle} or
\ref{e:apprangle}, is the obvious and unambiguous counterpart, as no
other star is nearby. This may be seen most directly in the Figures we
provide in Sect.\,\ref{s:figures}, in which filled circles indicate a
star from the Brahe Catalogue and open circles stars from the {\em
Hipparcos Catalogue}. Bigger symbols indicate brighter stars.  A
filled circle within an open circle is strongly indicative of a secure
counterpart.  Aries and Gemini are examples of constellations in which
all stars are thus securely identified (Figs.\,\ref{f:aries},
\ref{f:gemini}).  In some cases the offset between the positions is
large and indicates an error, but the identification is still secure
with the nearest counterpart. Examples are $\beta$\,Aur (near +5,+5 in
Fig.\,\ref{f:auriga}), and the three northern stars in Corona Borealis
(Fig.\,\ref{f:corbor}).

In a number of cases we consider not the nearest but another nearby
{\em Hipparcos} match to the star in the Brahe Catalogue as a secure
identification. There are two possible reasons for this: first, the
nearest {\em Hipparcos} star is much fainter than another unidentified
{\em Hipparcos} star at a slightly larger angular distance; second,
the nearest {\em Hipparcos} star is more plausibly identified with
another star in Brahe's Catalogue.  An example of the first reason is
K\,101, near $-$1,+28 in Fig.\,\ref{f:draco}.  K\,101 is 0\farcm7 from
HIP\,80309 ($V$=5.7), and 10\farcm5 from HIP\,80331 ($\eta$\,Dra,
$V$=2.7) and we consider HIP\,80331 the secure counterpart.  Because
we consider a pair of {\em Hipparcos} stars to be one object when they
are within 2\arcmin\ from one another, most identifications of Brahe
entries with a not-nearest {\em Hipparcos} star involve large
positional errors. For example, K\,646 lies about 61\arcmin\ from the
6.0 magnitude star HIP\,65545 and about 78\arcmin\ from the 3.4
magnitude star HIP\,66249 (near +4,+5 in Fig.\,\ref{f:virgo}). We
consider HIP\,66249 (= $\zeta$\,Vir) a secure identification even if
not the nearest Hipparcos star.  An example of the second reason may
be seen in the constellation Centaurus: Fig.\,\ref{f:centaurus} shows
convincingly that the position of the whole constellation is shifted
to lower longitude in the Brahe Catalogue. This implies that the
counterpart of the middle star on the left is not HIP\,67153 just
south of it, but HIP\,67669 almost a degree east of it. Another
example of the second reason is $\kappa$\,UMa which forms a pair with
$\iota$\,UMa just north of it (near $-$20,$-$7 in
Fig.\,\ref{f:ursamaior}).  $\kappa$\,UMa is closest to HIP\,44127, but
so is $\iota$\,UMa, and $\kappa$\,UMa may be securely identified with
HIP\,44471 just east of it.  A further example is illustrated in
Figure\,\ref{f:dracodetail}. K\,93 is closest to HIP\,86219 and K\,94
to the very close ($30\arcsec$) pair HIP\,86614+86620
($=\psi^1$\,Dra).  From the brightness of the stars, however, we
prefer to identify K\,93 with HIP\,86614+86620 and K\,94 with
HIP\,87728. Dreyer (1916) agrees in the identification of K\,93, but
identifies K\,94 with HIP\,89937 ($=\chi$\,Dra), relying more on the
description in Brahe's Star Catalogue of stars K\,93-95 as a
triangle. Rawlins (1993) corrects the position of K\,94, and validates
Dreyer's interpretation. All these examples illustrate the importance
of identifying stars in context, rather than individually.

In a number of cases we cannot find a secure counterpart. Some of these
are illustrated in Fig.\,\ref{f:umadetail}. K\,68, K\,72 and K\,74 may be 
identified with nearby {\em Hipparcos} stars, which we consider
probable rather than secure. K\,75 is at an equal distance of two
{\em Hipparcos} stars, and might be identified with either of them.
The remaining stars cannot be identified with any {\em Hipparcos} star,
either because no nearby matches are found, or -- in the case of K\,67 --
because the nearby star is identified with another star from the Brahe
Catalogue. Another constellation with stars we cannot confidently
identify is Ophiuchus (the group near +10,$-$16 in Fig.\,\ref{f:ophiuchus},
see also Fig.\,\ref{f:ophdetail}).
We discuss uncertain identifications in Sect.\,\ref{s:notes}.

Finally, in three cases we find that two entries in the Star
Catalogue are identical (K\,10=K\,252; K\,339=M\,701; K\,483=K\,470), 
and in other cases almost identical and probably referring to the 
same star. In two cases three entries appear to refer to the
same star. These multiple entries are listed in Table\,\ref{t:doubles}.

\begin{table}
\caption{Multiple entries in the Brahe Catalogue.\label{t:doubles}}
\begin{tabular}{l@{\hspace{0.1cm}}rrrrcl}
& \phantom{KK}K & K(M) & HIP\phantom{I} & $d_{1,2}$(\arcmin) & $\Delta M$ & see\\
 &  10 & 252 & 23734 &   0.0 & 0 \\
a &  14 &  64 & 51808 &  13.8 & 0 & Fig.\,\ref{f:ursaminor} near +6,+2  \\
 & 201 & 216 & 96441 &   7.8 & 0 & Fig.\,\ref{f:cygdetail} \\
 &     & 220 & 96441 &   1.0 & 2 & Fig.\,\ref{f:cygdetail} \\
 & 333 & 360 & 80883 &   6.6 & 1 & Fig.\,\ref{f:ophiuchus} near $-$6,+6 \\
 & 336 & 347 & 86284 &  46.3 & 0 & Fig.\,\ref{f:ophdetailb} \\
 &     & 364 & 86284 &  30.0 & 1 & Fig.\,\ref{f:ophdetailb} \\
 & 339 & M701 & 84012 &  0.0 & 0 \\
a & 341 & 353 & 84893 &  22.3 & 0 & Fig.\,\ref{f:ophdetail} near +8,$-$16 \\
 & 344 & 361 & 84880 &  11.2 & 1 & Fig.\,\ref{f:ophdetailb} \\
 & 345 & 362 & 86263 &   5.1 & 1 & Fig.\,\ref{f:ophdetailb} \\
 & 346 & 363 & 86565 &  16.5 & 1 & Fig.\,\ref{f:ophdetailb} \\
 & 470 & 483 &  --   &   0.0 & 0 \\
 & 584 & M94 & 39780 &   3.6 & 1 \\
 & 908 & M913 & 17593 &  0.4 & 0 \\
b & 13 & 297 & 15520 & -- & 0 \\
b & 738 & 739 & 108036 & -- & 0 \\
b & \multicolumn{2}{l}{205-7\phantom{777}219} & 102589 & -- & 2
\end{tabular}
\tablefoot{For each pair of K (or M) numbers the table lists the
{\em Hipparcos} identification, with the difference in positions and
magnitude of the Brahe entries and the approximate position in a
Figure in Sect.\,\ref{s:figures}. Effects of position corrections
by Rawlins (1993, esp.\ footnote 77): a: no longer repeated entry, b:
new repeated entry.}
\end{table}

We note that all three stars from \manuscript\ omitted in \kepler\ are
in Table\,\ref{t:doubles}, i.e.\ they
were presumably recognised by Kepler as double entries. We already
noted that one of the stars added in \kepler, viz.\ K\,483, is a
repeat entry. Thus, if we accept that all 15 entries in the second
column of Table\,\ref{t:doubles} refer to the corresponding star in
the first column, this implies that the 1004 entries in the \manuscript\
actually correspond to 990 stars; and that the
1004 entries in \kepler\ correspond to 992 stars. It
may be argued that some entries from the second column of
Table\,\ref{t:doubles} refer to a hitherto unidentified star. This can
be excluded in all cases where the angular separation between the
Catalogue entries is less than 2\arcmin, the resolution of the human
eye. On the basis further of the absence of sufficiently bright
candidate counterparts, as illustrated in the Figures mentioned in
Table\,\ref{t:doubles}, we consider it most likely that all entries
from the second column are in fact repeat entries.

While looking for identifications we occasionally encountered cases
where an emendation to the Brahe Star Catalogue appears to be warranted.
We describe these in Sect.\,\ref{s:emendations}.
\begin{table*}
\caption{First lines from the machine-readable tables \keplere\ and {\em
    Variants}. \label{t:machine}}
\begin{tabular}{rrrr@{\,}r@{\,}rr@{\quad}r@{\quad}lr@{\quad}l@{\quad}clrrrrrrrrc@{}c@{}c@{}c}
M\phantom{3} & B\phantom{3}&K & C & & & Z & G & M & G & M & 
& $V_B$ & HIP\phantom{P} & I & D & R & $V$ & 
$\Delta\lambda$ & $\Delta\beta$ & $\Delta$ \\
\hline
\\
 336 & 307 &  1 & 1 &=UMi&  1 & 3 & 23& 02.5& 66 &02.  & B & 2 & 11767 & 1 & 1 & 1 & 2.0 & $-$3.2 & 1.2 & 1.8\\
 337 & 308 &  2 & 1 &=UMi&  2 & 3 & 25& 36. & 69 &50.5 & B & 4 & 85822 & 1 & 1 & 1 & 4.3 & 1.6 & 3.4 & 3.5\\
 338 & 309 &  3 & 1 &=UMi&  3 & 4 & 03& 24. & 73 &50.  & B & 4 & 82080 & 1 & 1 & 1 & 4.2 & 8.0 & 2.3 & 3.2\\
 339 & 310 &  4 & 1 &=UMi&  4 & 4 & 21& 29. & 75 &00.  & B & 4 & 77055 & 1 & 1 & 1 & 4.3 & 15.2 & 4.5 & 6.0\\
 340 & 311 &  5 & 1 &=UMi&  5 & 4 & 24& 52. & 77 &38.5 & B & 5 & 79822 & 1 & 1 & 1 & 4.9 & $-$10.4 & 10.2 & 10.4\\
\ldots \\
 348 &   0 & 13 & 1 &=UMi& 13 & 3 & 18& 03. & 42 &56.  & B & 6 & 22783 & 3 & 1 & 4 & 4.3 &$-$158.1 & 26.0 & 118.2\\
 349 &   0 & 14 & 1 &=UMi& 14 & 4 & 21& 38. & 57 &55.  & B & 6 & 51808 & 6 & 1 & 1 & 4.9 & $-$62.3 & 38.7 & 50.7\\
 350 &   0 & 15 & 1 &=UMi& 15 & 3 & 21& 55. & 70 &42.  & B & 6 &     0 & 5 & 3 & 4 & 0.0 & 0.0 & 0.0 & 0.0\\
 351 &   0 & 16 & 1 &=UMi& 16 & 3 & 24& 31. & 69 &03.  & B & 6 & 85699 & 1 & 1 & 4 & 5.8 & 61.4 & 27.8 & 35.2\\
\hline
 344 &   0 &  9 & 1 &=UMi&  9 & 4 & 27& 20.5& 70 &18. & B & 6 & 69112 & 1 & 1 & 1 & 4.8 & $-$10.5 & $-$0.8 & 3.6 & .&.&M&.\\
 351 &   0 & 16 & 1 &=UMi& 16 & 3 & 24& 31. & 69 & 8. & B & 6 & 85699 & 1 & 1 & 1 & 5.8 & 61.4 & 22.8 & 31.4 & .&.&M&.\\
\ldots \\
 403 &   0 & 68 & 2 &=UMa& 48 & 5 &  3& 58. & 47 &55. & B & 6 &     0 & 5 & 0 & 1 & 0.0 & 0.0 & 0.0 & 0.0 & K&.&.&.\\
 413 & 344 & 78 & 3 &=Dra&  2 & 9 &  4& 14.5& 78 &14.5 & B & 4.& 85819 & 1 & 1 & 1 & 4.9 & 22.7 & $-$3.2 & 5.6 & .&P&M&.\\
\ldots \\
 586 & 478 & 238& 10& =Cas& 17& 1& 22 & 23.0 & 54 & 27.0 & B & 6 & 115990 & 1 & 3& 1 & 4.9&  15.9 & $-$24.3 & 26.0&  K&.&.&.\\
 586 & 478 & 238& 10& =Cas& 17& 2& 22 & 33.0 & 54 & 27.0 & B & 6 &   7078 & 1 & 3& 1 & 5.8& $-$17.1& $-$14.1& 17.3&  .&P&M&.\\
 586 & 478 & 238& 10& =Cas& 17& 1& 22 & 32.0 & 54 & 27.0 & B & 6 & 115990 & 1 & 3& 1 & 4.9&  6.9 & $-$24.3  & 24.6&  .&.&.&C
\end{tabular}
\tablefoot{For explanation of the columns for the basic
  table \keplere\ (above) and from the file with variants {\em
    Variants} see Sect.\,\ref{s:machine}.}
\end{table*}

\subsection{Identifications by Dreyer and Rawlins \label{s:dreyer}}

Unavoidably, deciding between secure and probable identifications or
between probable and possible identifications is sometimes
subjective. We therefore compare our identifications with those by
Dreyer (1916) and Rawlins (1993) in their editions of the Manuscript
version of the star catalogue by Brahe, thus obtaining independent
opinions.  

To compare the identifications by Dreyer with those by us, we convert
his identifications to a Hipparcos number.  In most cases,
\dreyer\ provides a Flamsteed number which is present in the
\textit{Bright Star Catalogue} (Hoffleit \&\ Warren 1991), allowing us
to obtain the \textit{Hipparcos} identification from the HR number via
the HD number. In some cases a Flamsteed number given by Dreyer is
not listed in the \textit{Bright Star Catalogue}. For these we first
check the \textit{SIMBAD} data base: for any Flamsteed number in this
data base the \textit{Hipparcos} number is provided. In seven cases we
have not found the Flamsteed number in the \textit{Bright Star
  Catalogue} or in the \textit{SIMBAD} database.  For these (54\,And,
51\,Lib, 31\,Mon, 2,4,9\,Crt and 35\,Cam) we consulted the Flamsteed
catalogue (Flamsteed 1725). To find the corresponding
\textit{Hipparcos} numbers we convert the positions of all
\textit{Hipparcos} entries brighter than $V=6.0$ to the equinox of the
Flamsteed catalogue 1690.0 (old style) = JD\,2338331 by first
correcting for proper motion and then precessing the coordinates. We
then find the \textit{Hipparcos} entry with the closest coordinates to
each of the seven Flamsteed stars. (In the constellation
Camelopardalis some entries only provide a latitude, omitting the
longitude; in counting entries to find number 35, we ignore these
entries.)  This leads to an unambiguous identification in all cases,
with positional differences ranging from 0\farcm3 to 1\farcm0.

In seventeen cases \dreyer\ gives an identification consisting of the
letter P followed by a roman and an arabic number. This refers to the
catalogue of Piazzi (1803; we use the 1814 reprint).\nocite{piazzi} To
find the corresponding \textit{Hipparcos} numbers we convert the
positions of all \textit{Hipparcos} entries brighter than $V=6.0$ to
the equinox of the Piazzi catalogue 1800.0 = JD\,2378497 and find the
nearest \textit{Hipparcos} entry to each of the 17 Piazzi stars.  This
leads to an unambiguous identification in all 17 cases with positional
differences ranging from 0\farcm06 to 0\farcm19.

In two instances \dreyer\ gives an identification consisting of a letter
G followed by a number (and in both cases by a question mark). This
refers to the catalogue by Groombridge (1838), which uses an equinox
1810.0 = JD\,2382149.\nocite{groombridge}
G\,3887 is 107\arcmin\ from HIP\,112519 (corrected for proper motion
and precession); however, G\,3928 is only 0\farcm03 from HIP\,112519.
G\,2807 has no {\em Hipparcos} counterpart with $V<6$, but HIP\,90182
($V$=6.2) lies at 11\farcm7 from it.

Some emendations to Dreyer's identifications that we consider probable
are discussed in Sect.\,\ref{e:emendreyer}.

To compare the identifications by Rawlins (1993) with those by us we
convert his {\em Bright Star Catalogue} (HR) number for each of his
identifications to an HD number using the \textit{Bright Star
  Catalogue} (Hoffleit \&\ Warren 1991), and use the HD number to find
the corresponding {\em Hipparcos} number.

\section{The machine-readable catalogues \label{s:machine}}

A large majority of the entries is identical in the three catalogues.
To save unnecessary duplication we produce the basic machine-readable
catalogue from \kepler\ to which we add the three
stars from \manuscript\ that Kepler omits. This basic
catalogue is referred to in the following as \keplere. A second
catalogue, referred to below as \variants, collects variants to the
entries in \keplere.  A third catalogue, {\em Names}, contains the
descriptions of the stars from \manuscript. 
% together with the identifications given by Dreyer (1916).

\subsection{The basic catalogue}

Kepler (1627) gives alternative positions for some stars, indicated
with {\em meus catalogus} or with {\em Piferus} or abbreviated
versions of these; these are omitted from \keplere\ (and given in
\variants). In \keplere\ the emendations discussed in 
Sect.\,\ref{s:emendapp} have been implemented.

The columns in \keplere\ give the following information.

Columns 1, 2 and 3 give the sequence numbers of the stars in the three
versions of the catalogue viz.\ M for the manuscript version (Brahe 1598),
P for the printed edition of Brahe (1602), and K for the Kepler (1627) edition.
\keplere\ follows the sequence of entries in Kepler (1627); three stars
from \manuscript\ omitted by Kepler (1627) are listed at the end of
\keplere. Thus \keplere\ contains 1007 entries. To obtain
the ordering of the stars/entries in the Manuscript (1598) or Brahe
(1603) edition, one simply orders the table on column 1 or 2,
respectively. A sequence number zero indicates that the entry is
absent in the corresponding catalogue.

Columns 4 and 5 indicate the constellation in which the star is listed
in Kepler (1627).  For convenience the constellation is indicated both
with its sequence number C in Kepler (1627; see
Table\,\ref{t:numbers}) and with the modern abbreviation (preceded by
the equal sign =). We use {\em Atn} and {\em Arg} as abbreviations for
{\em Antinous} and {\em Argo}, respectively, constellations which are
no longer in use.  Column 6 gives the sequence number of the star
within the constellation in \keplere.

Columns 6 -- 14 copy the information of the original catalogue:
columns 7, 8, 9 give the longitude ($Z$, $G$ and $M$), columns 10, 11,
12 the latitude ($G$, $M$ and A or B), and colums 13, 14 the magnitude
$V_B$ and its qualifier (blank, .\ or :).  The meaning of these numbers and
qualifiers is explained in Sect.\,\ref{s:description}.  Kepler (1627)
often omits the B or A for the latitude, implicitly setting it equal
to the value of the previous entry; \keplere\ always gives B or A
explicitly.  A magnitude indicated as `ne' in the original catalogue
is denoted 9 in \keplere.

Columns 15 -- 22 provide additional information from our analysis.
Columns 15 and 16 give the number of the star in the {\em Hipparcos
Catalogue} with which we identify the entry, and a flag I giving the
quality of the identification, as explained in Sect.\,\ref{s:identification}.
The meanings of the identification flags are given in Table\,\ref{t:idflags}.

\begin{table}[!]
\caption{Meaning of flags I classifying our identifications \label{t:idflags}}
\begin{tabular}{ll}
1 & nearest star, secure identification \\
2 & not nearest star, secure identification \\
3 & probable identification, not secure \\
  & because too far or too faint \\
4 & possible identification \\
  & other identification(s) also possible \\
5 & not identified \\
6 & double entry 
\end{tabular}
\end{table}

It should be noted that in general we limit our search for
counterparts to Hipparcos stars with $V<6.0$, and the term
nearest in our identification flag ignores stars fainter than this limit.
In the case of double entries it is not clear whether the star
listed in the first or that in the second column of Table\,\ref{t:doubles}
should be considered the repeat entry. In general we have the entry
closest to the {\em Hipparcos} star a flag 1 or 3, and the other
doubles entries a flag 6.

Columns 17 and 18 give flags D and R that compare our identification
with those by Dreyer (1916) and Rawlins (1993), respectively.  The
meanings of these flags are given in Table\,\ref{t:drflags}.

\begin{table}[!]
\caption{Meaning of flags D and R for identifications by Dreyer (1916)
  and Rawlins (1993) \label{t:drflags}}
\begin{tabular}{ll}
0 & no identification given by Dreyer/Rawlins \\
1 & identification by Dreyer/Rawlins identical to ours \\
2 & identification by Dreyer/Rawlins identical to other of \\
  & alternative identifications given by us  \\
3 & identification by Dreyer/Rawlins differs from ours  \\
4 & identification by Rawlins different due to corrected position \\
5 & catalogue entry called `utter fake' by Rawlins 
\end{tabular}
\end{table}

Column 19 gives the visual (Johnson) magnitude $V$ of the {\em
  Hipparcos} object given in column 15. Columns 20, 21 give the
differences in longitude $\Delta\lambda$ and latitude $\Delta\beta$
between the correct position (based on information from the {\em
  Hipparcos Catalogue}) and the catalogue entry in tabulated Minutes
$M$. If the catalogue entry for minutes as computed from the position
and proper motion in {\em Hipparcos Catalogue} is $M_\mathrm{H}$, and
$M$ is the value actually given in the Brahe Catalogue, then Columns
20 and 21 give $M_\mathrm{H}-M$.  Column 22 gives the difference
$\Delta$ between correct and tabulated position in \arcmin.

\subsection{Catalogue with variants}

The second catalogue \variants\ collects five types of
variants. First, it gives the original entries of Kepler (1627) for
all entries which are emended in \keplere\ to allow the reader to
judge the validity of our emendations and if required to restore the
original Kepler (1627) edition.  Next, \variants\ gives entries from
the 1598 manuscript version of the catalogue and from the 1602 edition
when these are different from the corresponding entry in
\keplere. 
Finally we give the alternative positions given by Kepler (1927) as
originating from {\em meus catalogus} or {\em Piferus}.

Each entry in \variants\ starts with the variant of an entry in
\keplere, and ends with a 4-character string indicating the origin of
the variant.  If the first character in this string is K the entry
gives the original values of the Kepler (1627) edition (for which
\keplere\ gives emended values). If the second character is P the
entry gives values from Brahe (1602) that differ from \keplere.  If
the third character is M the entry gives the values for the Manuscript
version of Brahe (1598) that differ from \keplere. This notation
allows us to give variants which are common between different versions
of the Star Catalogue in a single line.  If the fourth character of
the string is C or P the entry gives a variant given by Kepler (1627)
with the characterization {\em meus catalogus} or {\em Piferus},
respectively.

Note that the identification procedure is done independently for all
variant positions; when the variant position differs much from the
\keplere\ position, this may lead to a different identification.

\subsection{Catalogue with descriptions}

The third catalogue {\em Names} gives the descriptions of the stars as
given in the Manuscript version of the Star Catalogue (Brahe
1598). For convenience we add to this the sequence numbers M, B and K
of each entry in the various editions of the catalogue.

\section{Analysis and discussion}

\begin{figure}
% gemaakt met plotvar.f
\includegraphics[angle=270,width=\columnwidth]{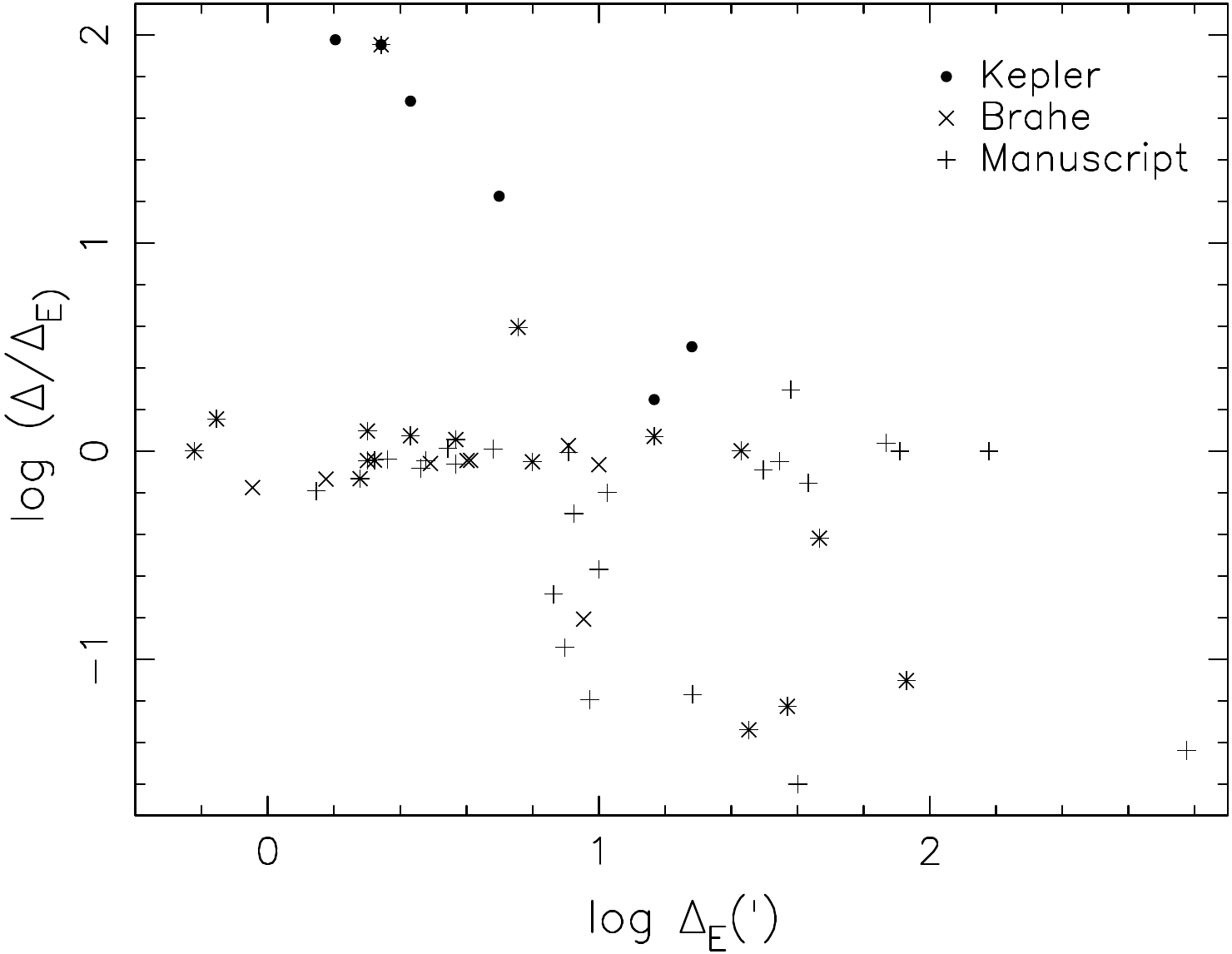}

\caption{Comparison of the accuracies of positions that are different
  in different versions of the Star Catalogue of Brahe. $\Delta_E$ is the
  positional accuracy of \keplere, $\Delta$ the positional accuracy of
  \kepler, \progym, or \manuscript. Note that symbols may be
  superposed when a variant position is common to several versions of
  the catalogue. Stars identified in one but not
  in the other catalogue are omitted.\label{f:variants}}
\end{figure}

It is beyond the purpose of this paper to provide a full analysis of the
Star Catalogue by Brahe and it different editions, but a few general
remarks may be made. 

\subsection{Identifications, emendations and the three versions}

In Table\,\ref{t:dreyer}, in the columns labelled {\em all}, we list
the numbers of our identifications of the entries in \keplere\ and
in \progym. It is seen that only 14 entries remain unidentified
in \keplere, and 5 in \progym. A large majority of stars is securely 
identified.

\begin{table}
\caption{Frequency of flags D of Dreyer (1916) identifications as a
  function of the flags I of our identifications \label{t:dreyer}}
\begin{tabular}{l|rrrrr|rrrrr}
        & \multicolumn{5}{c}{KeplerE} &  \multicolumn{5}{c}{Progymnasmata} \\
I\verb+\+D & 0 & 1 & 2 & 3 & all & 0 & 1 & 2 & 3 & all \\
\hline 
1 &  11 & 896 &   0 &  22 & 929 &   2 & 735 &   0 &  15 & 752 \\
2 &   0 &  12 &   0 &   2 &  14 &   0 &   8 &   0 &   1 &   9 \\
3 &  11 &   5 &   1 &   1 &  18 &   1 &   2 &   0 &   1 &   4 \\
4 &   3 &   4 &   8 &   2 &  17 &   0 &   4 &   3 &   0 &   7 \\
5 &  11 &   0 &   0 &   3 &  14 &   3 &   0 &   0 &   2 &   5 \\
6 &   2 &   8 &   0 &   2 &  12 &   0 &   0 &   0 &   0 &   0 \\
all &  38 & 925 & 9 & 32 & 1004 &  6 & 749 &   3 &  19 & 777 
\end{tabular}
\end{table}

The sixteen emendations that we apply to \kepler\ affect the
number of identifications. In particular, eight lead to secure
identifications (all with $\Delta\le3\farcm2$) of previously unidentified
stars. One other emendation leaves an unidentified star (K\,67)
unidentified, one gives a probable identification of a previously
unidentified star (K\,68), one improves the positional correspondence
with its {\em Hipparcos} counterpart (K\,801), and five lead to
different identifications with better positions. Note that with one
exception (K\,801) all the emendations that we make to \kepler\ are
taken from \progym\ and/or \manuscript.  In Figure\,\ref{f:variants}
we show the change in position caused by our emendations to \kepler,
and by different positions in \progym\ or \manuscript\ with respect to
\keplere.  Not surprisingly, all emendations lead to better
positions. Figure\,\ref{f:variants} shows that most differences
between \progym\ and \keplere\ are small, as are most differences
between \manuscript\ and \keplere.  Remarkably, the Figure also shows
that in all cases where the positions differ strongly between
\kepler\ and \progym\ and/or \manuscript, the positions in the older
catalogues are better.  25 of the 26 entries from \progym\ that differ
from the corresponding entries in \keplere\ have the same
identification in both versions, but K\,547 is identified in
\keplere\ but unidentified in \progym.  Similarly, 39 of 42 entries
from \manuscript\ that differ from the corresponding entries in
\keplere\ have the same identification in both versions, one (K\,251)
is identified in \keplere\ but not in \manuscript, and two (K\,64 and
K\,300) have a different identification in \keplere\ than in
\manuscript.

\subsection{Comparison with Dreyer (1916) and Rawlins (1993)}

In Table\,\ref{t:dreyer} we compare the identifications as found by us with
those given by Dreyer (1916). For both the Brahe (1602) and the emended
Kepler (1627) versions, we find that our identifications agree with the
earlier ones by Dreyer in most cases. We have identified a number of stars
not identified by Dreyer, in some cases prefer another one from several
plausible possibilities, and in some cases reject an identification by
Dreyer. 

The numbers in Table\,\ref{t:dreyer} should be read as indicative
rather than exact, due to unavoidable arbitrariness in some
classifications.  The pair K\,146/K\,147 is an example (see
Sect.\,\ref{s:notes}): we flagged our identifications as secure (I=1)
and Dreyer's as wrong (D=3), but could have chosen ours as one of
several possibilites (I=4) and Dreyer's as an alternative to our
choice (D=2).  Another example is the case of K\,218 in Cygnus. We
have assigned HIP\,106062 as its counterpart, because the closer and
brighter counterpart has been assigned to K\,440, a star in
Pegasus. This is a reasonable choice {\em if} we assume that Kepler
was aware of the proximity of K\,440 to K\,218. If such was not the
case, we could follow Dreyer and consider K\,218 as a repeated entry
for K\,440, and our identification flag would be I=6 rather than I=2;
and the flag for Dreyer's identification D=1 rather than D=3.

The sixteen emendations that we apply to \kepler\ also affect the
numbers in Table\,\ref{t:dreyer}.

Comparison of our identifications with those by Rawlins (1993)
must be made with some care, because his identifications refer to
an ideal version of the catalogue, that Brahe might have produced
given the time, whereas our identifications refer to the catalogue
in the versions edited by Brahe and Kepler. Thus Rawlins identifies the
stars that Brahe actually observed, whereas we identify the stars closest
to the catalogue positions.

In Table\,\ref{t:rawlins} we compare the identifications as found by
us with those given by Rawlins for the emended Kepler edition.  The
three entries in \keplere\ that do not occur in \manuscript\ are not
discussed by Rawlins; all other entries are identified. In 937 cases
our identification agrees with the one by Rawlins. We include in this
four identifications (of K\,583, K\,718, K\,120, and K\,411) given by
Rawlins (1993) that refer to one of a close pair of stars, whereas our
identification refers to the other star of the pair. In each of these
cases the pair is not separable with the naked eye, with a separation
$<2$\arcmin, and our identification refers to the star that is
brighter in the {\em Hipparcos Catalogue}. In 911 cases the
identifications given by Dreyer (1916), Rawlins (1993) and us all
agree.  In one case the identification given by Rawlins refers to one
of two possibilities considered by us.  In 53 cases Rawlins finds a
different identification because he has corrected the catalogue
position.  For 4 entries, all in Ophiuchus, Rawlins concludes that
Brahe invented positions without having observed them: they are `utter
fakes' (see Fig.\,\ref{f:ophdetail}).

\begin{table}
\caption{Frequency of flags R of Rawlins (1993) identifications
 as a function of the flags I of our identifications for \keplere.
\label{t:rawlins}}
\begin{tabular}{l|rrrrrrrr}
I\verb+\+R & 0 & 1 & 2 & 3 & 4 & 5 & all \\
\hline 
1 &   2 & 900 &   0 &   5 &  22 &   0 & 929 \\
2 &   0 &  11 &   0 &   0 &   3 &   0 &  14 \\
3 &   0 &   6 &   0 &   0 &   8 &   4 &  18 \\
4 &   0 &   9 &   1 &   0 &   7 &   0 &  17 \\
5 &   0 &   0 &   0 &   1 &  13 &   0 &  14 \\
6 &   1 &  11 &   0 &   0 &   0 &   0 &  12 \\
all &  3 & 937 &  1 &   6 &  53 &   4 & 1004
\end{tabular}
\end{table}

This leaves 6 entries where our identification is different from that
given by Rawlins. Five of these concern pairs of stars, with
separations varying from 3\farcm8 to 10\farcm5, in which our suggested
counterpart is closer to the catalogue position than the counterpart
given by Rawlins. An example is shown in Fig.\,\ref{f:oridetail}.  In
four cases (K\,183, K\,209, K\,671, K\,870) our counterpart is the
brighter star of the pair, in one case (K\,804) only slightly fainter
than the other star. In some cases, e.g. K\,175, K\,183 and K\,209,
Rawlins combines two stars separated by 8-11\arcmin\ into
one counterpart; in such cases we may choose the brighter and/or closer
star as the counterpart, or leave the entry unidentified.

The corrections applied by Rawlins (1993) also affect the number of
repeated entries, as indicated in Table\,\ref{t:doubles}.

\subsection{Accuracy}

Table\,\ref{t:dreyer} shows that 14 stars remain unidentified in our
emended Kepler catalogue. Two entries in \keplere\ have a secure identification,
but no counterpart from the {\em Hipparcos Catalogue}: K\,267 is SN\,1572
and K\,577 is Praesepe.
To identify the fourteen unidentified stars would require that
one accepts either a fainter counterpart, or a larger positional offset.
It is necessary to note that such acceptance increases the probability
of chance coincidences, i.e.\ of spurious identifications. That this
is a serious problem may be concluded from the fact that we classified
as `secure' four identifications of entries in Kepler (1627) that were
identified with {\em other} counterparts after our emendation was applied.
Thirteen of our unidentified entries are corrected by Rawlins (1993)  
to new positions, that allow him to identify them. One of our unidentified
entries (K\,175) is identified by Rawlins as the combined light of
HIP\,87045 ($V$=6.47) and HIP\,87119 ($V$=6.83) two stars separated by
8\farcm8.

\begin{figure}
\includegraphics[angle=270,width=\columnwidth]{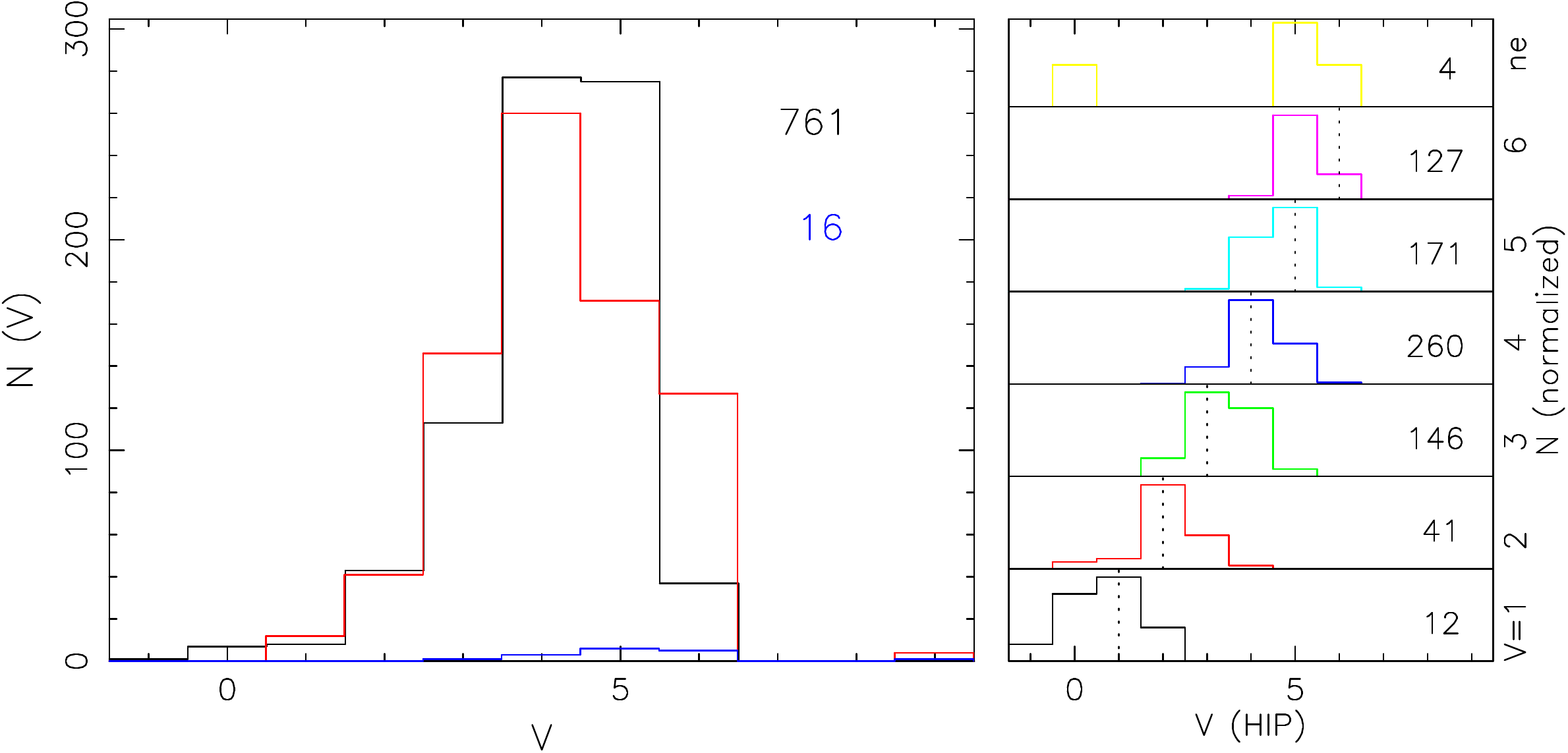}

\includegraphics[angle=270,width=\columnwidth]{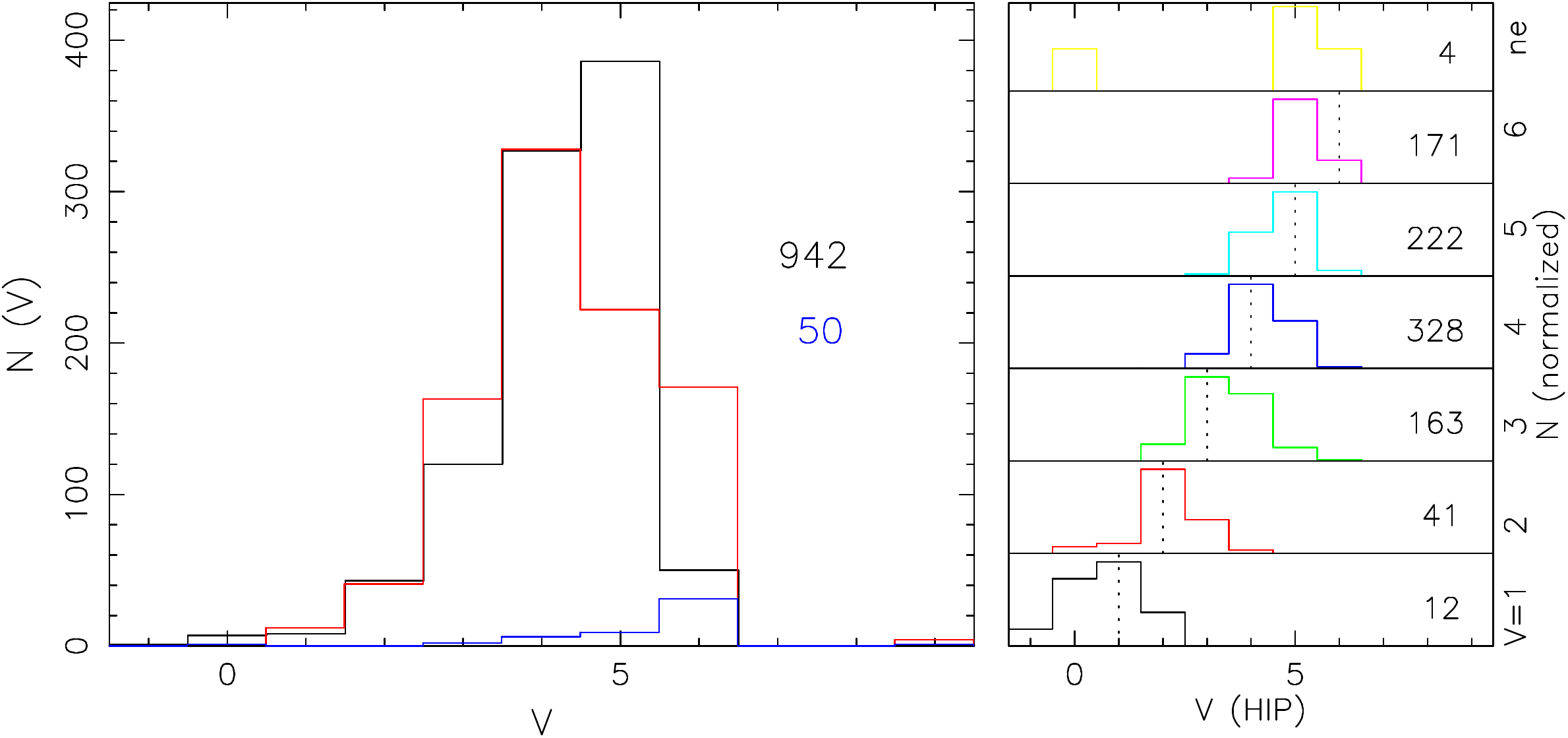}

\caption{Distribution of the magnitudes in \progym\ (above) and
  \keplere\ (below). In the large frames the histograms indicate the
  magnitudes according to Brahe for stars which we have securely
  identified (red; flags 1-2 in Table\,\ref{t:idflags}) or not
  securely identified (blue, flags 3-5), and the magnitudes from the
  {\em Hipparcos} catalogue for securely identified stars (black). The
  numbers of securely and not-securely identified stars are indicated.
  The small frames give the {\em Hipparcos} magnitude distributions
  for securely identified stars for each magnitude according to Brahe
  separately.  The number of securely identified stars at each (Brahe)
  magnitude is indicated.
 \label{f:magnitudes}}
\end{figure}

In Figure\,\ref{f:magnitudes} we compare the magnitude distributions
of the stars in \keplere\ to those of their securely identified
counterparts.  In our opinion, a difference between an {\em Hipparcos}
magnitude and the magnitude assigned by Brahe cannot be called an
error, since the magnitudes for Brahe correspond to a classification
rather than a measurement. It is striking that Brahe's classification
in general correlates well with the modern magnitude; only his
magnitude 6 corresponds to mostly brighter stars in {\em Hipparcos}.
The number of securely identified stars peaks at modern magnitudes 4
and 5, and rapidly drops for magnitude 6. This lends support to our
hesitance in accepting stars with modern magnitudes $V>6$ as feasible
counterparts.

\begin{figure}
% made with plotall.f
\includegraphics[angle=270,width=0.48\columnwidth]{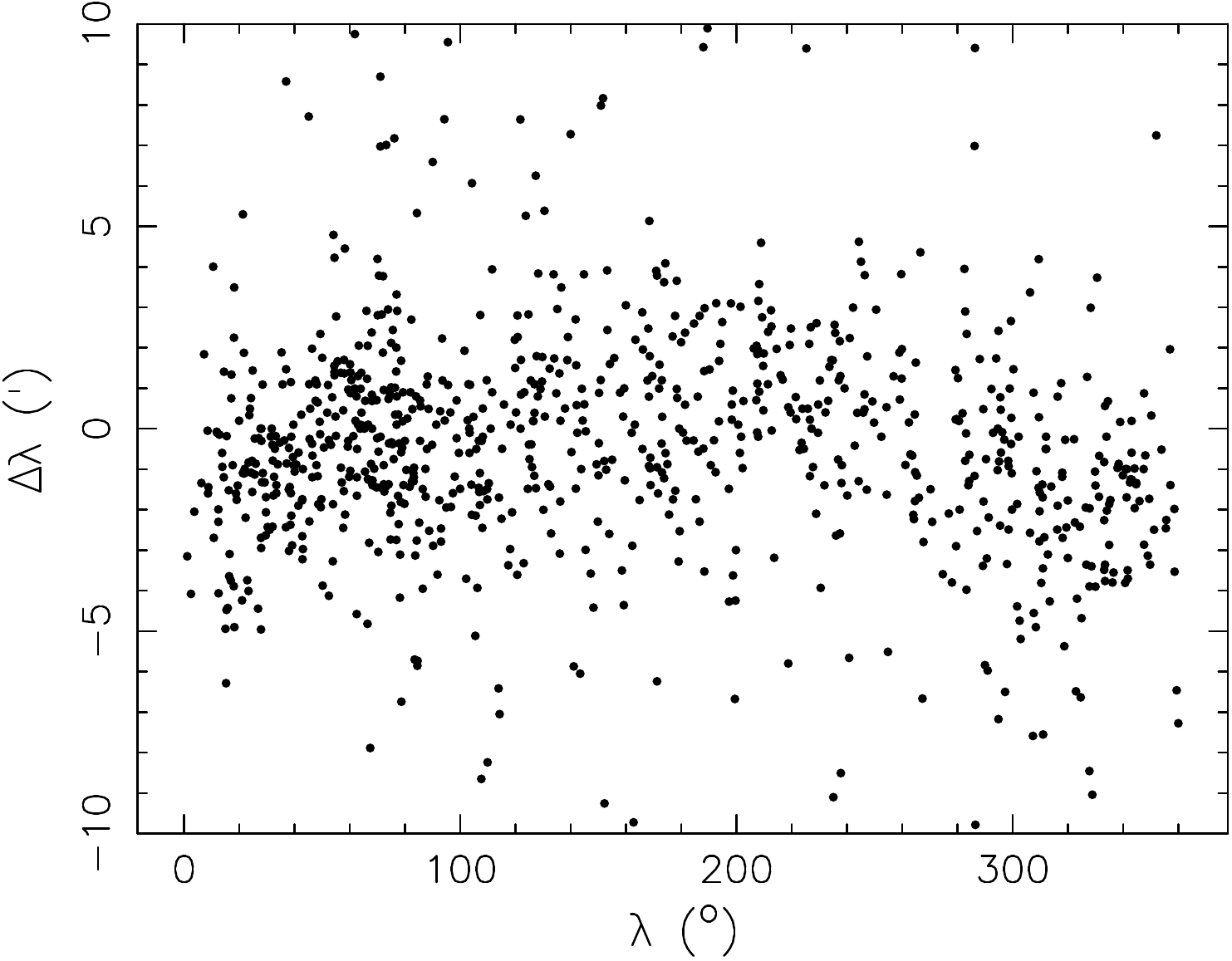}
\includegraphics[angle=270,width=0.48\columnwidth]{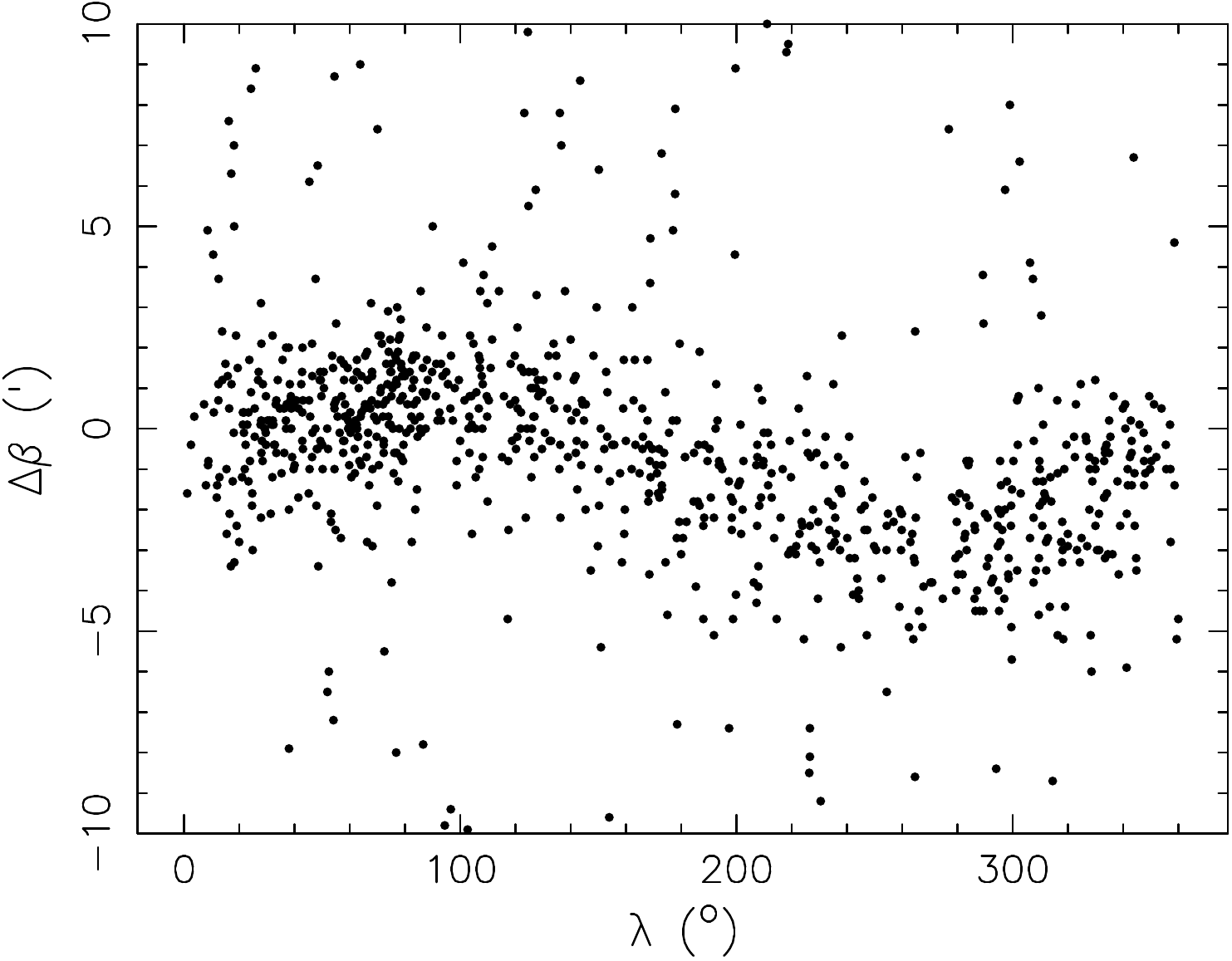}

\includegraphics[angle=270,width=\columnwidth]{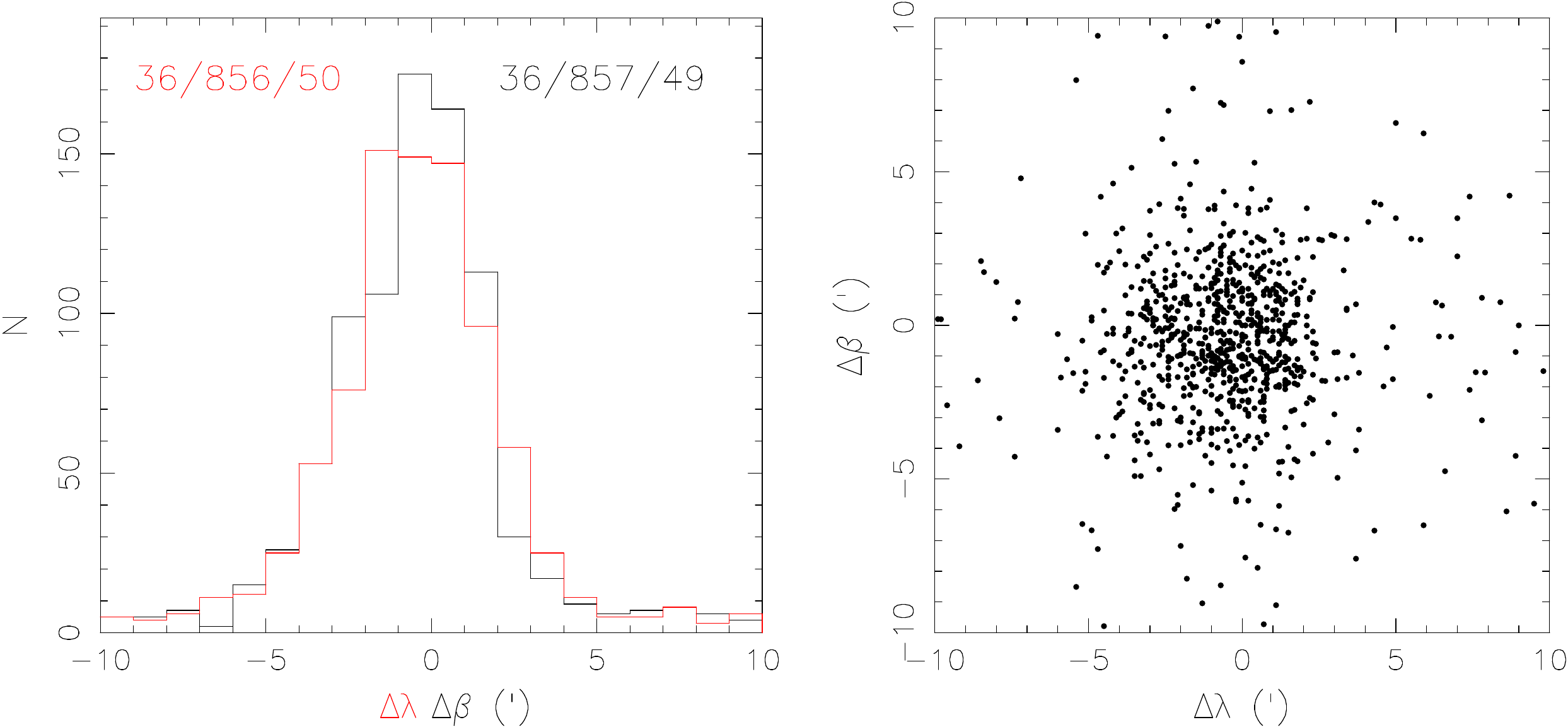}

\caption{Above: Correlations of the differences in longitude
  $\Delta\lambda\equiv(\lambda_\mathrm{H}-\lambda)\cos\beta$ and
  latitude $\Delta\beta\equiv\beta_\mathrm{H}-\beta$ of the entries
  in \keplere\ and their secure {\em Hipparcos} counterparts
  (converted to 1601) with longitude. Below left: Distributions of
  $\Delta\lambda$ and $\Delta\beta$. The numbers of sources with
  $\Delta\lambda,\Delta\beta<-10\arcmin$, of sources included in the
  histogram ($-10\arcmin<\Delta\lambda,\Delta\beta<10\arcmin$), and of
  sources with $\Delta\lambda,\Delta\beta>10\arcmin$ are
  indicated. Below right: correlation between $\Delta\lambda$ and
  $\Delta\beta$. \label{f:dlongdlat}}
\end{figure}

In Figure\,\ref{f:dlongdlat} we show the offsets between the position
in \keplere\ and the position derived from the {\em Hipparcos} data,
for longitudes and latitudes separately. We showed in
Sect.\,\ref{s:conversion} that errors in conversion of the modern data
to the positions in 1601 are negligible, so that the offsets describe
the errors in the position as given by Brahe (or Kepler).  If the
errors were fully random, we might expect their distributions to be
Gaussian, but this is not the case: gaussians that fit the peak of the
distribution ($|\Delta\lambda|,|\Delta\beta|<5\arcmin$, say) have a
width $\sigma\simeq2\arcmin$ and predict much smaller numbers at
$|\Delta\lambda|,|\Delta\beta|>5\arcmin$ than observed.  The excess in
the wings of the distributions with respect to a gaussian description
is presumably due to the correlations and buildup of errors when the
position of a star is determined by measurement with respect to
another star which already has a positional error.  However, for large
errors, $>10\arcmin$ say, the possibility of a copying error must be
considered, as illustrated by the differences between the different
versions of the catalogue (see Fig.\,\ref{f:variants}), and as proven
for many cases by Rawlins (1993).

\begin{figure}
% made with plotdis.f
\includegraphics[angle=270,width=\columnwidth]{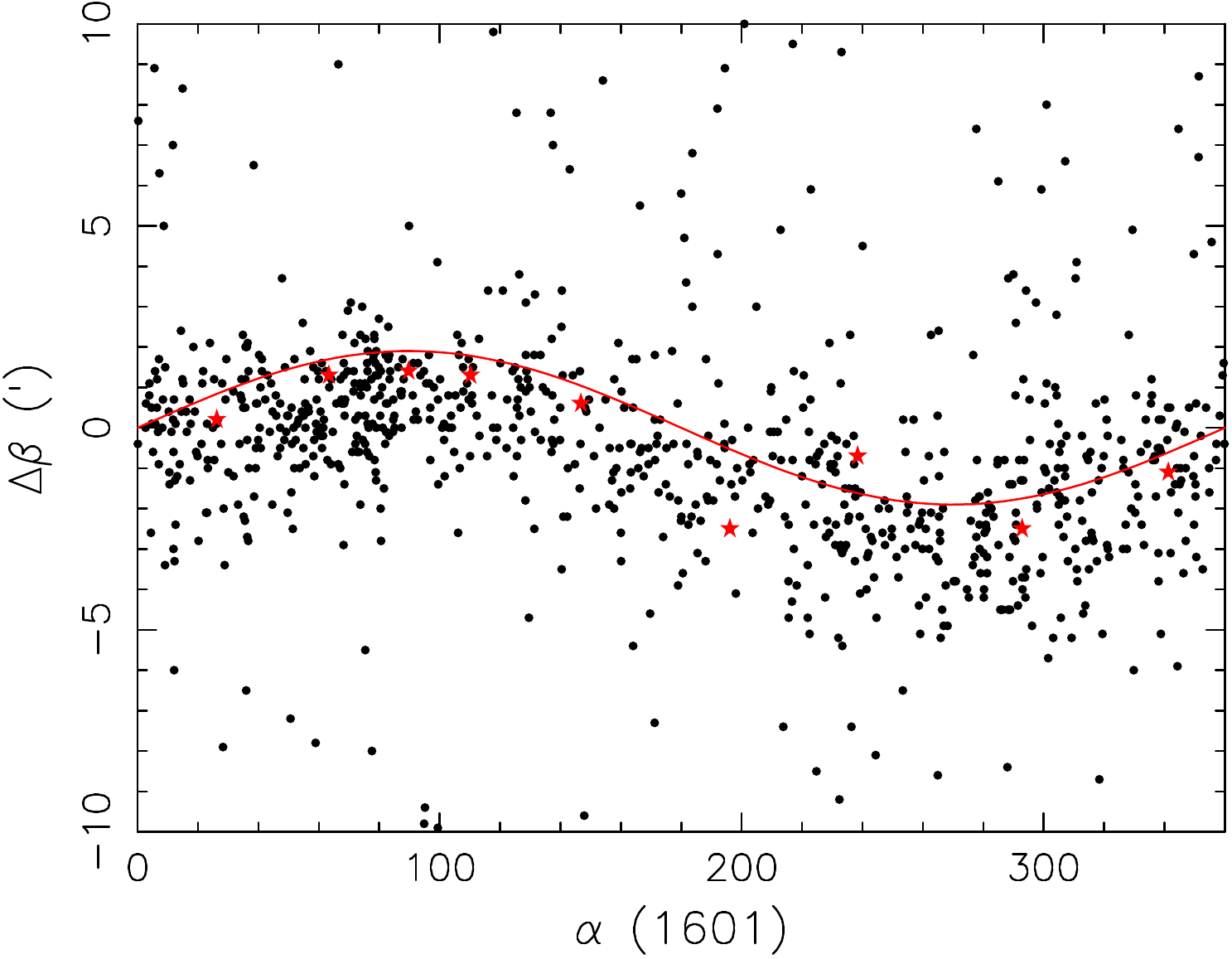}

\includegraphics[angle=270,width=\columnwidth]{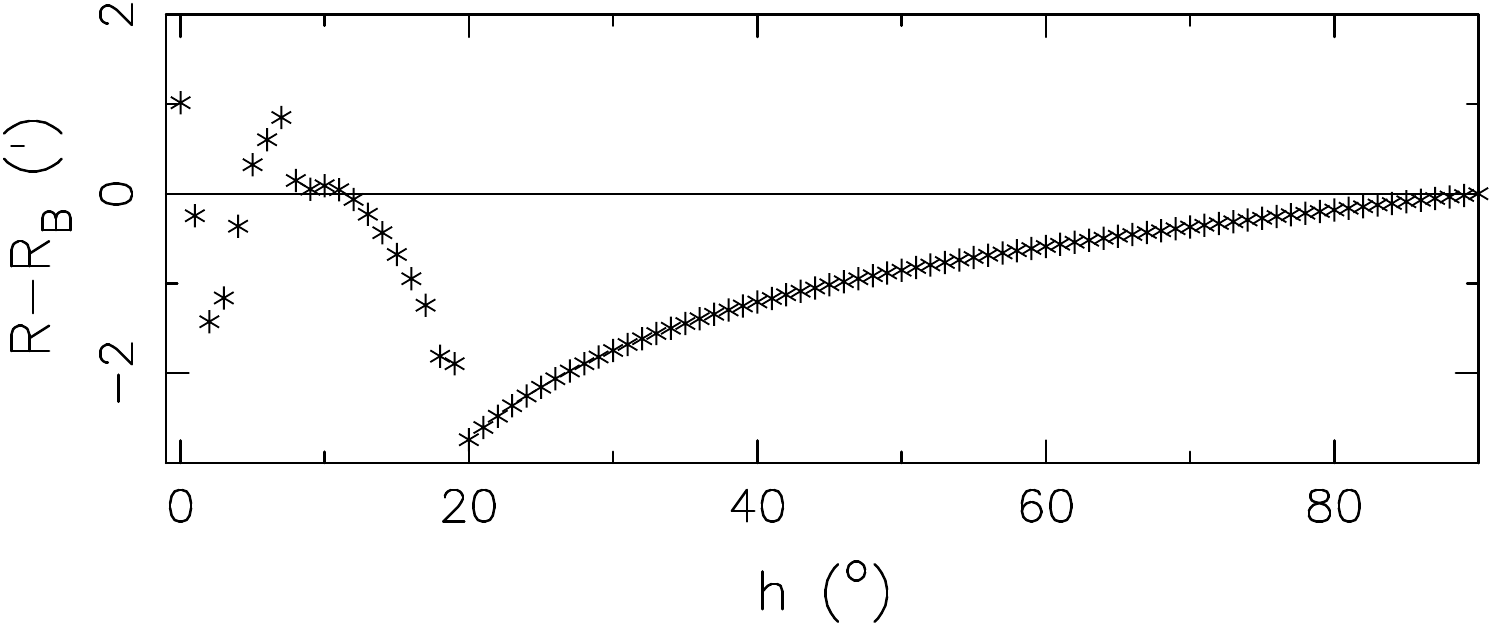}

\caption{Above: Latitude errors $\Delta\beta$ in \keplere\ as a
function of right ascension. The nine standard stars of Brahe are
indicated with a red star; the red line indicates 1\farcm9
$\sin\alpha$ (see Eq.\,\ref{e:alphadb}). Below: the difference between
modern values for diffraction and the values used by Brahe (zero at
$h>20$) for stars, as a function of altitude $h$ (Saemundsson, T.
1986; Meeus 1998; Brahe 1602, p.287). \label{f:alphadb}}
\end{figure}

The average offsets $\Delta\lambda$ and $\Delta\beta$ are not zero,
but for Gaussian fits to the central part of the distributions are
both around $-0\farcm5$. Together with the small but systematic
dependence of the average offsets on longitude, also shown in
Figure\,\ref{f:dlongdlat}, this suggests that a small part of the
errors may be due to small errors in the position of the zero point in
longitude and in the value of the obliquity that Brahe used.  Brahe
used an obliquity $\epsilon_B$=23\fdg525 (Brahe 1602, p.18 and p.208)
whereas the correct value for
1601 according to modern determinations (see
Sect.\,\ref{s:conversion}) was $\epsilon$=23\fdg491. For small
declinations $\delta$, the resulting error in latitude
$\Delta\beta\equiv\beta-\beta_B$ due to the error
$\Delta\epsilon\equiv\epsilon-\epsilon_B$ after converting equatorial
to ecliptic coordinates may be written
\begin{equation}
\cos\beta\,\Delta\beta \simeq -sin\alpha\cos\epsilon\,\Delta\epsilon=
1\farcm9\sin\alpha
\label{e:alphadb}\end{equation}
This dependence on right ascension is clearly seen in Brahe's
data, as illustrated in Fig.\,\ref{f:alphadb}.

Brahe also assumed that refraction at altitudes above 20$^\circ$ is
negligible, whereas a modern estimate would give 0\farcm7 at the
altitude of the equatorial pole for Hven (latitude almost 56$^\circ$;
the actual refraction depends somewhat on weather circumstances).
Declination measurements with respect to this pole would thus all be
offset by $-$0\farcm7, and this would lead to a systematic offset in
$\beta$. Even though the real situation would be more complicated,
involving differences in refraction errors between stars measured at
different altitudes (see Fig.\,\ref{f:alphadb}), we think that this
offset largely explains the overall offset in $\beta$ seen in
Figs.\,\ref{f:dlongdlat} and \ref{f:alphadb}, which averages
$-$0\farcm5, and is already present in the positions of Brahe's nine
standard stars (Fig.\,\ref{f:alphadb}; see also Dreyer 1890, p.387).

\begin{figure}
% made with plotdis.f
\includegraphics[angle=270,width=\columnwidth]{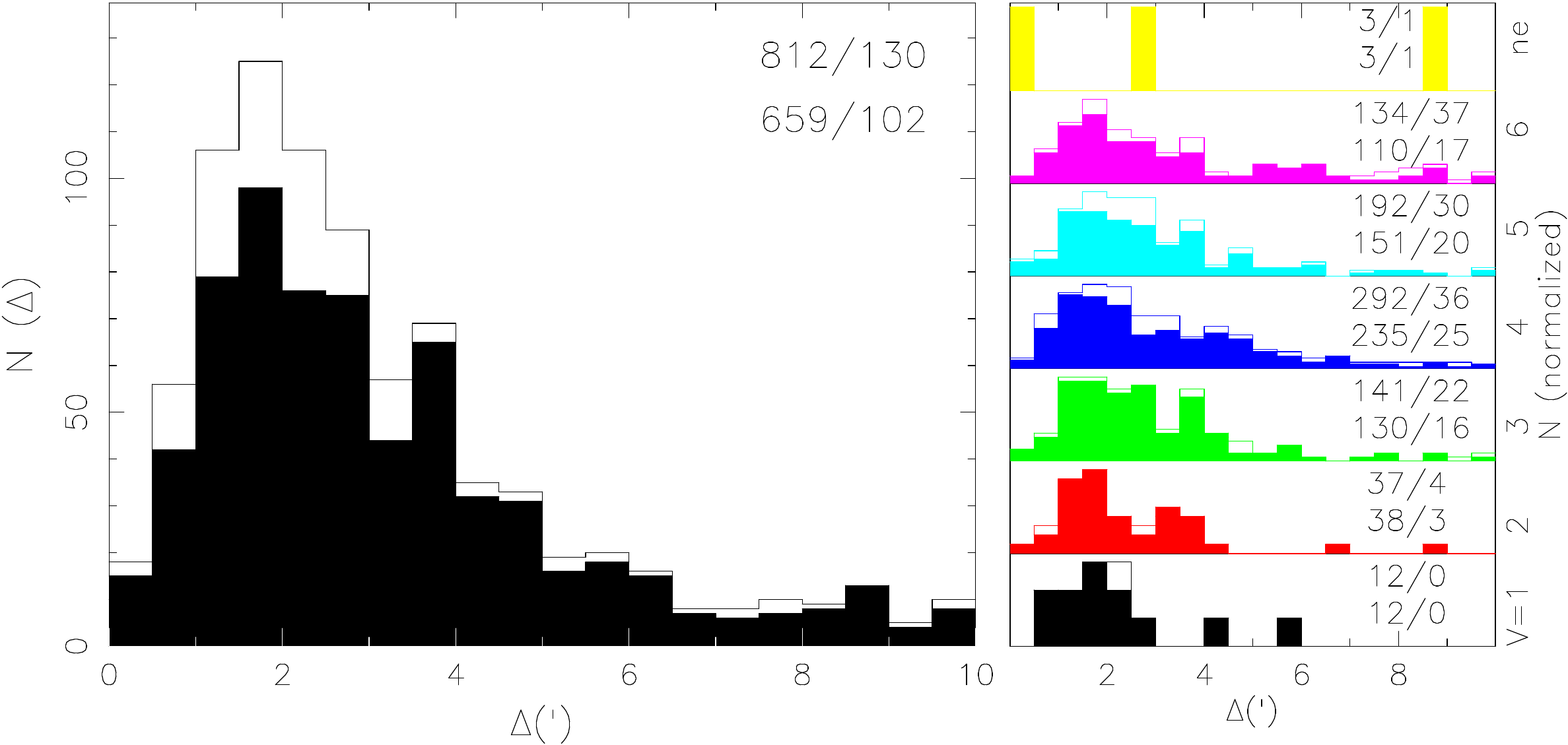}

\caption{Distribution of the position errors $\Delta$ in \progym\ (solid histograms)
and in \keplere\ (open histograms) for all securely identified stars
(left) and for the securely identified stars at each Brahe magnitude
separately (right). The numbers indicate the number of stars included
in the plot (i.e.\ with $\Delta<10\arcmin$) and those excluded 
($\Delta\ge10\arcmin$) \label{f:delta}}
\end{figure}

If $\Delta\lambda$ and $\Delta\beta$ were distributed as Gaussians
centered on zero, the distribution of the total position errors  
$\Delta\equiv\sqrt{\Delta\lambda^2+\Delta\beta^2}$ could be written
\begin{eqnarray}
N f(\Delta\lambda)f(\Delta\beta)d\Delta\lambda d\Delta\beta & = &
{N\over2\pi\sigma^2}e^{-0.5\Delta\lambda^2/\sigma^2}
e^{-0.5\Delta\beta^2/\sigma^2}d\Delta\lambda d\Delta\beta
\nonumber \\
\equiv N f(\Delta)d\Delta & = & 
N{\Delta\over\sigma}e^{-0.5(\Delta/\sigma)^2}d{\Delta\over\sigma} 
\label{e:delta}
\end{eqnarray}
The maximum of this function is at $\Delta=\sigma$.  In
Figure\,\ref{f:delta} we show the distribution of $\Delta$.  The
observed error distribution is similar to Eq.\,\ref{e:delta}, but has
more measurements at large $\Delta$ than expected from
Eq.\,\ref{e:delta} with $\sigma=2$ because the observed distributions
of $\Delta\lambda$ and $\Delta\beta$ are not centered on zero (a small
effect) and show excesses at higher offsets with respect to a Gaussian
(the dominant effect).

Using Kolmogorow-Smirnov tests to compute the probability that the
distributions of the errors $\Delta$ in the range $\Delta<10\arcmin$
for $V=n$ is the same as for $V=n+1$ we find no significant
differences. However, the distribution of $\Delta$ for $V\le2$ is
significantly different from that for $3\le V\le5$ (probability for
being identical is 0.003).  Comparing the differences between the
distributions of $\Delta$ over the full range of $\Delta$, i.e.\
including large offsets, we find a significant difference between
$V=5$ and $V=6$ (probability for being identical is 0.001).  We
conclude that the positions of stars with Brahe magnitudes 1 and 2 are
better than those of fainter stars; and that a larger fraction of the
positions of stars with magnitude 6 is wrong by more than 10\arcmin\
than for positions of fainter stars.

\begin{acknowledgements}
We thank Dr. Gertie Geertsema for help in photographing \kepler.
This research has made use of the SIMBAD database, operated at CDS, Strasbourg, France,
and was supported by the Netherlands Organisation for Scientific Research under grant
614.000.425.

\end{acknowledgements}

%\Online

\begin{appendix}
\section{Emendations \label{s:emendations}}

We have emended some obvious errors in Kepler (1627). Positional
errors are obvious if they place a star well outside the
constellation; in general we emend them only if the correct version
appears in the Brahe (1603) version. In the following,
we abbreviate \meuscat\ for an alternative
position labelled {\em Meus Catalogus} in Kepler (1627).
In this Section reference to Dreyer is short for Dreyer (1916) and
reference to Rawlins for Rawlins (1993).

\subsection{Applied emendations \label{s:emendapp}}

We list errors in Kepler (1627) that we emended.

\noindent K\,67,68 have $Z$ = \cancer\ and \leo\ for their longitudes in
\kepler, and \leo\ and \virgo, respectively, in \manuscript. Because
the description {\em Tertia} (the third) for K\,69 couples it to that
of K\,68 {\em Secunda earundem} (the second of the same [sc.\ stars])
we apply the correction.  Whether or not we apply the emendation, we
don't find close counterparts for these stars.

\noindent K\,145 has \libra\ in \kepler, but \scorpio\ in \manuscript.
Because K\,145 is described as following K\,144 ({\em hanc sequens}),
we follow \manuscript.

\noindent K\,238 has $Z$ = \aries\ in \kepler\ for the longitude, but
\taurus\ in \progym\ and \manuscript. Because the star is described
together with K\,237 and K\,239 as in the bench (scabellum) of
the throne on which Cassiopeia sits, it should be near these two stars.

\noindent K\.378 has 29$^\circ$ in the latitude in \kepler, too far from
the other stars of Sagitta. We emend to the value 39$^\circ$ of 
\progym\ and \manuscript.

\noindent K\,499,500 have southern latitude A in \kepler, but
northern latitude B in \progym\ and \manuscript.

\noindent K\,530 has southern latitude A in \kepler, both
\manuscript\ and \progym\ give B.

\noindent \kepler\ gives $Z$ of K\,539 as \taurus; both
\manuscript\ and \progym\ give \gemini.

\noindent The magnitudes of K\,535-545 are erroneously replaced in
\kepler\ with those of K\,536-546, and the magnitude of K\,546 is left
empty. This is due to an erroneous shift by one line of the magnitude
column, starting with a {\em meus cat} alternative position for K\,535.

\noindent K\,547. \manuscript\ has B for the latitude of K\,547,
whereas \progym\ and \kepler\ give A. B is clearly correct, as it fits
the description of the star {\em Quae est inter binas praec.\ in
  $\square$ colli} (which is between the two leading stars in the
square at the neck) and lies within 0\farcm3\ of HIP\,19171.

\noindent K\,685, 686 are A(ustralis) in \kepler\ but B(orealis) in
\manuscript. The southern positions give no
near counterparts, whereas the northern positions do. The descriptions
for K\,685 {\em Superior et Orientalior} (North and East, sc.\ of
K\,684) and K\,686 {\em Quae hanc sequitur} (Which follows this one)
imply positions north of K\,685, and thus B. 

\noindent K\,801. \manuscript, \progym\ and \kepler\ all have
B  for the latitude of this star, close to K\,802, and with no
near counterpart. A latitude A gives an excellent match
both to the otherwise unidentified o\,Psc, and to the description
{\em In lineo boreo a connexu praecedens} (in the northern line leading
the connection). 

\noindent K\,934. \manuscript\ and \progym\ have $G$=11 in the
longitude, \kepler\ has $G$=19.  The descripton {\em Quae in fronte ad
  dextram aurem} (Which is in the forehead at the right ear) is not
compatible with the position in \kepler.

\noindent K\,972, K\,973. \manuscript\ and \progym\ have \leo\ for
longitude, \kepler\ has \virgo.

\subsection{Not-applied emendations}

We list errors in Kepler (1627) which we decided not to emend.

\noindent K\,59-61 all three have magnitude 5 in \manuscript, and 
3 in \kepler. Their counterparts indicate that the \manuscript\ magnitudes
are correct.

\noindent K\,534  has the same position in \manuscript, \progym, and 
\meuscat\ which differs from that in \kepler. The \manuscript\
position give a better match to Electra (see Fig.\,\ref{f:tauruspl}).

\noindent K\,581. \manuscript\ has B for the latitude of K\,581,
whereas \progym\ and \kepler\ give A. The manuscript position gives a
better match with HIP\,42911 (= $\delta$\,Cnc) at 0\farcm9.

\subsection{Emendations to Dreyer \label{e:emendreyer}}

In a number of cases the Flamsteed number is accompanied by an
(extended) Bayer identification that is different from the one given
in the 5th edition of the Bright Star Catalogue (Hoffleit \&\ Warren
1991). For example 35\,Lib (K\, 679= HIP\,76126) is listed as
$\zeta^1$ by Dreyer and as $\zeta^4$\,Lib in the Bright Star
Catalogue. Similarly with $\pi$\,Ori (K\,860-865). Since we use only
Flamsteed numbers to identify the counterpart we do not emend such
differences.

\noindent K\,580 Dreyer gives Flamsteed no.\,41 for $\gamma$\,Cnc;
correct is 43.

\noindent K\,646.
Dreyer gives Flamsteed no.\,70 for $\zeta$\,Vir; correct is 79.

\noindent K\,990.
Dreyer gives Flamsteed no.\,23 for $\theta$\,Crt. Correct is 21.

\noindent K\,263 and K\,264. Dreyer gives no identification for K\,263 and
P.IX.37 = HIP\,47193 for K\,264. From the position HIP\,47193 is a better match
for K\,263, which when accepted leaves K\,264 unidentified by Dreyer.
\end{appendix}

%-----------------------------------
\begin{appendix}
\section{Notes on individual identifications \label{s:notes}}

In this Section reference to Dreyer is short for Dreyer (1916) and
reference to Rawlins for Rawlins (1993).

\begin{figure}
\centerline{\includegraphics[angle=270,width=0.9\columnwidth]{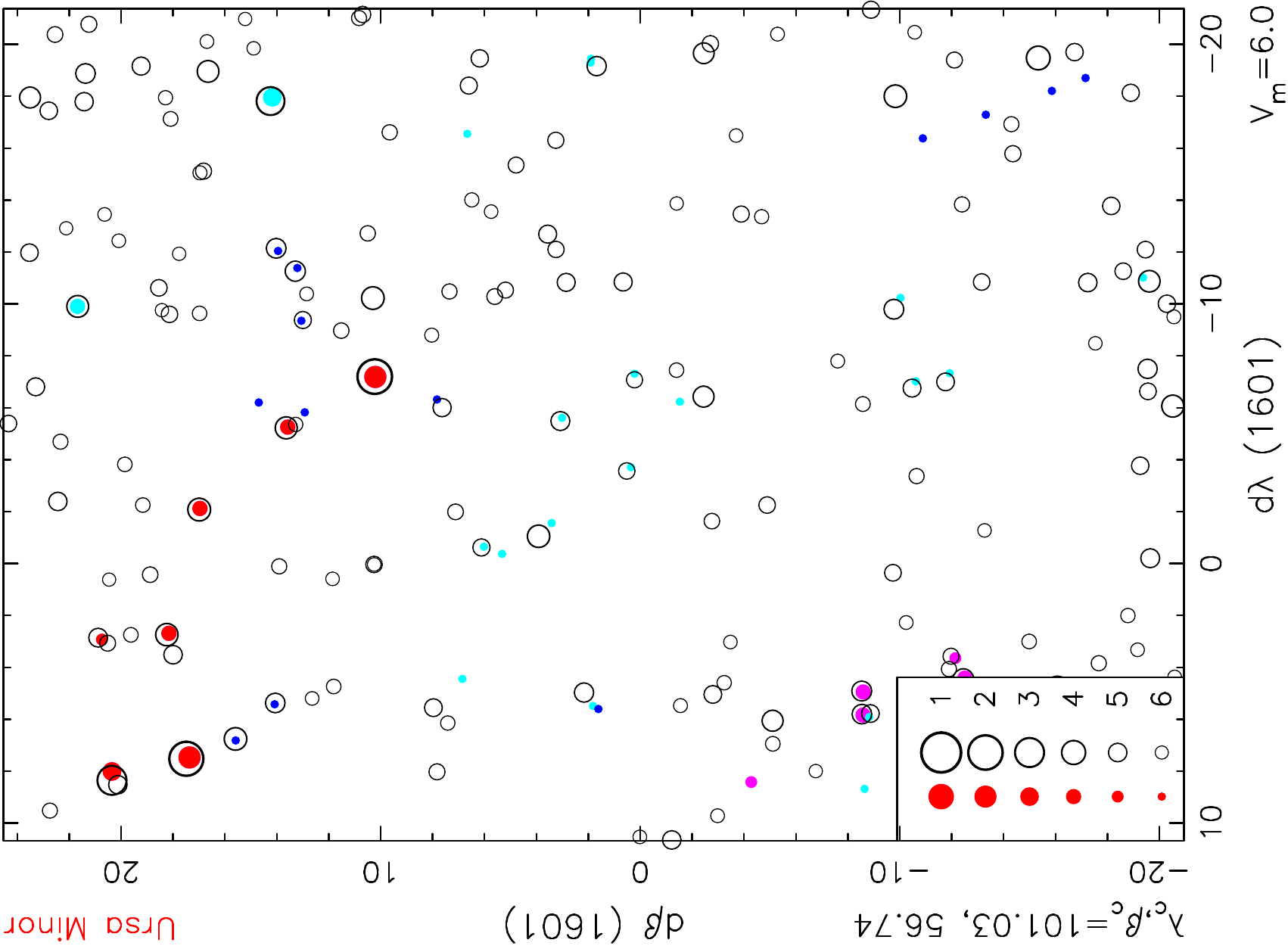}}
\caption{Ursa Minor. The stars additional to the 7 stars given in \progym\
include four stars in Camelopardalis (below right) with no close counterparts,
and one star (near +5.5,1.5) also listed in Ursa Maior. We agree with Dreyer
in not finding a suitable identification for K\,15 (near $-$6,14.5).
 \label{f:ursaminor}}
\end{figure}

\noindent K\,10-13 lie in a straight line in Fig.\,\ref{f:ursaminor}
(lower right corner), without close counterparts. All four
identifications we choose lie to the northwest. K\,12 lies rather closer to
HIP\,24914, 63\farcm4\ to the southeast, and this star is an alternative
identification. The counterpart suggested for K\,12 by Dreyer, HIP\,24348 is
too faint and still not very close ($V$=6.5, $\Delta$=45\farcm3).

\noindent K\,14, near +5.5,+1.5 in Fig.\,\ref{f:ursaminor}, is very
close (14\arcmin) to K\,64 in Ursa Maior, and refers to the same star.

\noindent K\,15, near $-$6,+15 in Fig.\,\ref{f:ursaminor}, is
tentatively identified by Dreyer with Groombridge 2708, possibly
identical to HIP\,90182 at 11\farcm7 from the Groombridge position. We
consider HIP\,90182 too faint, at $V=6.2$.

\noindent K\,18, near $-$12,+14 in Fig.\,\ref{f:ursaminor}, is
tentatively identified by Dreyer with Groombridge 3887; however,
HIP\,112519, our counterpart for K\,18, corresponds to Groombridge
3928.

\begin{figure}
\includegraphics[angle=270,width=\columnwidth]{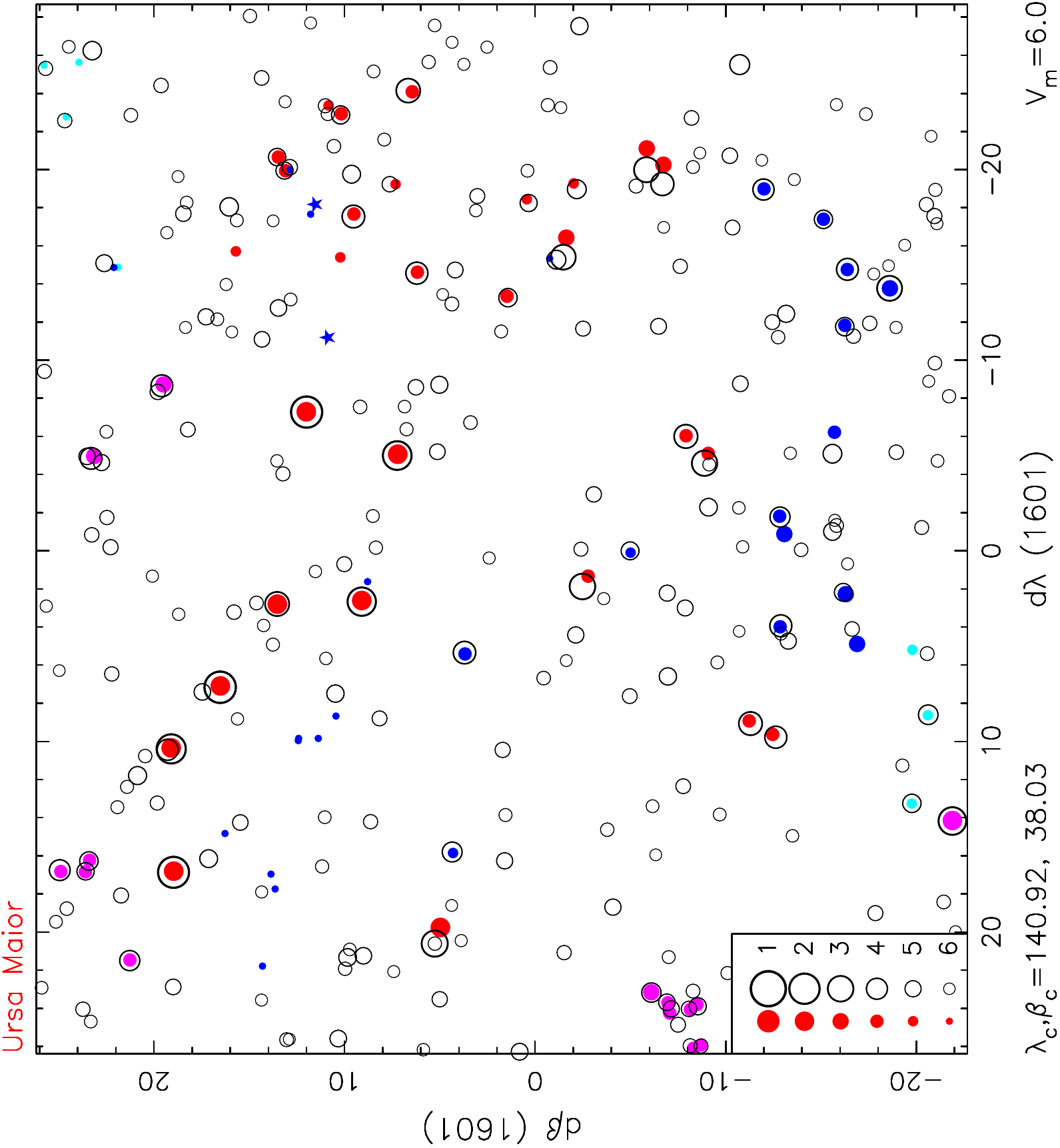}
\caption{Ursa Maior. K\,67 and K\,88 at the positions given in \kepler\ are
indicated $\star$. See also Fig.\,\ref{f:umadetail}.
 \label{f:ursamaior}}
\end{figure}

\noindent K\,26, near $-$16,+16 in Fig.\,\ref{f:ursamaior},
has no near counterpart. It lies between two faint
stars HIP\,50685 ($V$=5.9, $\Delta$=111\arcmin\ East) and HIP\,47594 ($V$=5.7,
$\Delta$=102\arcmin\ West), but we prefer the brighter star slightly further West.
Dreyer agrees.

\noindent K\,29, near $-$15,+10 in Fig.\,\ref{f:ursamaior},
is not identified by us. We consider HIP\,47911, the
counterpart given by Dreyer, too faint ($V$=6.6) and far ($\Delta$=46\farcm7)
to be viable.

\begin{figure}
\includegraphics[angle=270,width=\columnwidth]{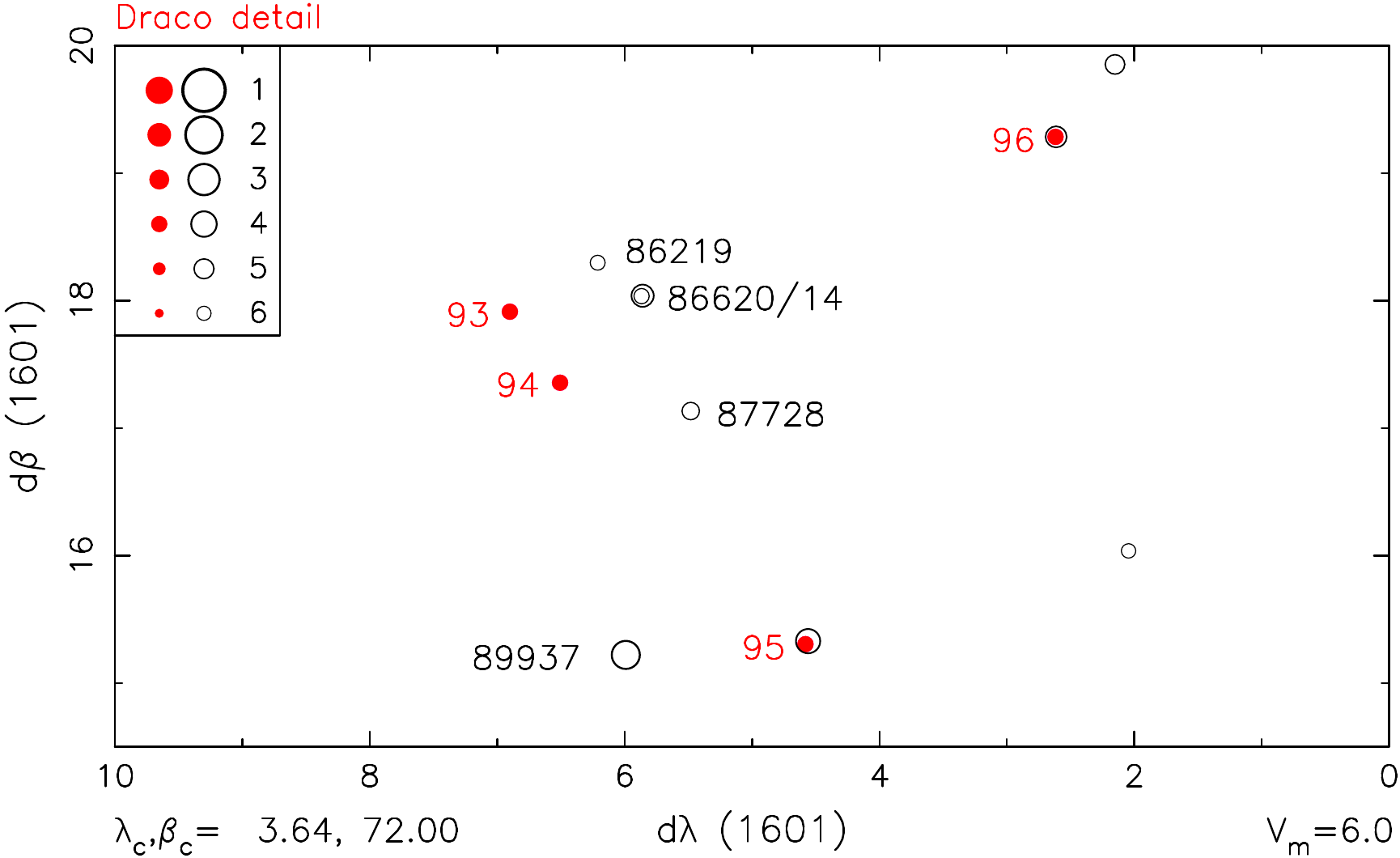}
\caption{Draco detail.
\noindent K\,93,94. We identify K\,93 with the close pair
HIP\,86614/86620 ($=\psi^1$\,Dra) and K\,94 with HIP\,87728.  For
K\,94 Dreyer (1916) and Rawlins (1993) prefer HIP\,89937
($=\chi$\,Dra) as counterpart.  . \label{f:dracodetail}}
\end{figure}

\noindent K\,93,94. We identify K\,93 with the close pair
HIP\,86614/86620 ($=\psi^1$\,Dra) and K\,94 with HIP\,87728.
For K\,94 Dreyer prefers HIP\,89937 ($=\chi$\,Dra) as counterpart.
See Fig.\,\ref{f:dracodetail}

\begin{figure}
\includegraphics[angle=270,width=\columnwidth]{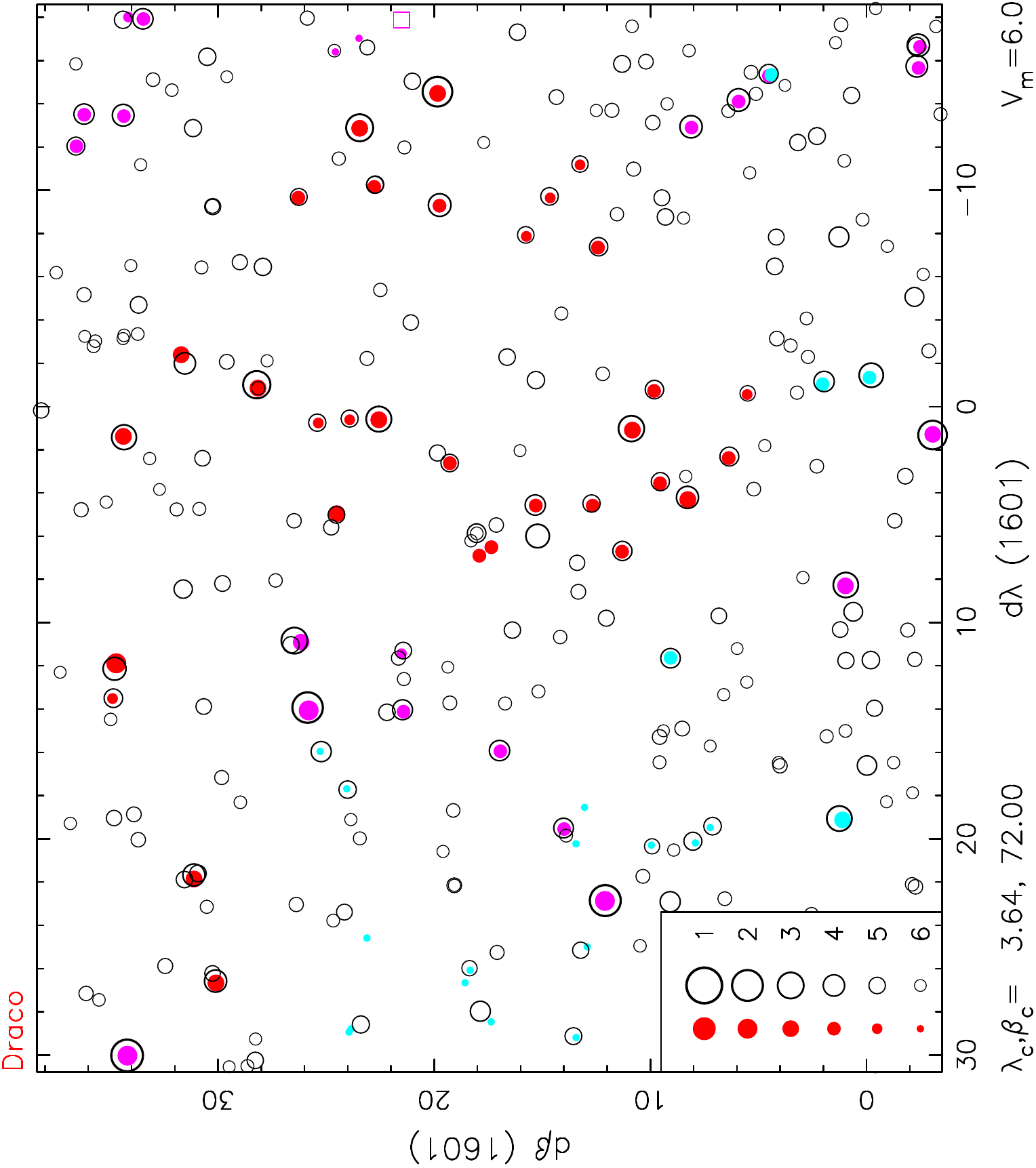}
\caption{Draco. \label{f:draco}}
\end{figure}

\noindent K\,101, near $-$1,+28 in Fig.\,\ref{f:draco},
lies very close to HIP\,80309 ($V$=5.7, $\Delta$=0\farcm7),
but its much brighter neighbour is our preferred counterpart.

\begin{figure}
\includegraphics[angle=270,width=\columnwidth]{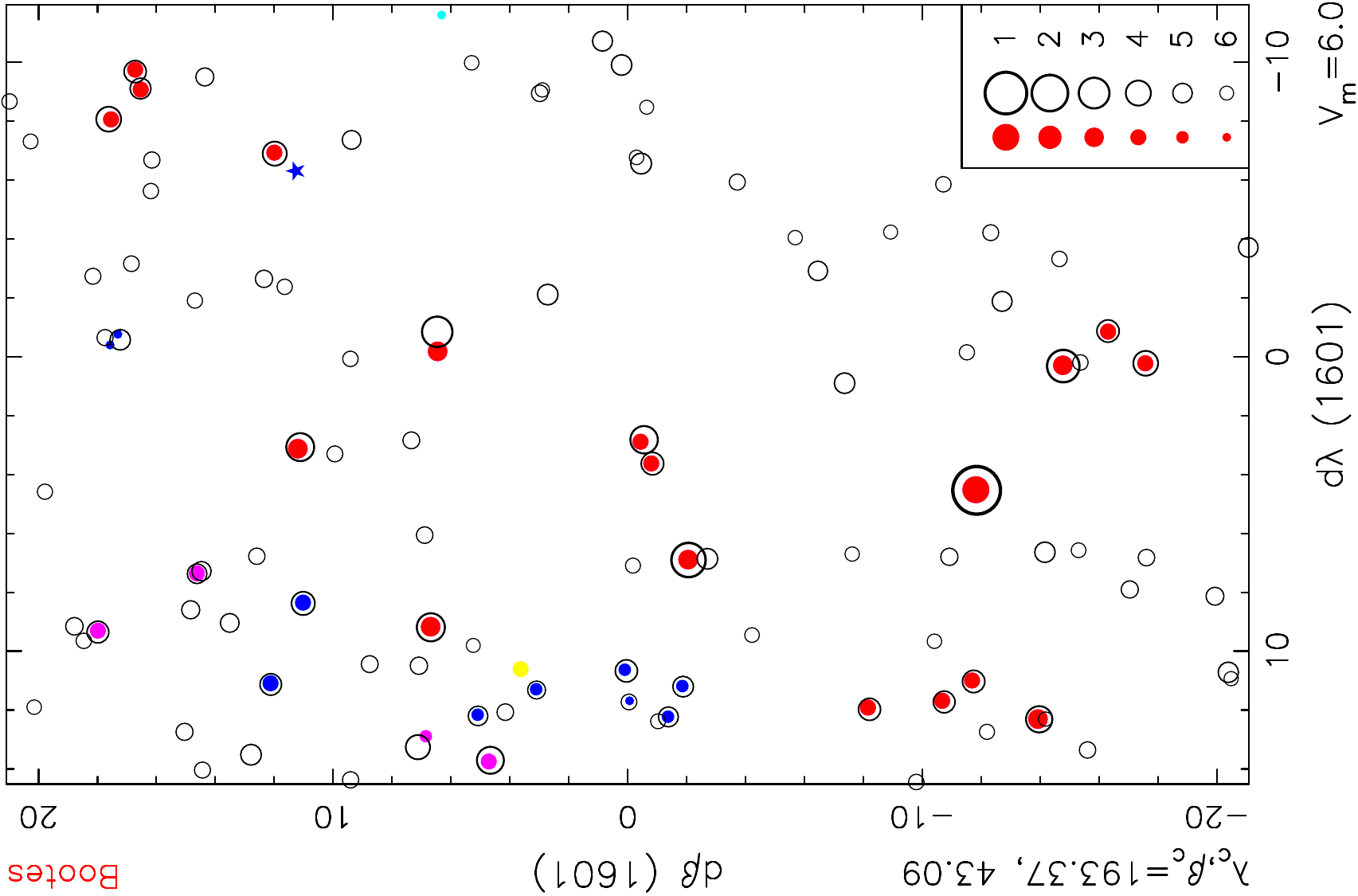}
\caption{Bootes. The position of K\,145 in \kepler\ is indicated $\star$.
\label{f:bootes}}
\end{figure}

\noindent K\,146 and K\,147, near $-$0.5,+17.5 in Fig.\,\ref{f:bootes}. 
Our identification for K\,146 is that of K\,147
in Dreyer, and vice versa, because we identify with the nearest object,
whereas Dreyer follows order in longitude. 

\begin{figure}
\includegraphics[angle=270,width=\columnwidth]{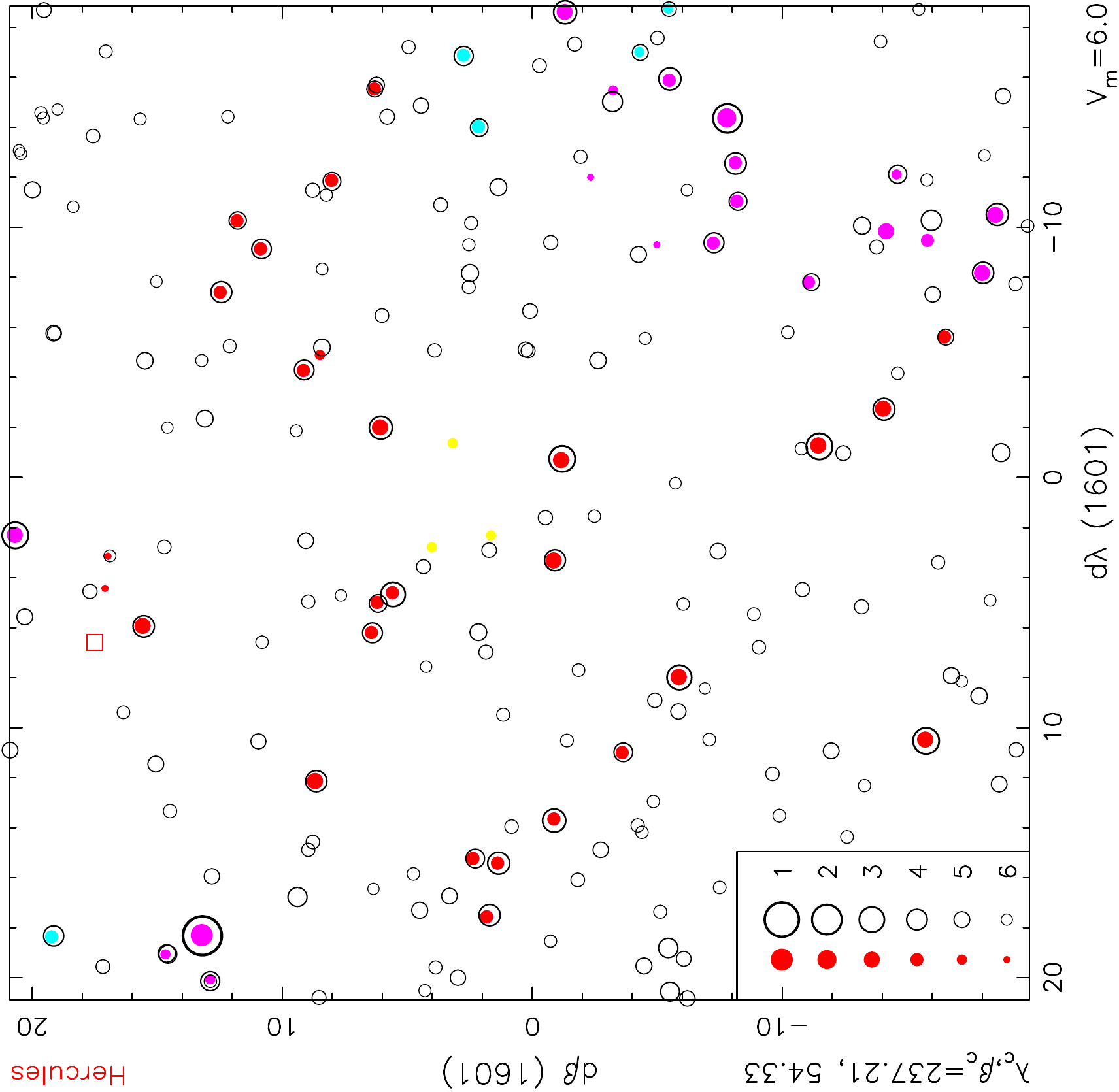}
\caption{Hercules. The star from {\em Secunda Classis} near $-$1.4,3.2 may be
identified with M\,13
. \label{f:hercules}}
\end{figure}

\noindent K\,175, near +7,+18 in Fig.\,\ref{f:hercules}, is listed in
the Catalogue as nebulous, and is identified by Dreyer with
HIP\,87280, a Be star. Although variable, HIP\,87280 is too faint
($V$=6.8) and too far ($\Delta$=50\arcmin) to be a viable
counterpart. Rawlins identifies K\,175 with the combined light of HIP\,87045
($V$=6.47) and HIP\,87119 ($V$=6.83) two stars separated by 8\farcm8.
We leave K\,175 unidentified.

\noindent K\,183, near $-$16,+6 in Fig.\,\ref{f:hercules}, is
identified by Dreyer with HIP\,86534, but this star is at a distance
$\Delta$=77\arcmin, whereas our identification, which corresponds to
53\,Boo, lies at 1\farcm2 from K\,183. Rawlins identifies this entry
with the combined light of 52\,Boo (= HIP\,75973, $V$=5.0) and
53\,Boo, separated by 10\farcm5.

\begin{figure}
\centerline{\includegraphics[angle=270,width=0.9\columnwidth]{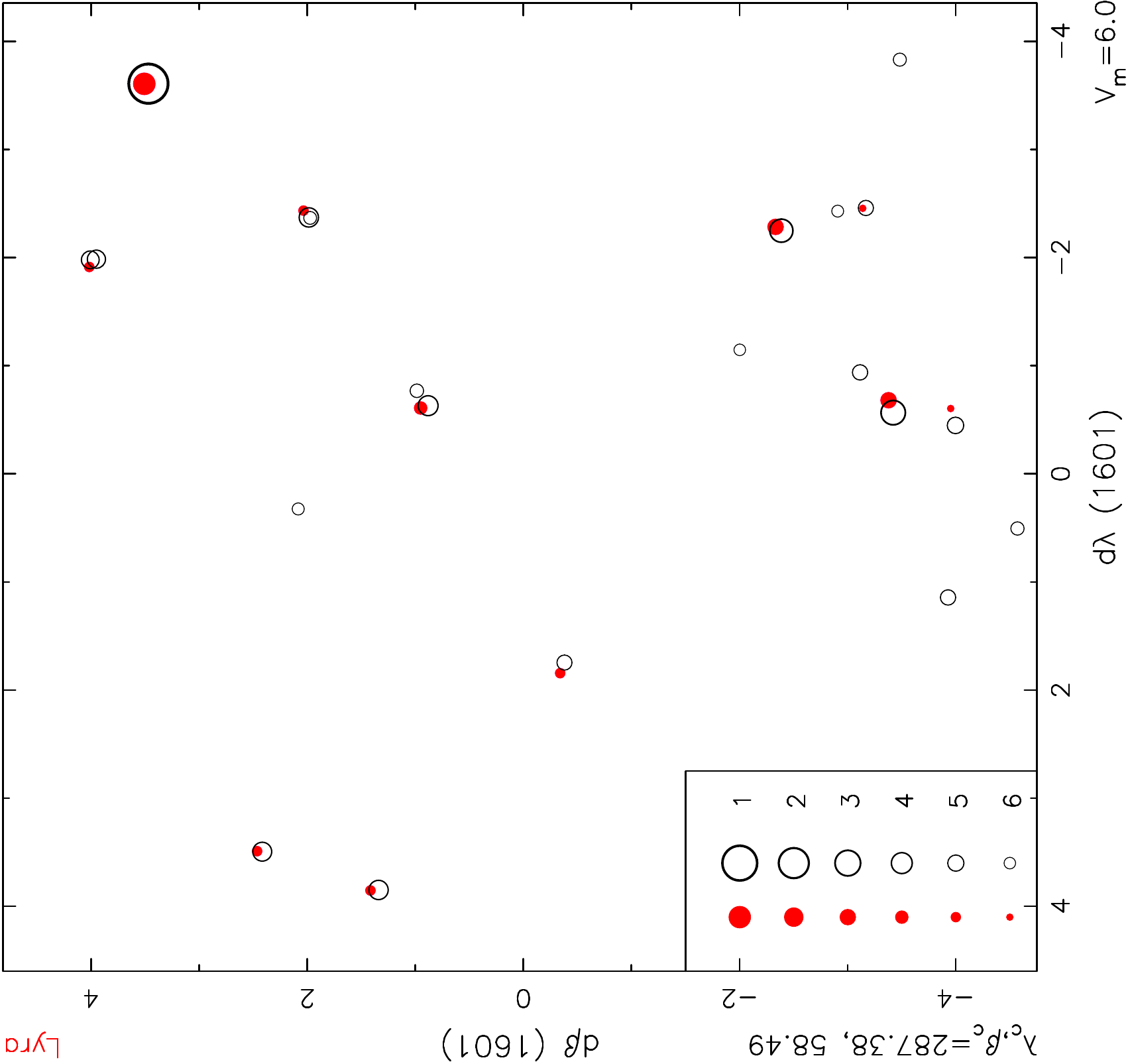}}
\caption{Lyra
. \label{f:lyra}}
\end{figure}

\noindent K\,191, near $-$2.5,$-$3 in Fig.\,\ref{f:lyra}, is
identified by Dreyer with HIP\,92398 (=$\nu^1$\,Lyr), which is both
fainter and further ($V$=5.9, $\Delta$=14\arcmin) as our
identification HIP\,92405 (=$\nu^2$\,Lyr).

\begin{figure}
\includegraphics[angle=270,width=\columnwidth]{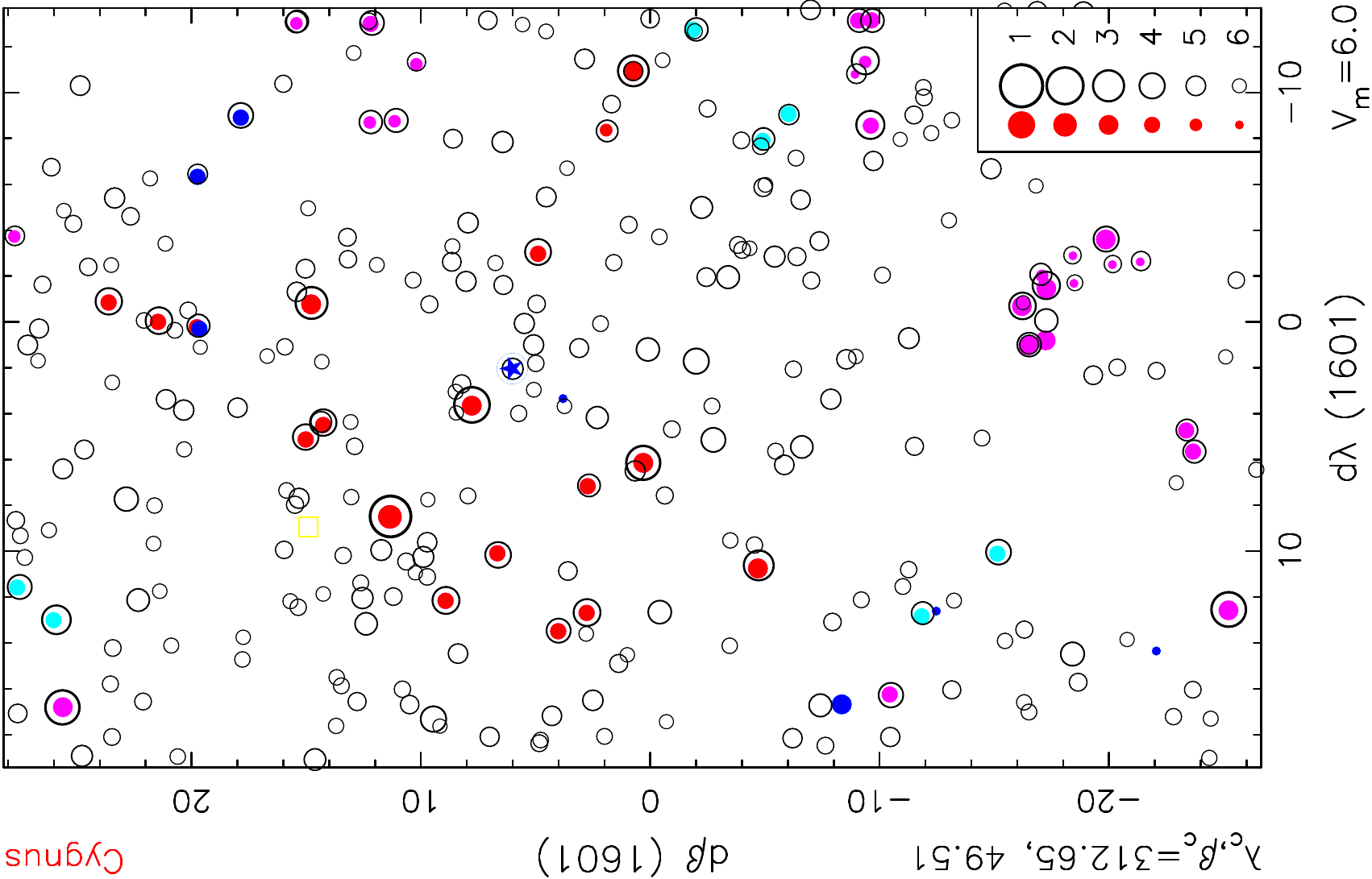}
\caption{Cygnus. Nova\,1600 aka P\,Cygni = HIP\,100044 is indicated with
$\star$. \label{f:cygnus}}
\end{figure}

\noindent K\,209, near +4.5,+14 in Fig.\,\ref{f:cygnus}, is identified
by Dreyer with HIP\,99639, both fainter and further ($V$=4.8,
$\Delta$=9\farcm5) than our suggested counterpart HIP\,99675.  Rawlins
identifies this entry with the combined light of these two stars.

\begin{figure}
\includegraphics[angle=270,width=\columnwidth]{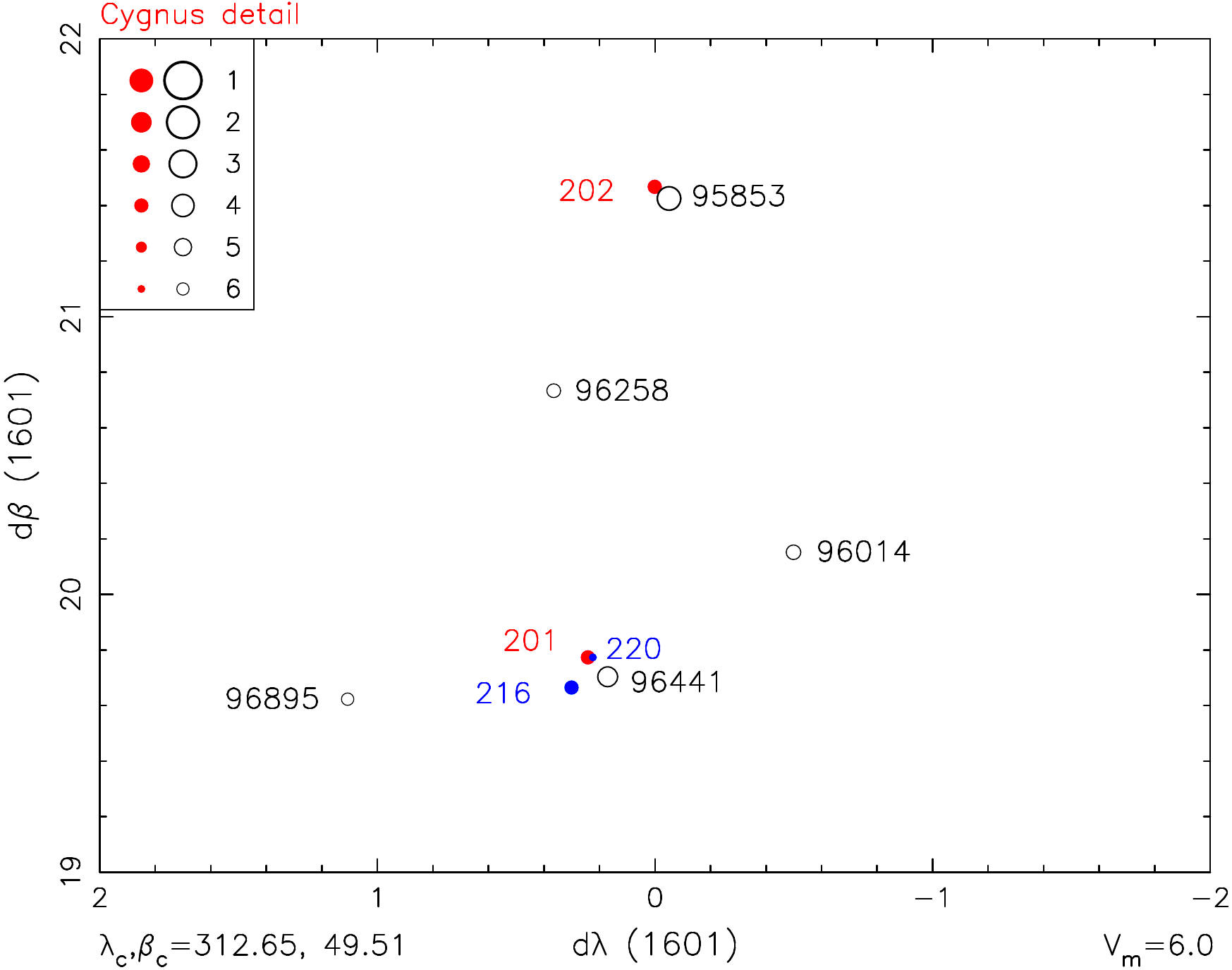}
\caption{Detail of Cygnus illustrating that K\,201, K\,216 and K\,220
are repeated entries for the same star. \label{f:cygdetail}}
\end{figure}

\noindent K\,216 is at 7\farcm8 from K\,201; K\,220 is at
1\farcm0 from K\,201 (Fig.\,\ref{f:cygdetail}).  We think that
these three entries refer to the same star, HIP\,96441.  Dreyer
identifies K\,201 as we do, but identifies K\,216 and K\,220 both with
HIP\,96895, but this star is 54\arcmin\ from K\,220.

\noindent K\,218, near +13,$-$12 in Fig.\,\ref{f:cygnus}. 
HIP\,106140, the counterpart suggested by Dreyer, is brighter and
closer than our suggested counterpart, but it is already taken by K\,440.
 
\noindent K\,220: see K\,216.

\begin{figure}
\includegraphics[angle=270,width=\columnwidth]{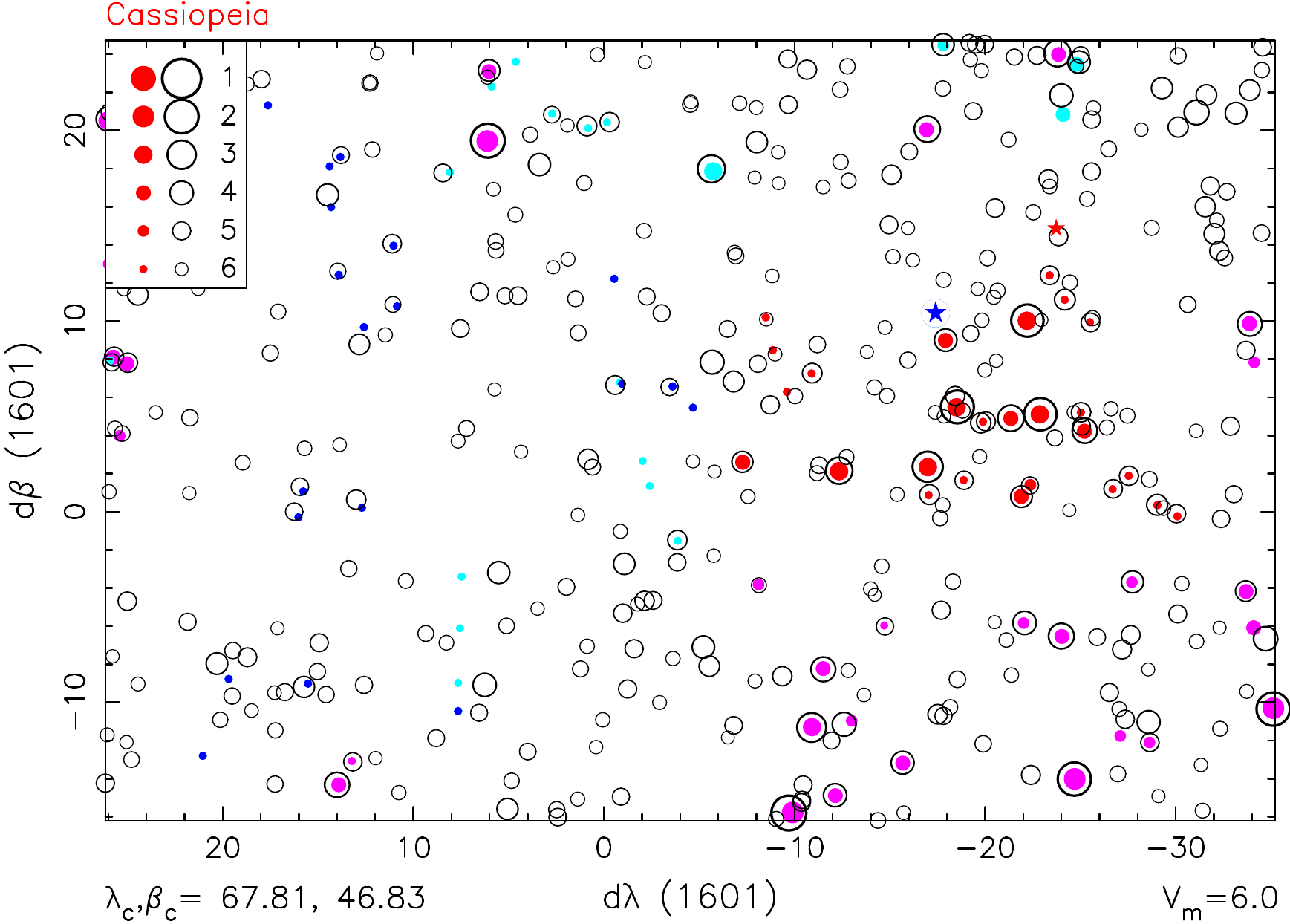}
\caption{Cassiopeia. The blue $\star$ indicates Nova 1572. The red
star indicates the position of K\,238 in Kepler.
. \label{f:cassiopeia}}
\end{figure}

\begin{figure}
\includegraphics[angle=270,width=\columnwidth]{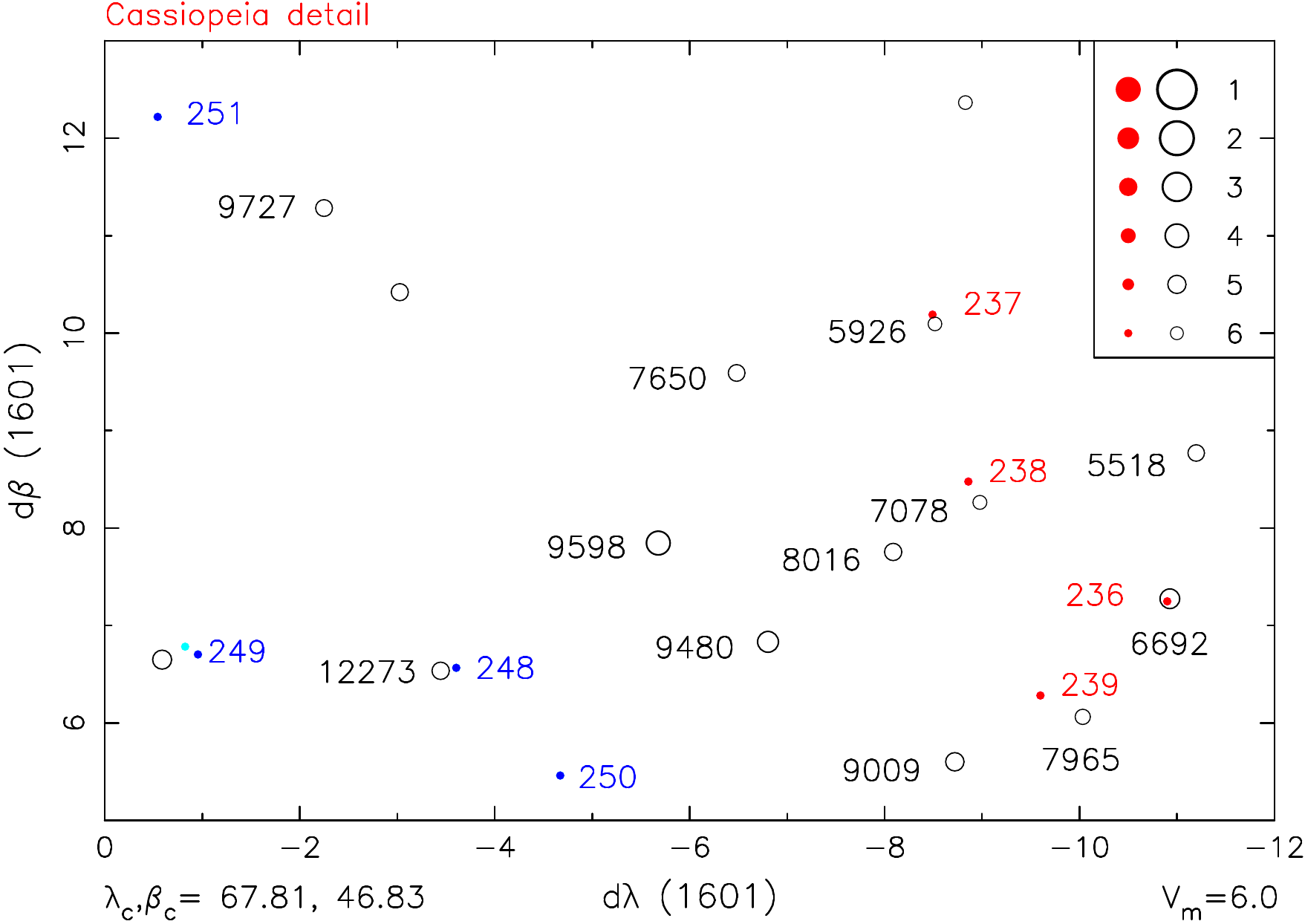}
\caption{Cassiopeia detail, illustrating the identifications by Dreyer 
(1916) and Rawlins (1993) of K\,237-239 with HIP\,9598, HIP\,9480 and HIP\,9009.
. \label{f:casdetail}}
\end{figure}

\noindent K\,237-239. Notwithstanding implied position errors of
224\arcmin, 161\arcmin, 68\arcmin, it is tempting to agree with
Hevelius, Dreyer and Rawlins and identify these three stars with
HIP\,9598, HIP\,9480 and HIP\,9009, respectively, since these
counterparts are not matched by other entries in Brahe's catalogue,
even though rather brighter than the counterparts we choose because of
their positional matches, HIP\,5926, HIP\,7078 and HIP\,7965.  See
Figs.\,\ref{f:cassiopeia},\ref{f:casdetail}.  Rawlins indeed corrects
the positions. Note that the Brahe catalogue give magnitudes 6 to
K\,237-239.

\noindent K\,249 lies very close to K\,300 (Fig.\,\ref{f:casdetail}),
but we do not consider this to be a double entry: see K\,297-300.

\noindent K\,254, near +20,$-$9 in Fig.\,\ref{f:cassiopeia}, 
lies between HIP\,33449 and HIP\,32489. We choose the
HIP\,33449 as a counterpart as it is brighter; Dreyer chooses HIP\,32489,
which is fainter but closer ($V$=5.3, $\Delta$=55\farcm3)

\noindent K\,264,265, near +14,+18.5 in Fig.\,\ref{f:cassiopeia}, form
a pair close to the pair HIP\,51384 and HIP\,51502. If we follow
Dreyer and identify K\,265 with HIP\,51502 (at 6\farcm0), K\,264 is
without counterpart. We have chosen to identify the pair, accepting
the larger offsets. Neither solution is very satisfactory.

\noindent K\,267 is SN\,1572. The positional error is rather large,
about 15\arcmin, mainly in longitude.

\noindent K\,294. Rawlins makes a large correction to its position and
identifies it with HIP\,19949 ($V$=5.2).

\begin{figure}
\includegraphics[angle=270,width=\columnwidth]{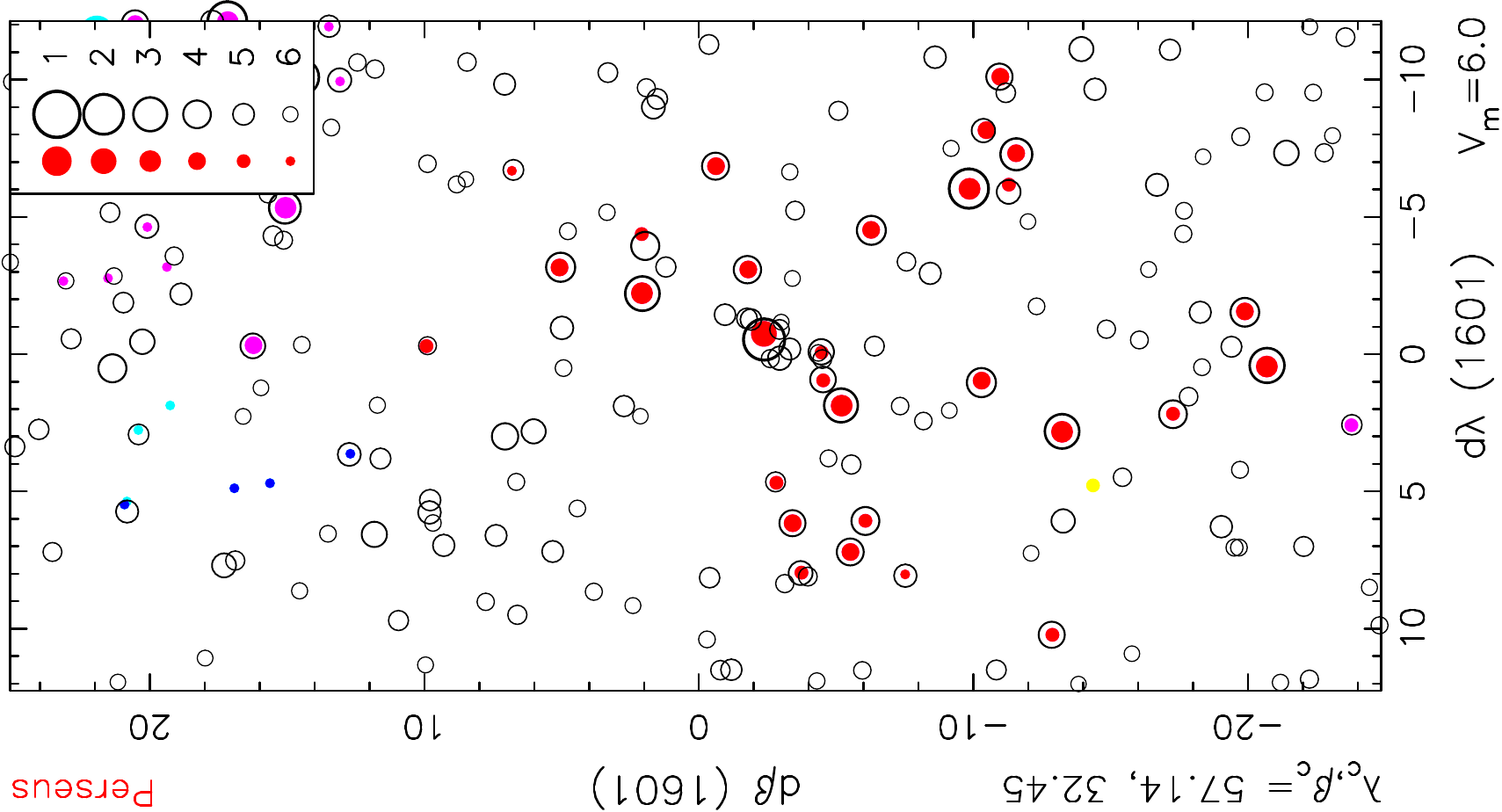}
\caption{Perseus. The additional stars (in blue) appear shifted from a
better match about 10$^\circ$ south. \label{f:perseus}}
\end{figure}

\noindent K\,297-300 are the four additional stars (indicated blue)
top left in Fig.\,\ref{f:perseus}. K\,297 and K\,300 are close to
HIP\,15520 ($\Delta$=2\farcm8) and HIP\,14862 ($\Delta$=16\farcm3),
respectively, whereas K\,298,299 have no nearby counterparts. This
is the choice made by Dreyer.  Looking at the Figure we have the
distinct impression that the correct position of this group of four
stars is about 10$^\circ$ (mainly towards the south) from the
catalogued positions. We give our identifications accordingly. Rawlins
identifies K\,297 with HIP\,15520 and corrects the positions of
K\,298-300 to identify K298 with HIP\,16281, K\,299 with HIP\,16292
and K\,300 with HIP\,16228.

\begin{figure}
\centerline{\includegraphics[angle=270,width=0.8\columnwidth]{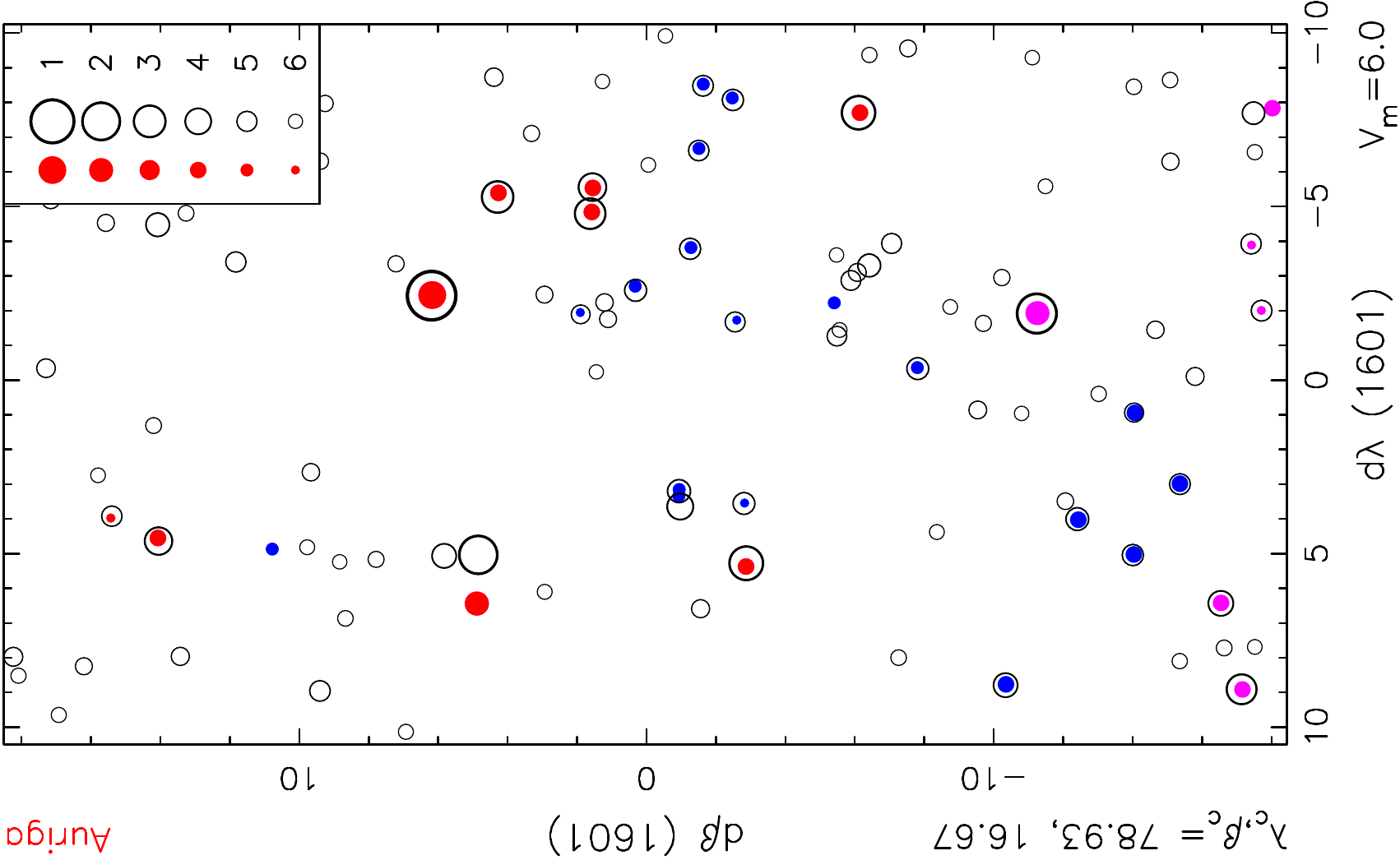}}
\caption{Auriga. \label{f:auriga}}
\end{figure}

\noindent K\,317. The \kepler\ position is closest to the counterpart
of K\,316, so that we choose HIP\,27673 as counterpart (near +3,$-$1
in Fig.\,\ref{f:auriga}, the two entries of the Brahe Catalogue
-- in blue -- are merged).  For the position of \meuscat\ HIP\,27673 is
in fact the star closest to K\,317.

\noindent K\,319, near $-$2,$-$5 in Fig.\,\ref{f:auriga},
lies between HIP\,25471 ($V$=5.9, $\Delta$=48\farcm1) and
our preferred counterpart HIP\,24879, brighter and at the same distance
($V$=5.1, $\Delta$=48\farcm5). Dreyer takes HIP\,25541 as counterpart ($V$=5.1,
$\Delta$=58\arcmin).

\begin{figure}
\includegraphics[angle=270,width=\columnwidth]{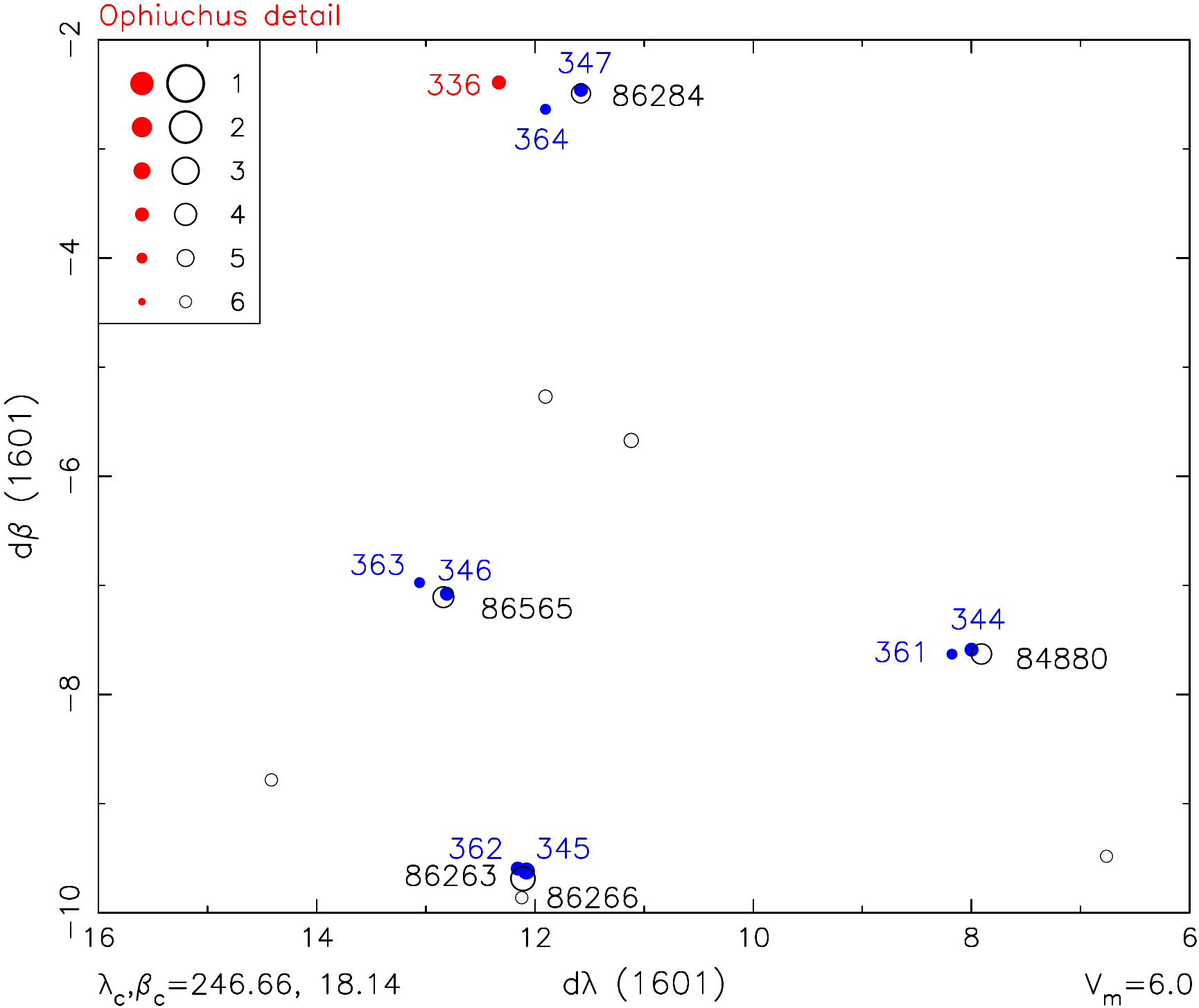}
\caption{Ophiuchus detail. \label{f:ophdetailb}}
\end{figure}

\noindent K\,336, K\,347 and K\,364 have virtually the same position, and
all correspond to HIP\,86284 (= $\mu$\,Oph). See Fig.\,\ref{f:ophdetailb}. 

\noindent K\,341: see K\,353.

\noindent K\,344-347. We agree wih Dreyer in considering K\,361 to
K\,364 as repeat entries for K\,344 to K\,347. Alternatively, one may
identify the pair K\,345/K\,362 with the pair
HIP\,86263/HIP\,86266. See Fig.\,\ref{f:ophdetailb}.

\noindent K\,347: see K\,336.

\noindent K\,353-364. Kepler annotates these entries with: {\em Desunt in
meo seqq.\ ad finem. Vide Classem secund.} (The following until the end
are absent in mine, see Second Class). Kepler refers to the
stars in the catalogue of Ptolemaios that Brahe omitted in his
catalogue as the Second Class, and lists these separately, on pages
following his edition of the Brahe Catalogue.

\begin{figure}
\includegraphics[angle=270,width=\columnwidth]{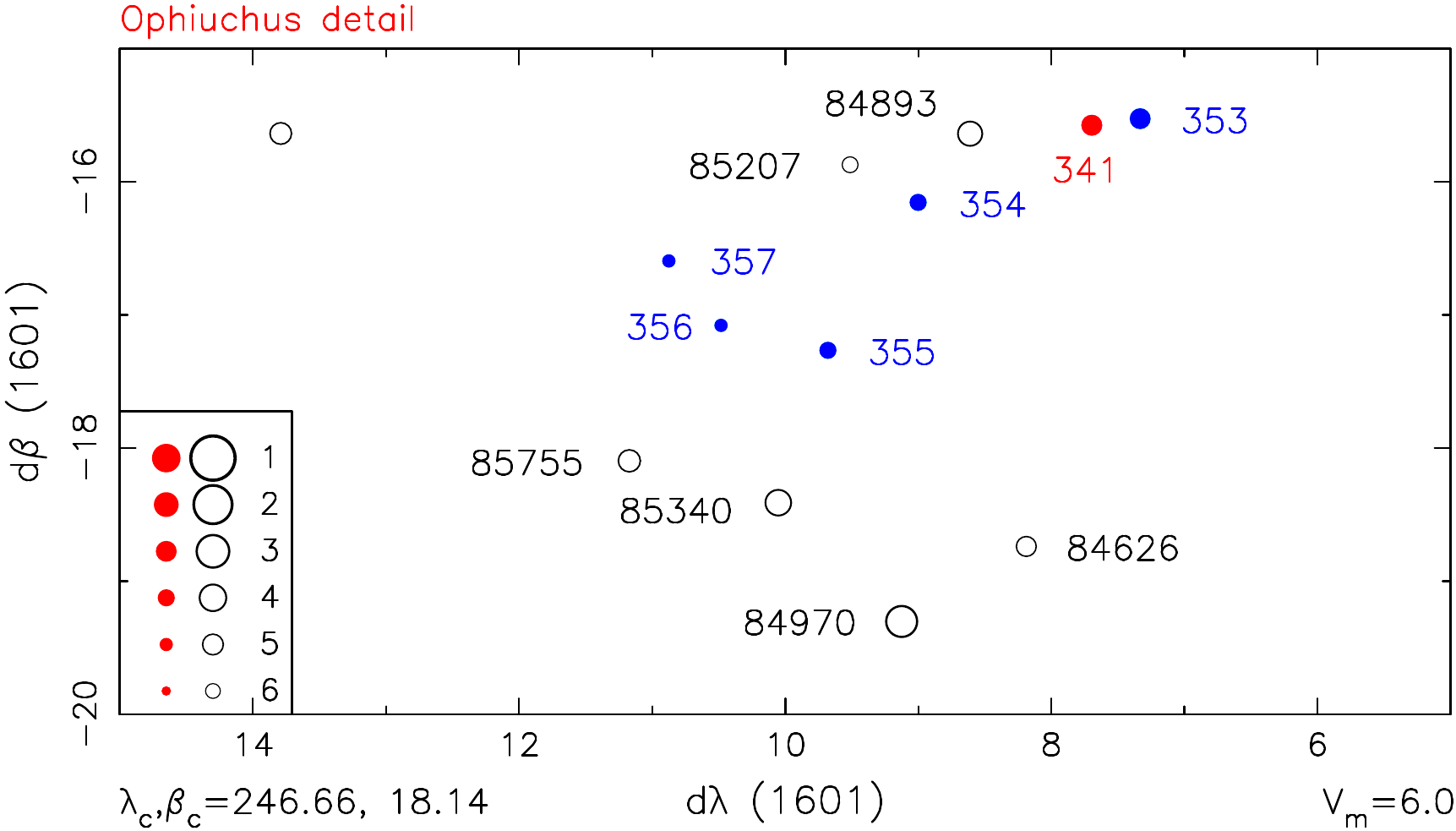}
\caption{Ophiuchus detail. \label{f:ophdetail}}
\end{figure}

\noindent K\,353-357 are illustrated in Fig.\,\ref{f:ophdetail}. Our
interpretation has K\,353 as a repeat entry for K\,341; and identifies
K\,354 with HIP\,85207 and K\,355-357 with HIP\,84970, HIP\,85340 and
HIP\,86755, respectively. An alternative possiblity would be to
identify K\,341 with HIP\,85207 and K\,354 with HIP\,84626, but we
consider this less likely. Dreyer gives no identifications for
K\,353-357. Rawlins argues that K\,354-357 are {\em utter fakes}.

\begin{figure}
\includegraphics[angle=270,width=\columnwidth]{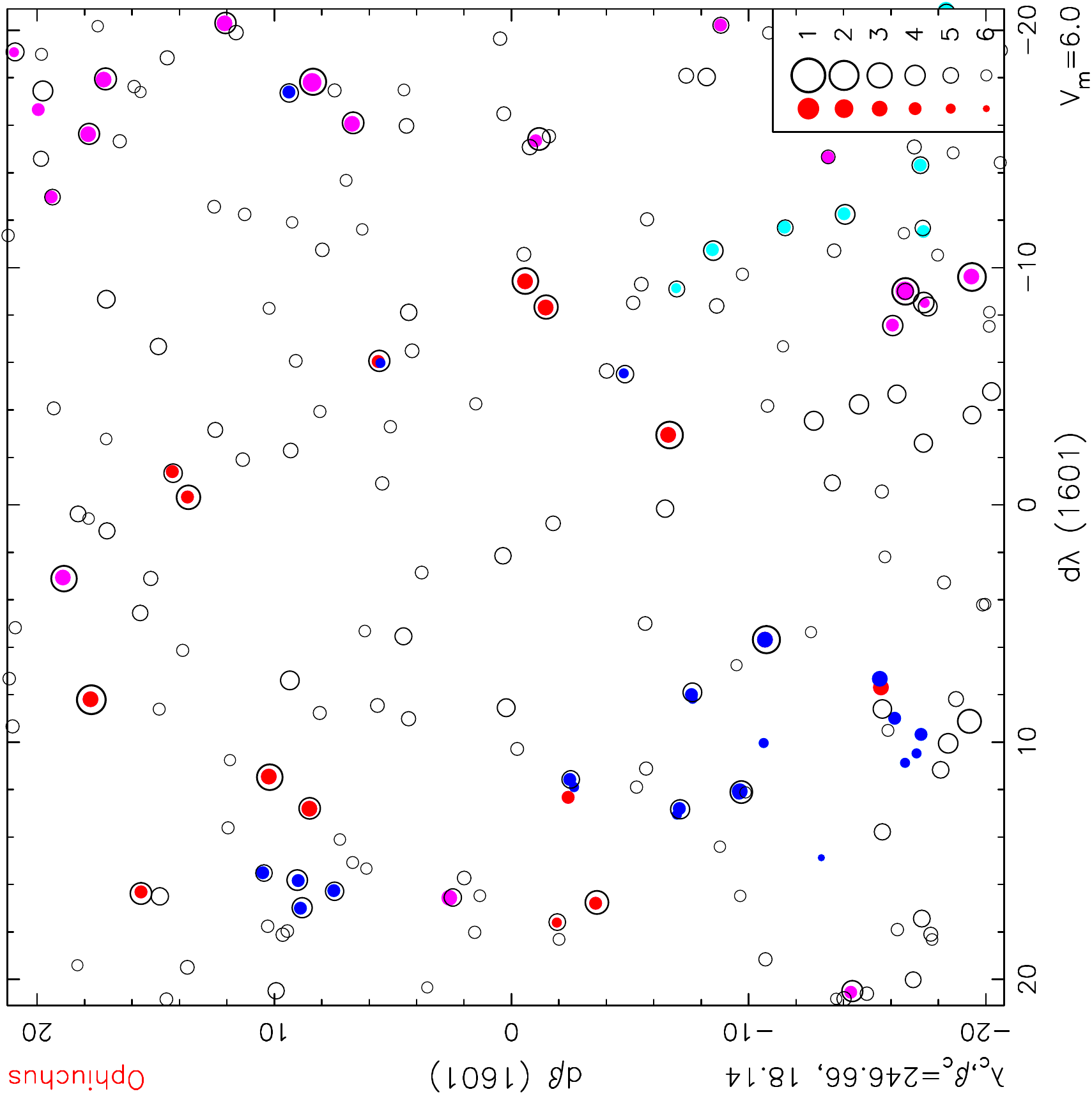}
\caption{Ophiuchus. \label{f:ophiuchus}}
\end{figure}

\noindent K\,360-364 are all repeat entries (see Table\,\ref{t:doubles})
K\,360 is near $-6$,+6 in Fig.\,\ref{f:ophiuchus}. For K\,361-364
see Fig.\,\ref{f:ophdetailb}. One may consider the pair K\,362 - K\,345
to correspond to the pair HIP\,86266 - HIP\,86263, but we consider this less
likely, because the orientation is not right and because the two entries
preceding K\,362 and the two following are all repeat entries.

\noindent K\,470 is a rather confusing entry. It is repeated exactly
as K\,483, apparently unnoticed by Kepler; \manuscript\ gives it
at the location of K\,470 only, and \progym\ gives it at the location of
K\,483 only. It has no nearby identification. \kepler\ annotates K\,470
{\em Meus solus. Forte eadem} (Mine only. Accidentally in the same place).
This may refer to its absence at this location in \progym\ and suggests
that Kepler thought it as almost at the same coordinates as K\,469. Did
he consider 28 for $M$ in longitude as an error for 18?

\begin{figure}
\centerline{\includegraphics[angle=270,width=\columnwidth]{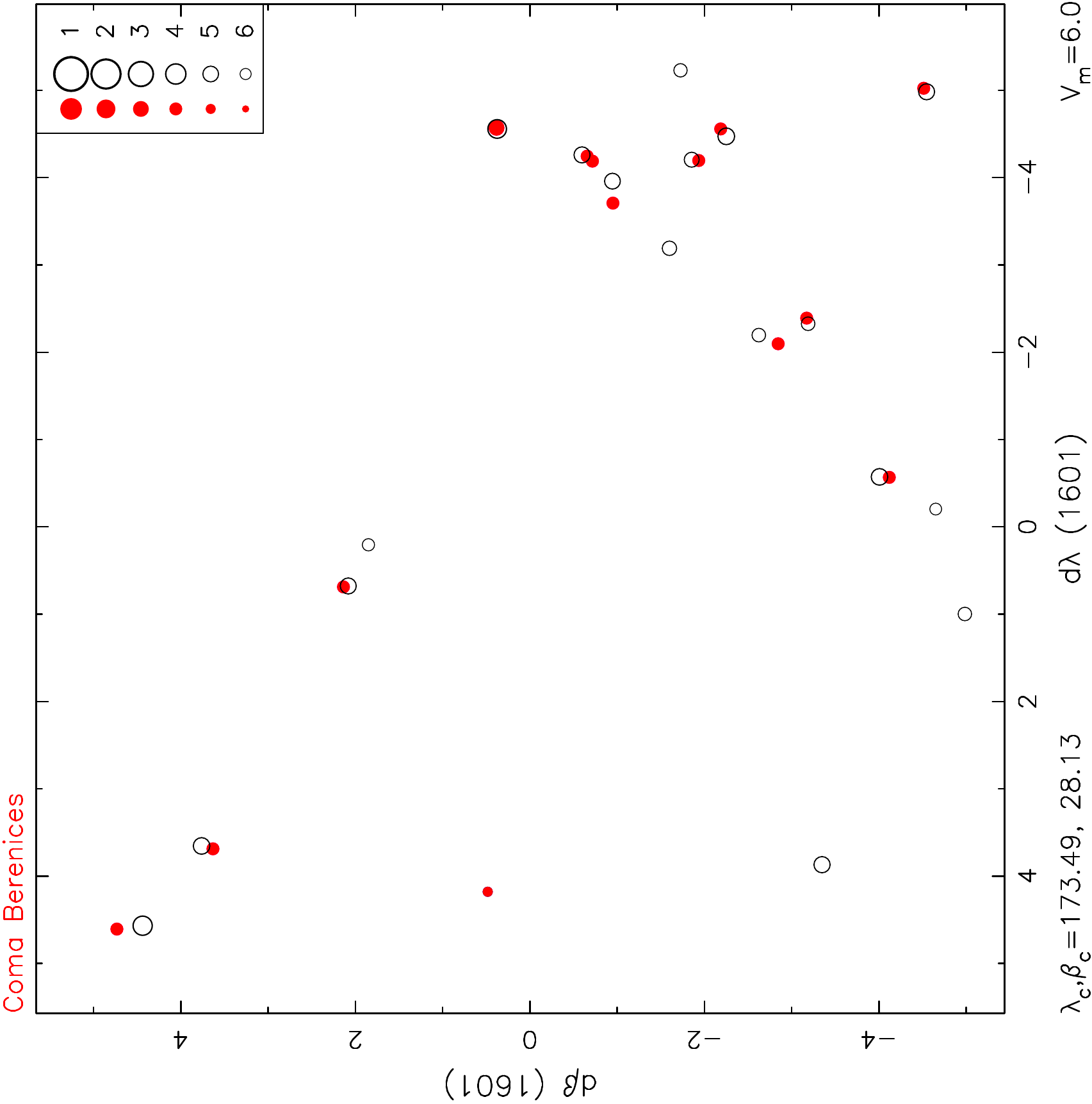}}
\caption{Coma Berenices \label{f:comaberenice}}
\end{figure}

\noindent K\,471-473. From the positions in the catalogue, K\,471 and
K\,472 both lie closest to HIP\,60697 (near $-$4.2,$-$0.7 in 
Fig.\,\ref{f:comaberenice}) and K\,473 is close to HIP\,60746, leaving
HIP\,60904 (near $-$3.2,$-$1.6) without catalogue counterpart.
The catalogue describes K\,471-473 as following one another,
hence we prefer to identify K\,472 with HIP\,60746 and K\,473 with
HIP\,60904. Dreyer agrees.

\noindent K\,484 = HIP\,8832 corresponds to HR\,545+546 
($\gamma^1+\gamma^2$\,Ari). 

\begin{figure}
\centerline{\includegraphics[angle=270,width=0.9\columnwidth]{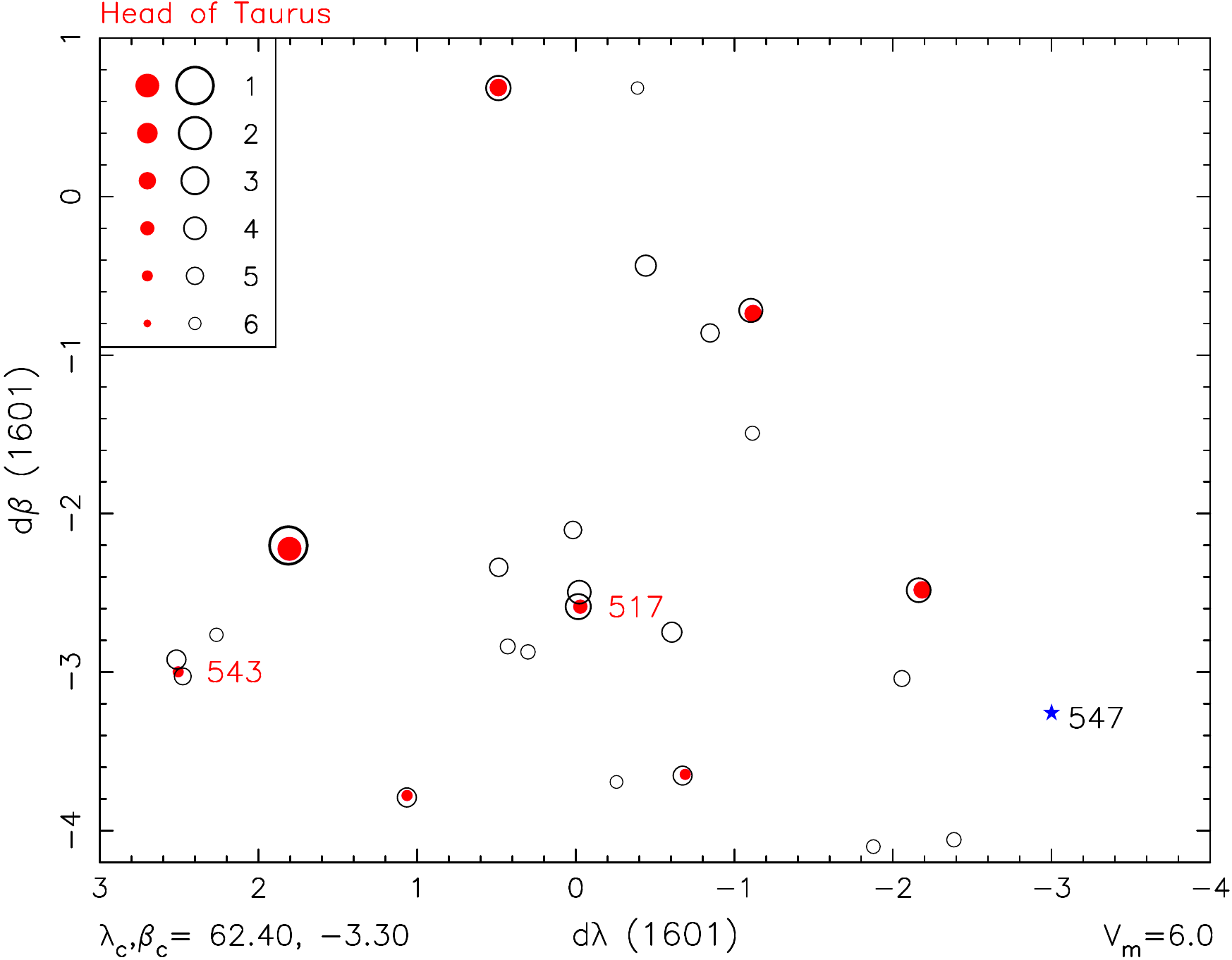}}
\caption{Taurus: enlarged detail of the head, showing K\,517 to be best
identified with $\theta^2$ the southern star of the close pair $\theta^1$/$\theta^2$,
and K\,543 to be between two posible {\em Hipparcos} counterparts.
The \kepler\ position of K\,547, indicated $\star$, has no good identification.
\label{f:taurush}}
\end{figure}

\noindent K\,517 is identified with the northern of the pair
$\theta^{1,2}$ (HIP\,20885 and HIP\,20894) by Dreyer, but actually
is closer to the southern star. See Figure\,\ref{f:taurush}.

\begin{figure}
\centerline{\includegraphics[angle=270,width=0.9\columnwidth]{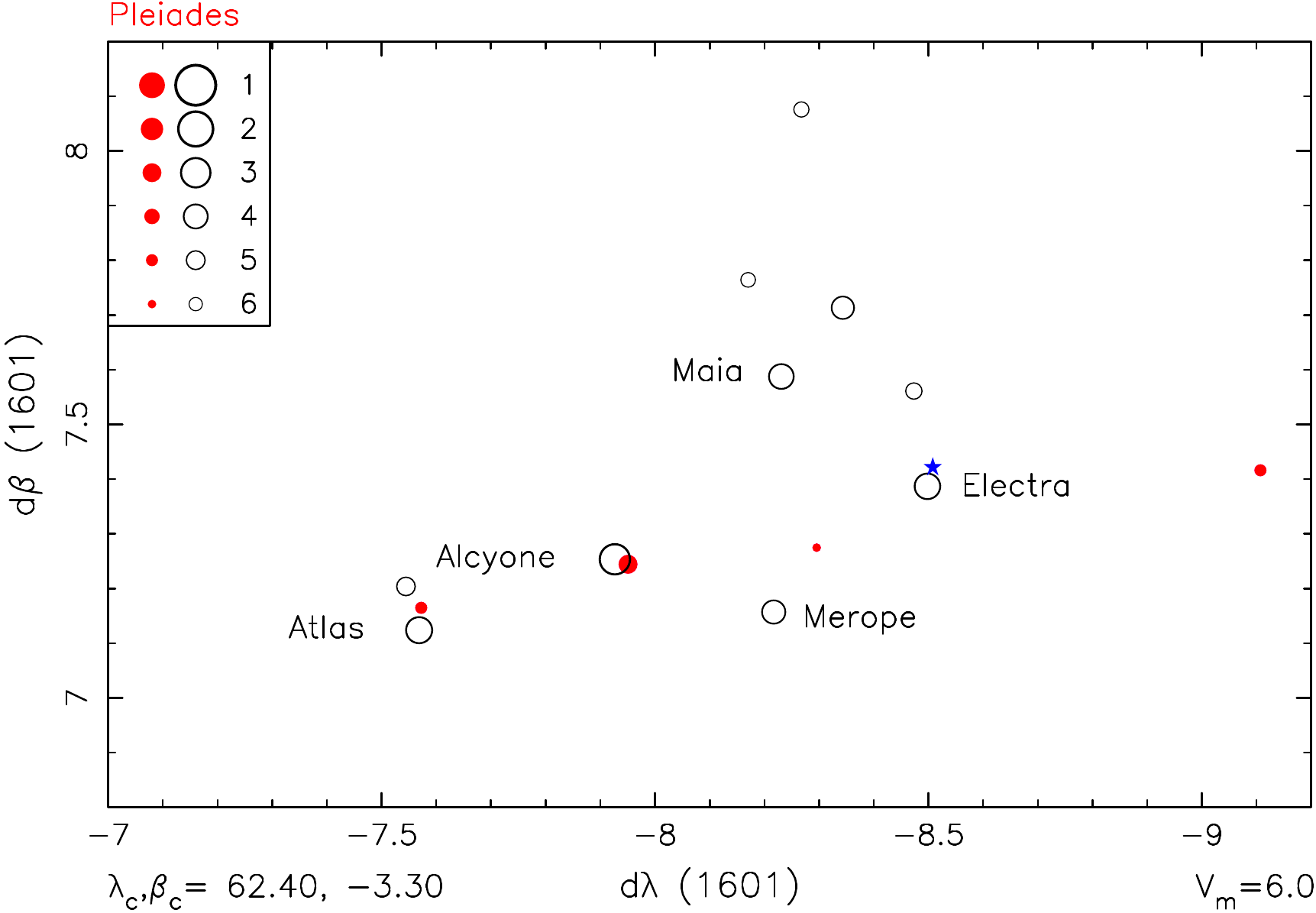}}
\caption{Taurus: enlarged detail of the Pleiades. The position of
  K\,534 in \manuscript\ and \progym\ is indicated with a blue $\star$, and
  shows that this star should be identified with Electra.
  \label{f:tauruspl}}
\end{figure}

\noindent K\,534 is Electra in the Pleiades, as shown by
its position in \manuscript\ (see Fig.\,\ref{f:tauruspl}).

\noindent K\,543 lies between HIP\,21683 and HIP\,21673, somewhat
closer to the latter (Fig.\,\ref{f:taurush}). We list the brighter star
as the counterpart. HIP\,21673 is fainter ($V$$=$5.1) but closer
(2\farcm4), and is listed by Dreyer as the counterpart.

\begin{figure}
\includegraphics[angle=270,width=\columnwidth]{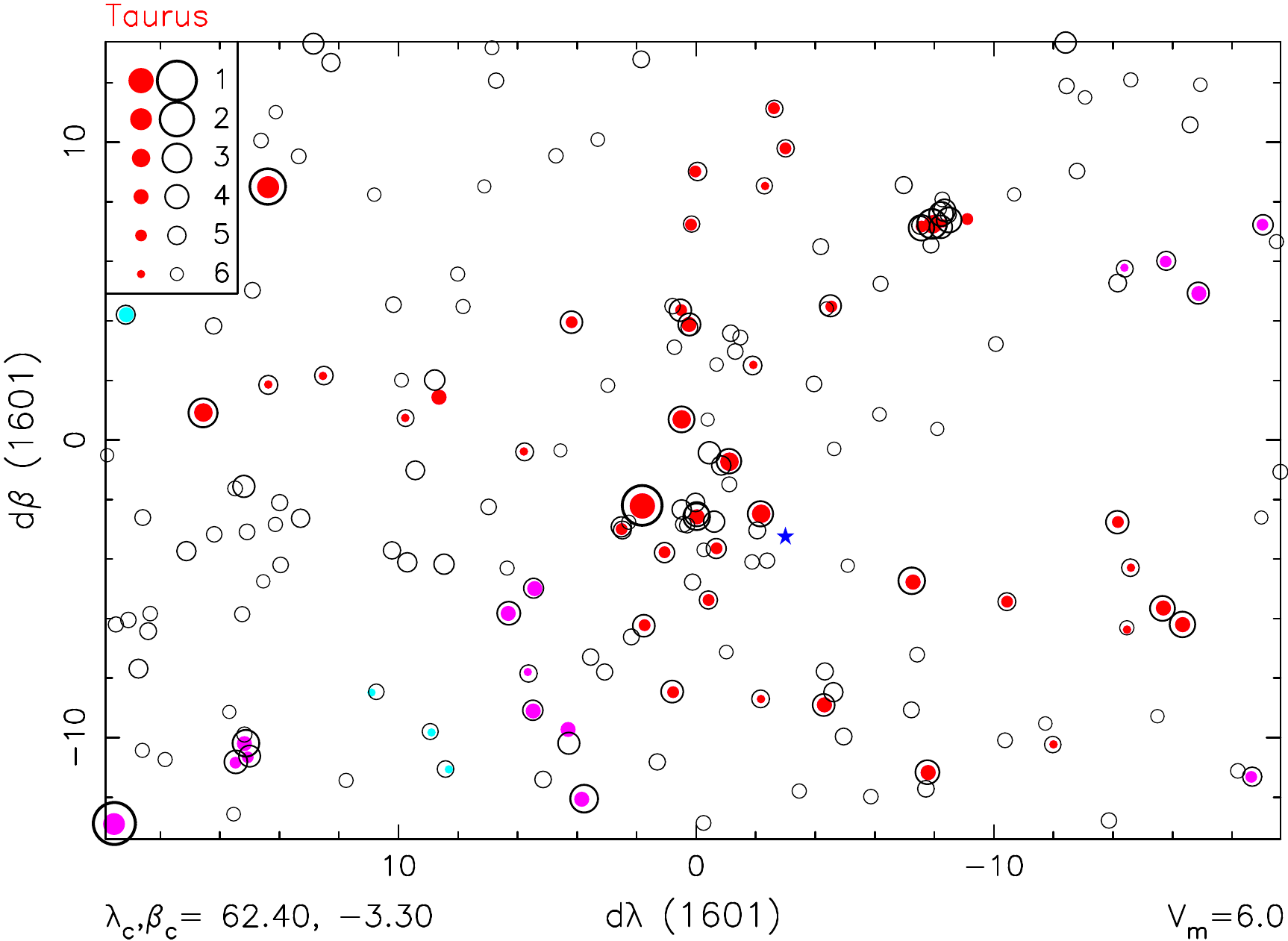}
\caption{Taurus. The \kepler\ position of K\,547, indicated $\star$,
  has no good identification. The emended position at about
  $-3,+10$ identifies it with HIP\,19171.\label{f:taurus}}
\end{figure}

\noindent K\,547. The position in \manuscript\ gives
excellent agreement with HIP\,19171 (at 0\farcm3), which is the
identification also given by Dreyer. The position given in \progym\ and
\kepler\ does not lead to a good identification. See Fig.\,\ref{f:taurus}. 

\begin{figure}
\includegraphics[angle=270,width=\columnwidth]{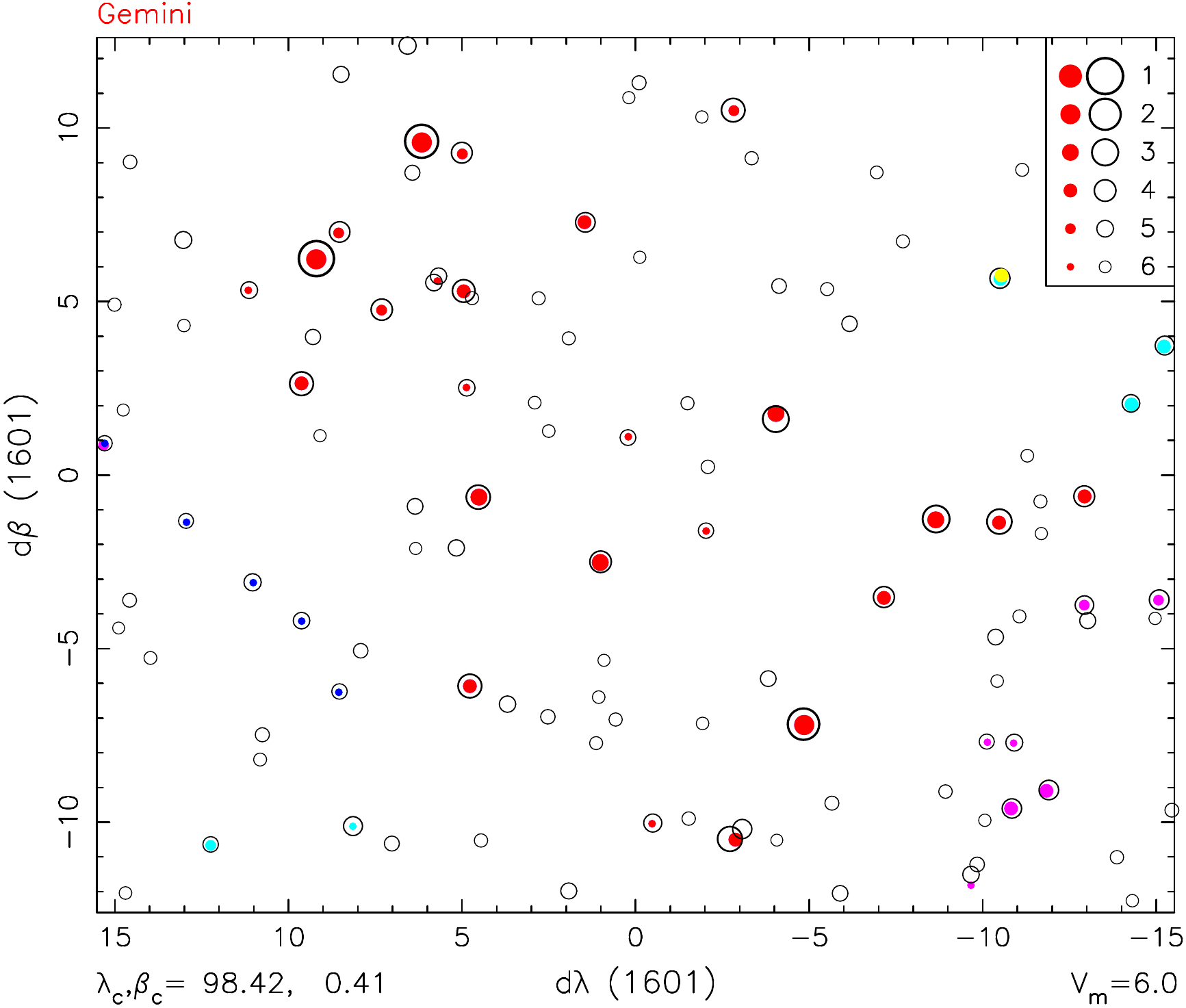}
\caption{Gemini.\label{f:gemini}}
\end{figure}

\noindent K\,556, near +5.5,+5.5 in Fig.\,\ref{f:gemini}, lies between
HIP\,36393 (at 8\farcm9) and HIP\,36492.  Dreyer gives HIP\,36393 as
identification, we choose the closer star.

\noindent K\,577 is the open star cluster Praesepe; Dreyer identifies
it with the brightest star (at $V=6.3$) in Praesepe, $\epsilon$\,Cnc =
HIP\,42556

\begin{figure}
\includegraphics[angle=270,width=\columnwidth]{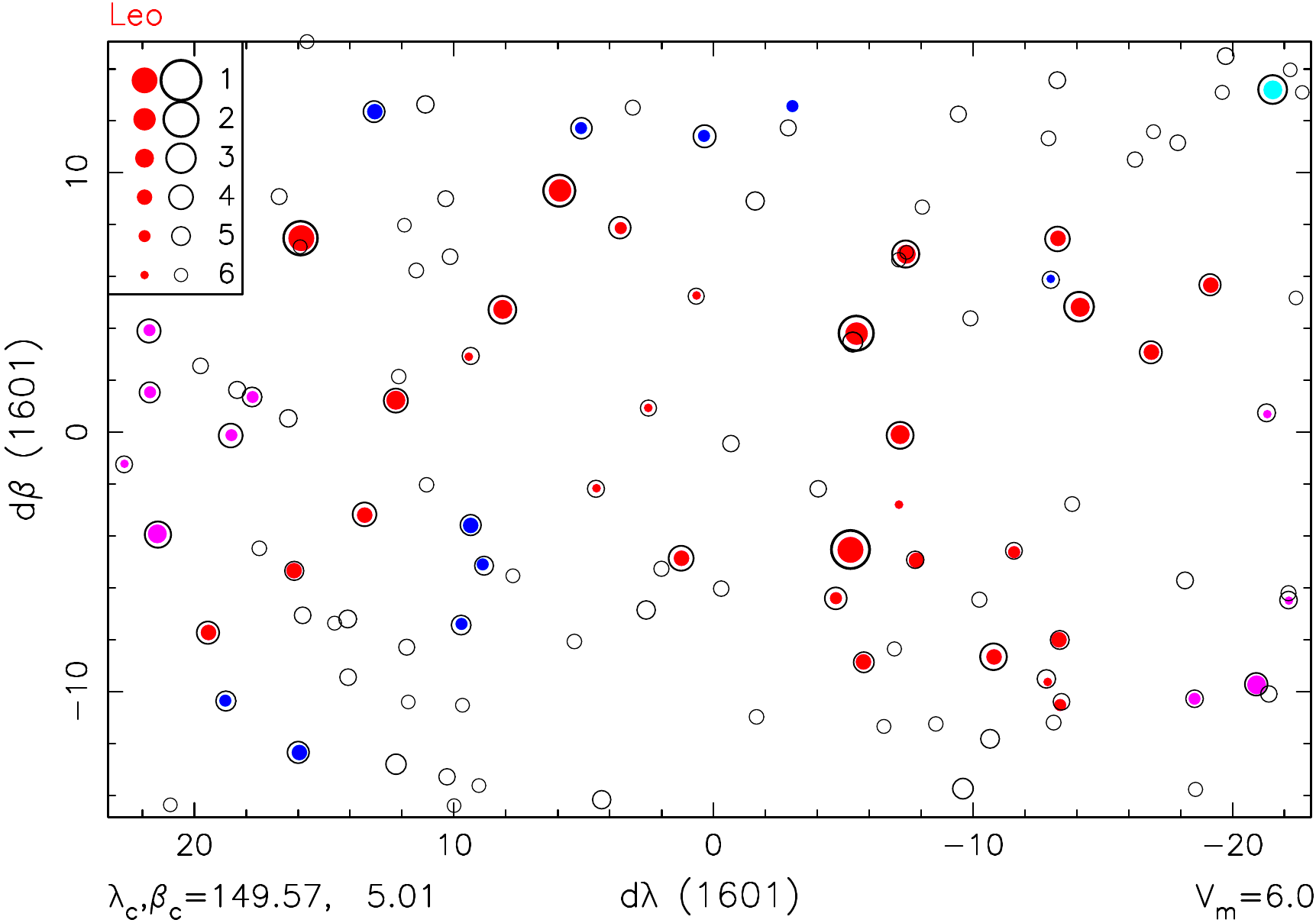}
\caption{Leo. \label{f:leo}}
\end{figure}

\noindent K\,607, near $-$7,$-$3 in Fig.\,\ref{f:leo}, is identified
with HIP\,51585 (46\,Leo, near $-$0.5,$-$0.5) by Dreyer; too far to be
believable. HIP\,50333, near $-$4,$-$2, is closer.

\noindent K\,625, near $-$3.0,+12 in Fig.\,\ref{f:leo}, is identified
with 7.3 mag star HIP\,52240 (39\,LMi) by Dreyer, too faint to be
believable. We think our identification with HIP\,52422 is secure, as
this is the only sufficiently bright star near K\,625 .

\begin{figure}
\includegraphics[angle=270,width=\columnwidth]{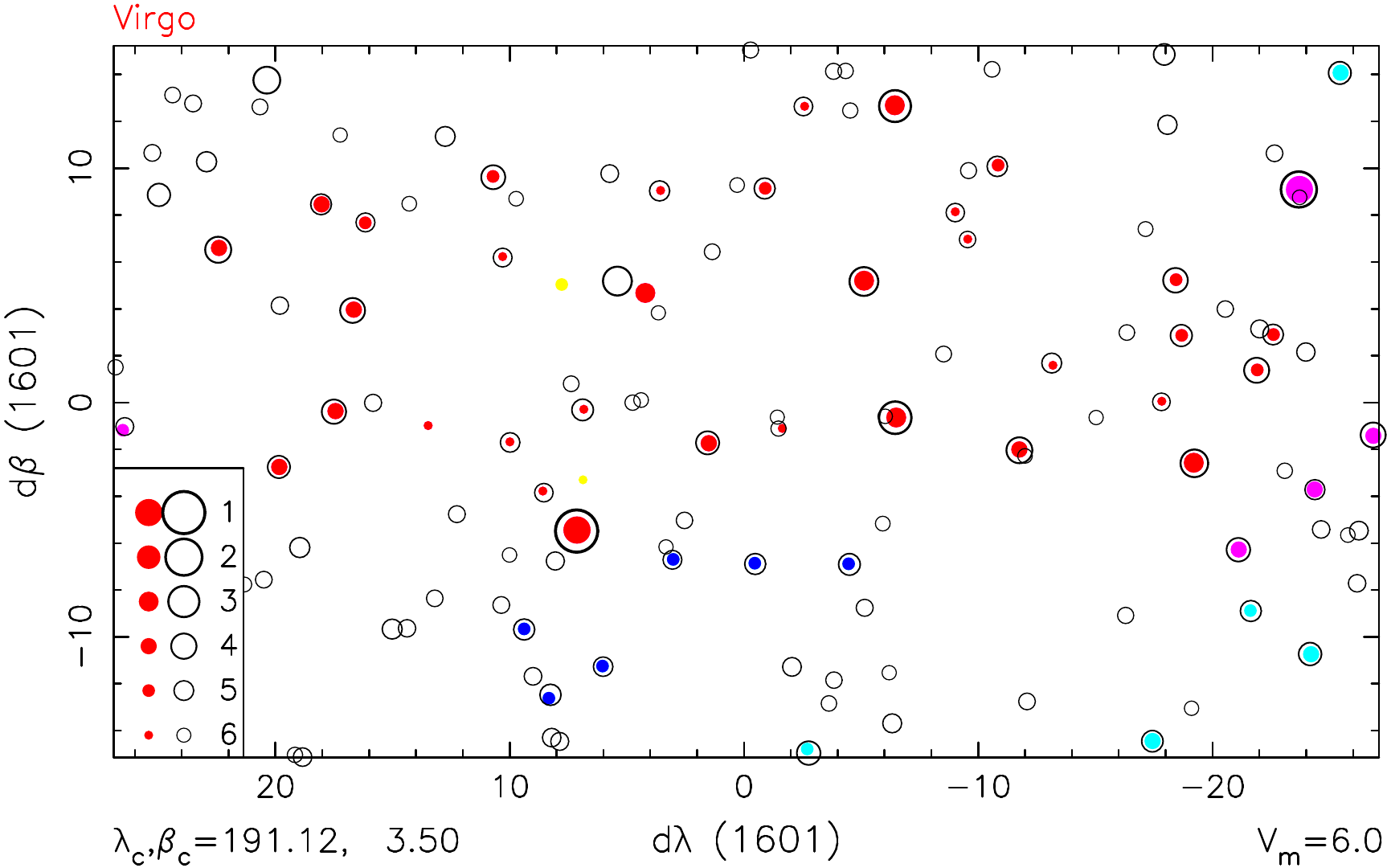}
\caption{Virgo. \label{f:virgo}}
\end{figure}

\noindent K\,646, near +4,+5 in Fig.\,\ref{f:virgo}, is securely
identified with HIP\,66249 ($\zeta$\,Vir) even though the rather
fainter HIP\,65545 is somewhat closer at 61\farcm2.

\noindent K\,650, near +13.5,$-$1.0 in Fig.\,\ref{f:virgo}, is possibly
identified with HIP\,68940, as no other sufficiently bright star is
nearer. Dreyer gives no identification.

\begin{figure}
\centerline{\includegraphics[angle=270,width=0.9\columnwidth]{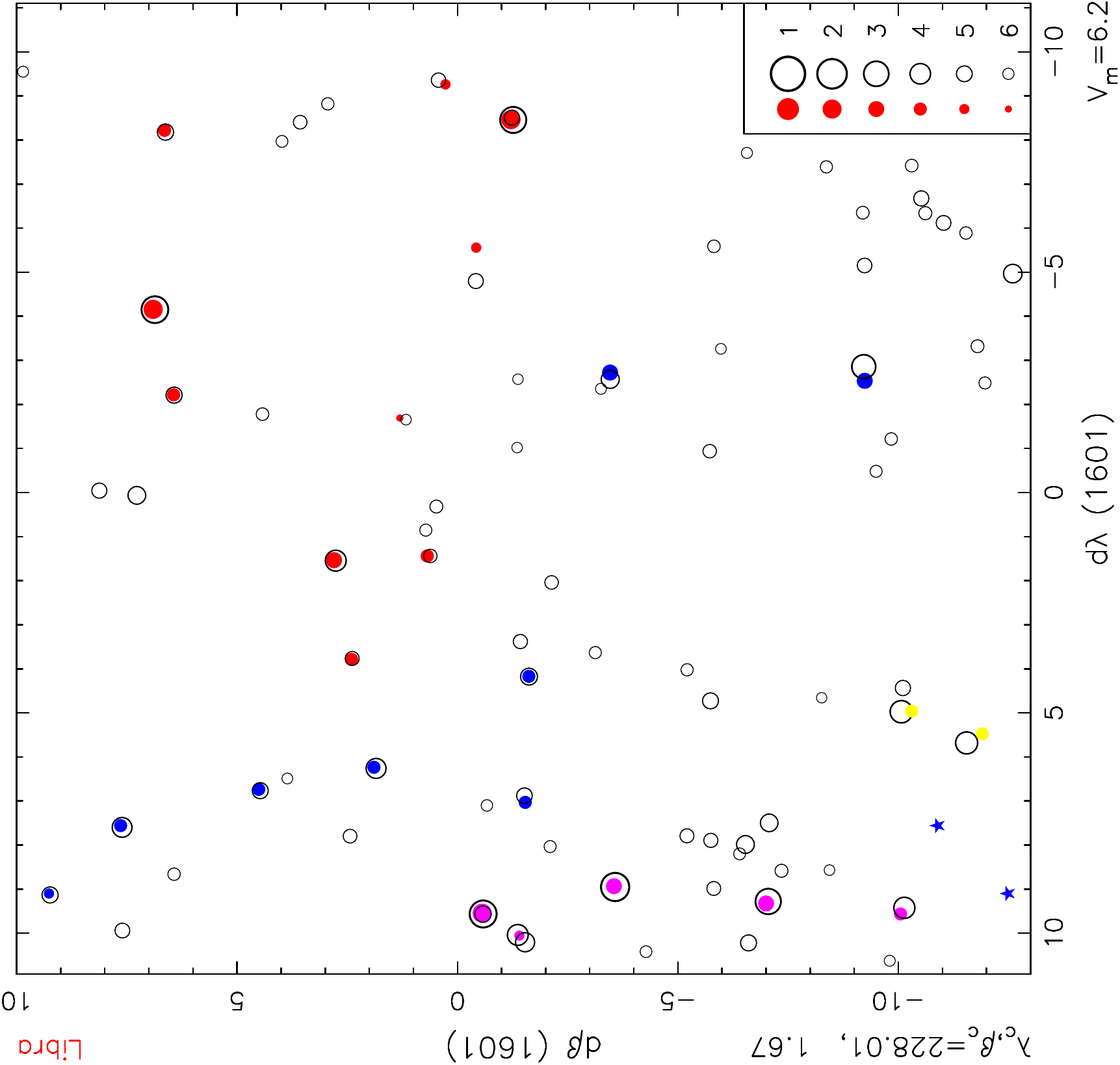}}
\caption{Libra. The position as given in \kepler\ for K\,685,686 
are indicated with $\star$ and have no near counterparts.
The emended positions are in the upper left corner. \label{f:libra}}
\end{figure}

\noindent K\,676, near $-$2,+1 in Fig.\,\ref{f:libra}.  We accept a
counterpart with $V=6.1$ as no brighter stars are nearby; the
identification by Dreyer, HIP\,75294, is fainter and further ($V$=6.5,
$d=33'$).

\noindent K\,687,688, near $-$2.5,$-$9 and $-$2.5,$-$3.5 in 
Fig.\,\ref{f:libra}, have better positions in \manuscript,
i.e.\ closer to the counterparts, than in \kepler.

\begin{figure}
\centerline{\includegraphics[angle=270,width=0.9\columnwidth]{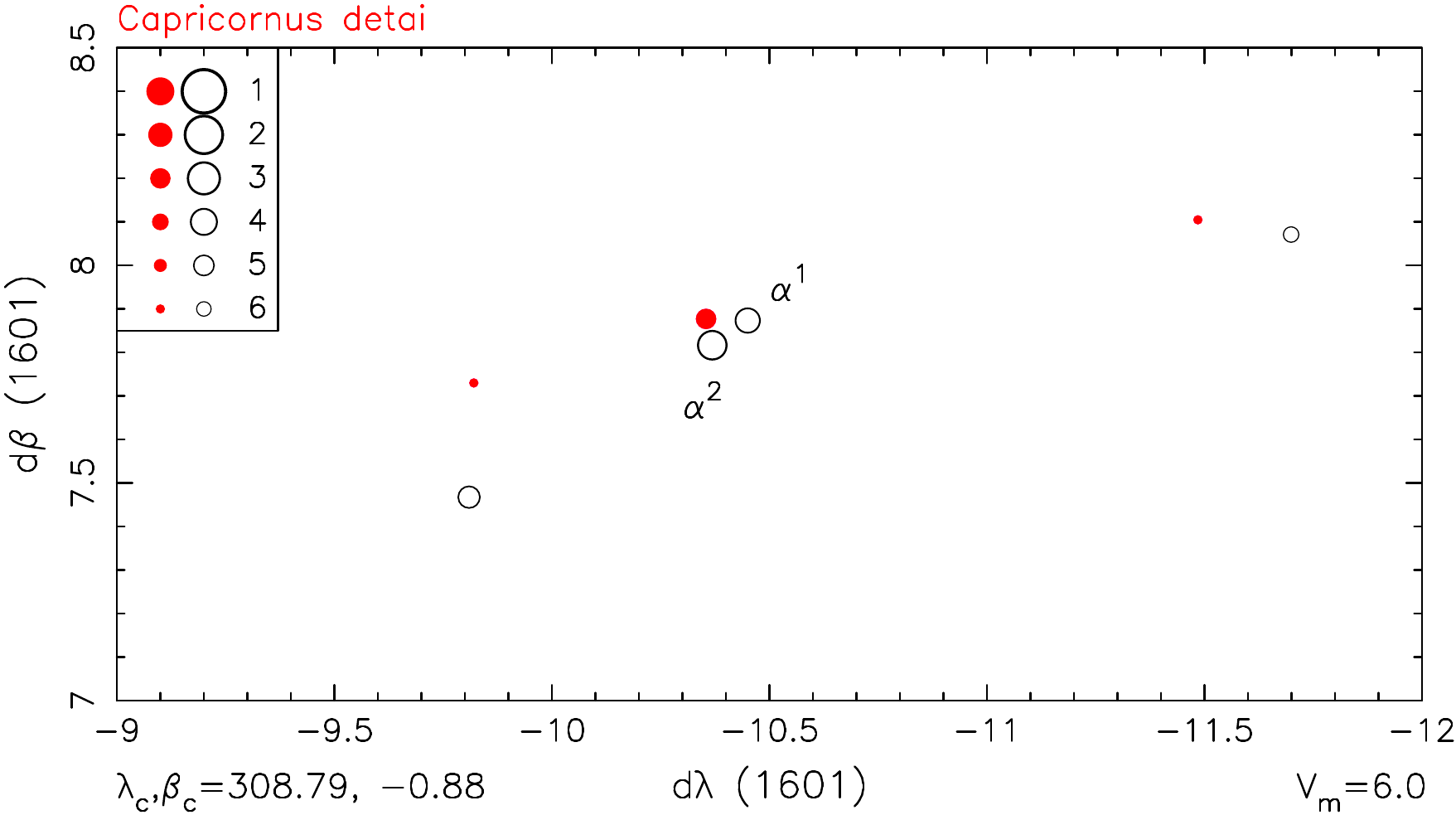}}
\caption{Capricornus detail, showing identification of K\,713 with
$\alpha^2$\,Cap. \label{f:capdetail}}
\end{figure}

\noindent K\,713 is closer to $\alpha^2$, the brighter and southern
star of the pair $\alpha^1$,$\alpha^2$ (Fig.\,\ref{f:capdetail}). Dreyer
gives $\alpha^1$ as identification.

\begin{figure}
\includegraphics[angle=270,width=\columnwidth]{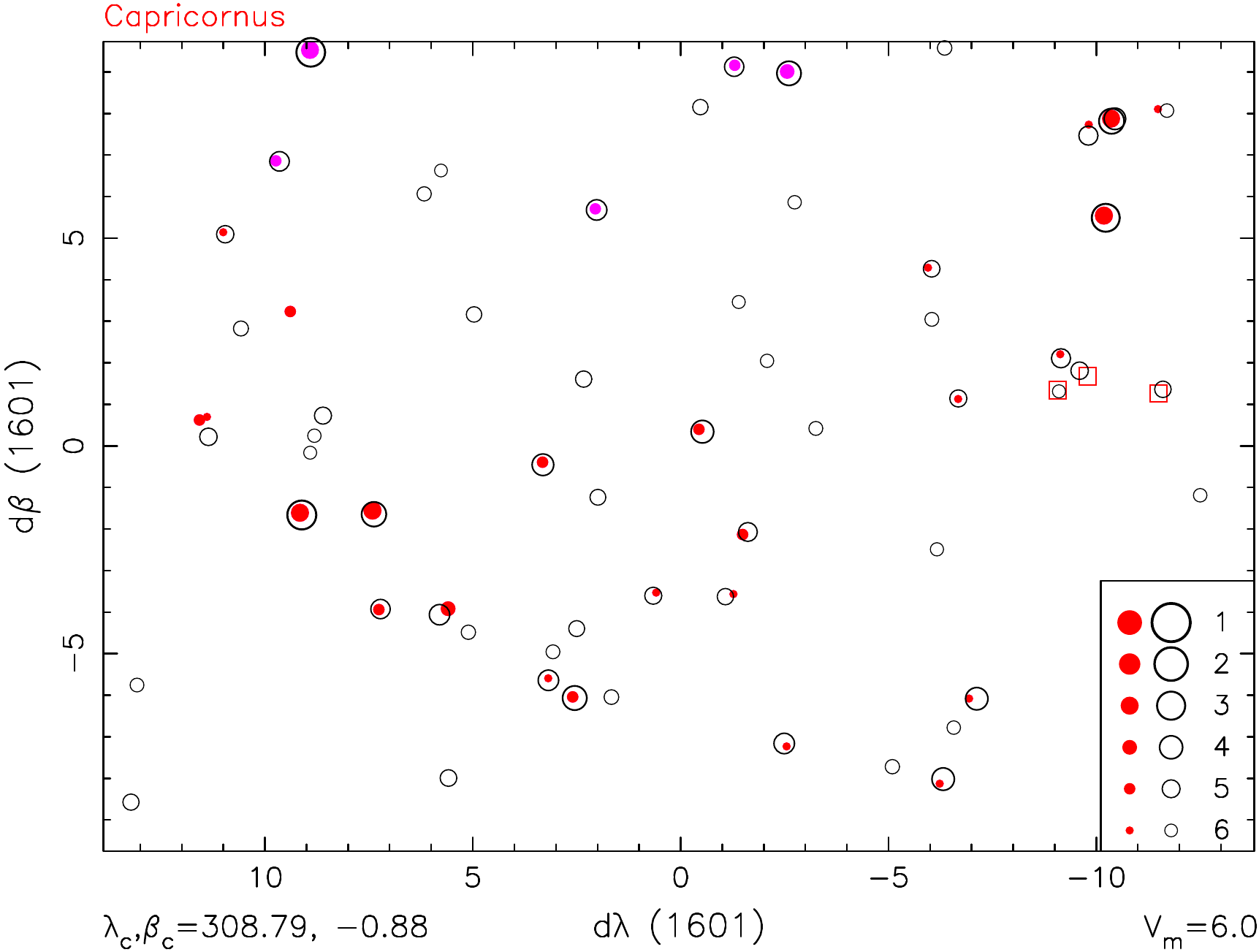}
\caption{Capricornus. Three stars listed as nebulous by Brahe,
indicated with $\square$, are identified with ordinary stars.
\label{f:capricornus}}
\end{figure}

\noindent K\,717, 718 and 720 are listed `ne', but are well identified
with stars (see Fig.\,\ref{f:capricornus}).

\noindent K\,737, near +9.5,+3 in Fig.\,\ref{f:capricornus}, is rather
far from its identification, but no star with $V$$<$6 is closer.

\noindent K\,738, near +11.5,+0.5 in Fig.\,\ref{f:capricornus},
is identified (by us and Dreyer) with HIP\,108036, which leaves
K\,739, right next to it, without counterpart, unless we are willing to accept a
$V$=6.3 star. 

\begin{figure}
\includegraphics[angle=270,width=\columnwidth]{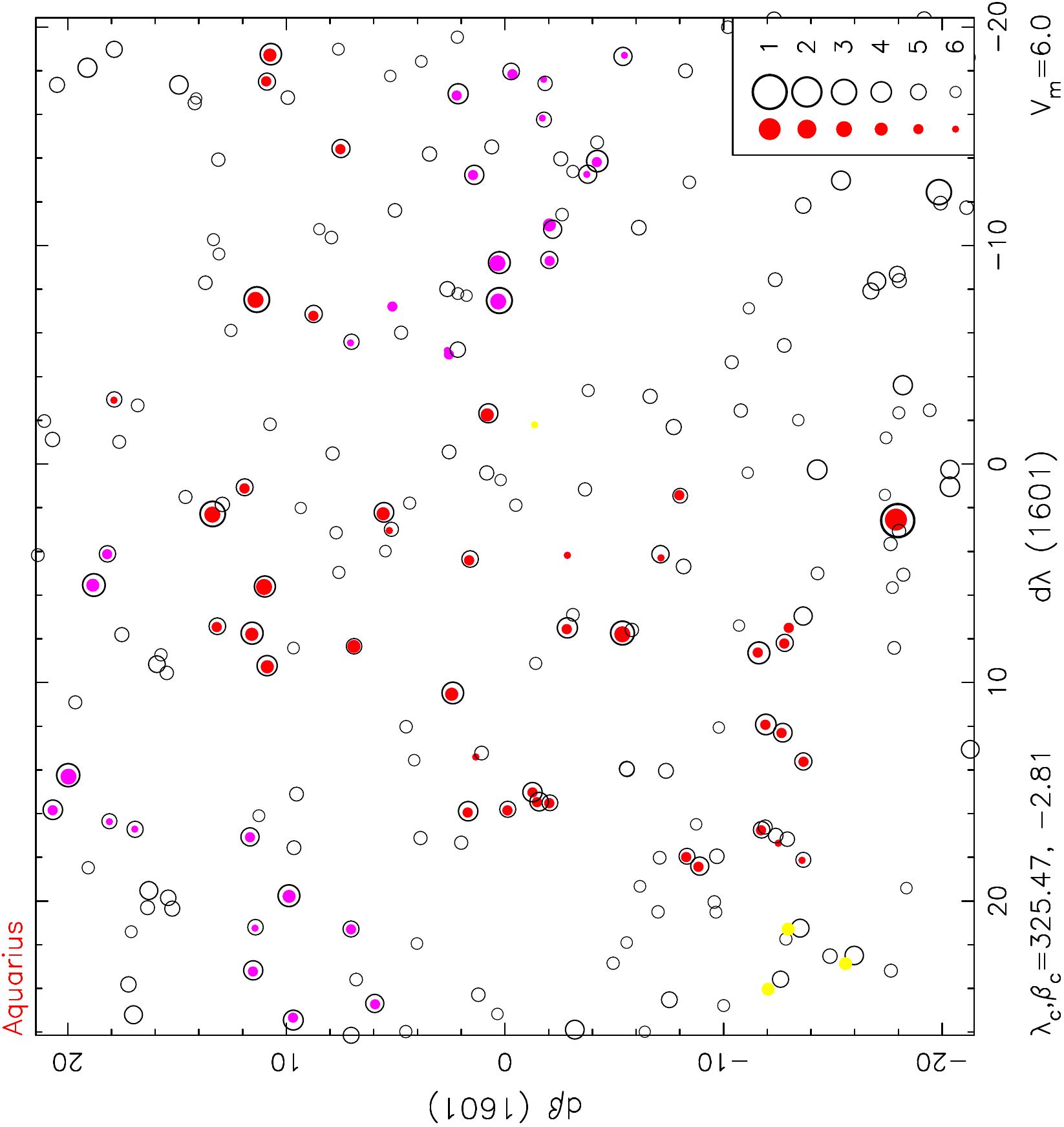}
\caption{Aquarius. \label{f:aquarius}}
\end{figure}

\noindent K\,759, near +4,$-$3 in Fig.\,\ref{f:aquarius}) is
tentatively identified by us with HIP\,112542 (near +7,$-$3) and by
Dreyer with HIP\,110778 (near +1,$-$4).

\noindent K\,772, near +17,$-$12 in Fig.\,\ref{f:aquarius}). Our
identication is both brighter and closer than the identification given
by Dreyer HIP\,116889 ($\Delta$=14\farcm2).

\begin{figure}
\includegraphics[angle=270,width=\columnwidth]{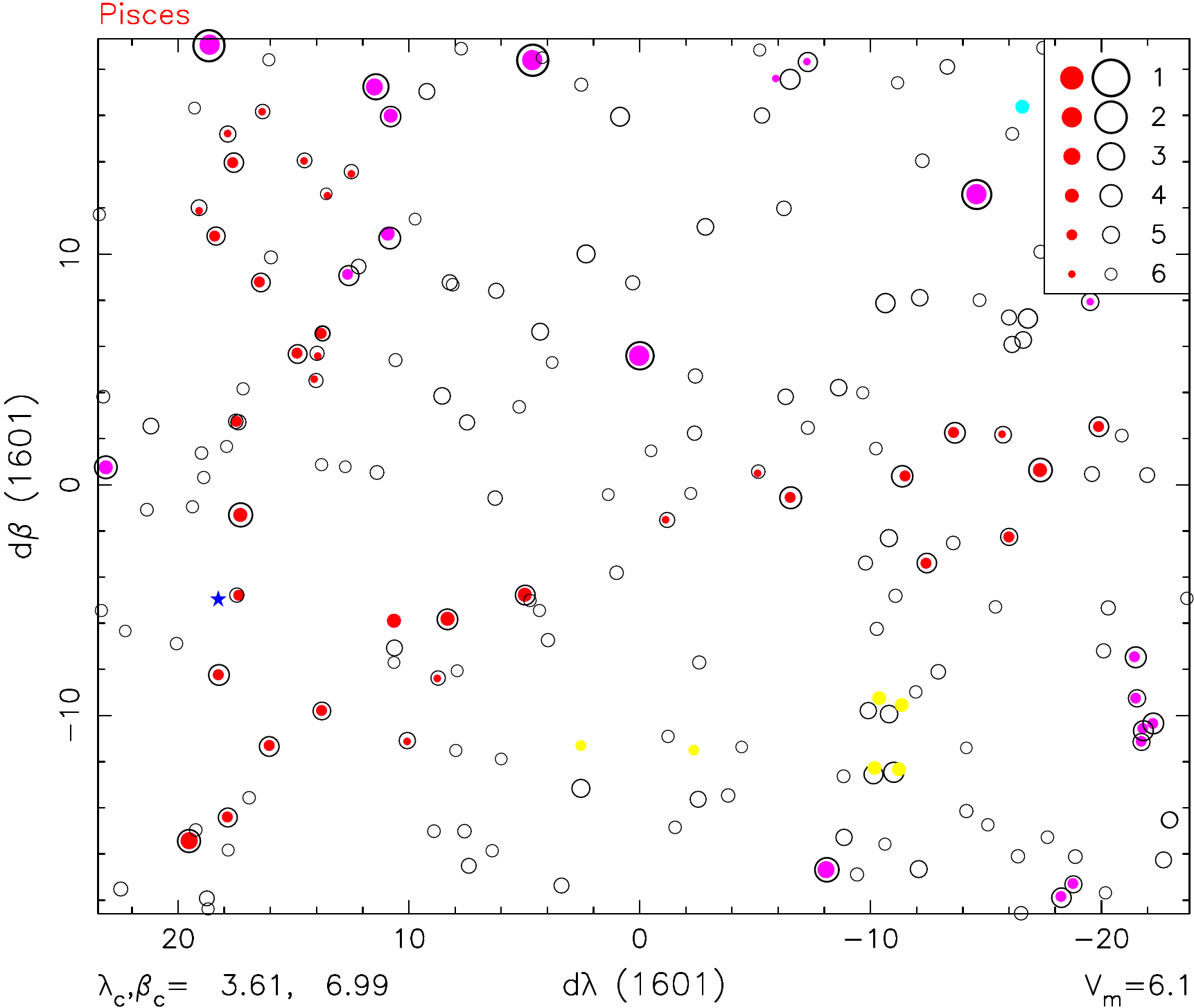}
\caption{Pisces. The blue $\star$ indicates the position of K\,801 
according to \kepler, the emended position gives a much better
coincidence with o\,Psc, about 3$^\circ$ to the south.\label{f:pisces}}
\end{figure}

\noindent K\,804, near +17,+3 in Fig.\,\ref{f:pisces}).  Our
identification is closer than, but slightly fainter than the
identification given by Dreyer and Rawlins, $\rho$\,Psc = HIP\,6706
($\Delta$=5\farcm6).

\noindent K\,808, near +13,+12 in Fig.\,\ref{f:pisces}). Our
counterpart has $V$=6.1.

\begin{figure}
\includegraphics[angle=270,width=\columnwidth]{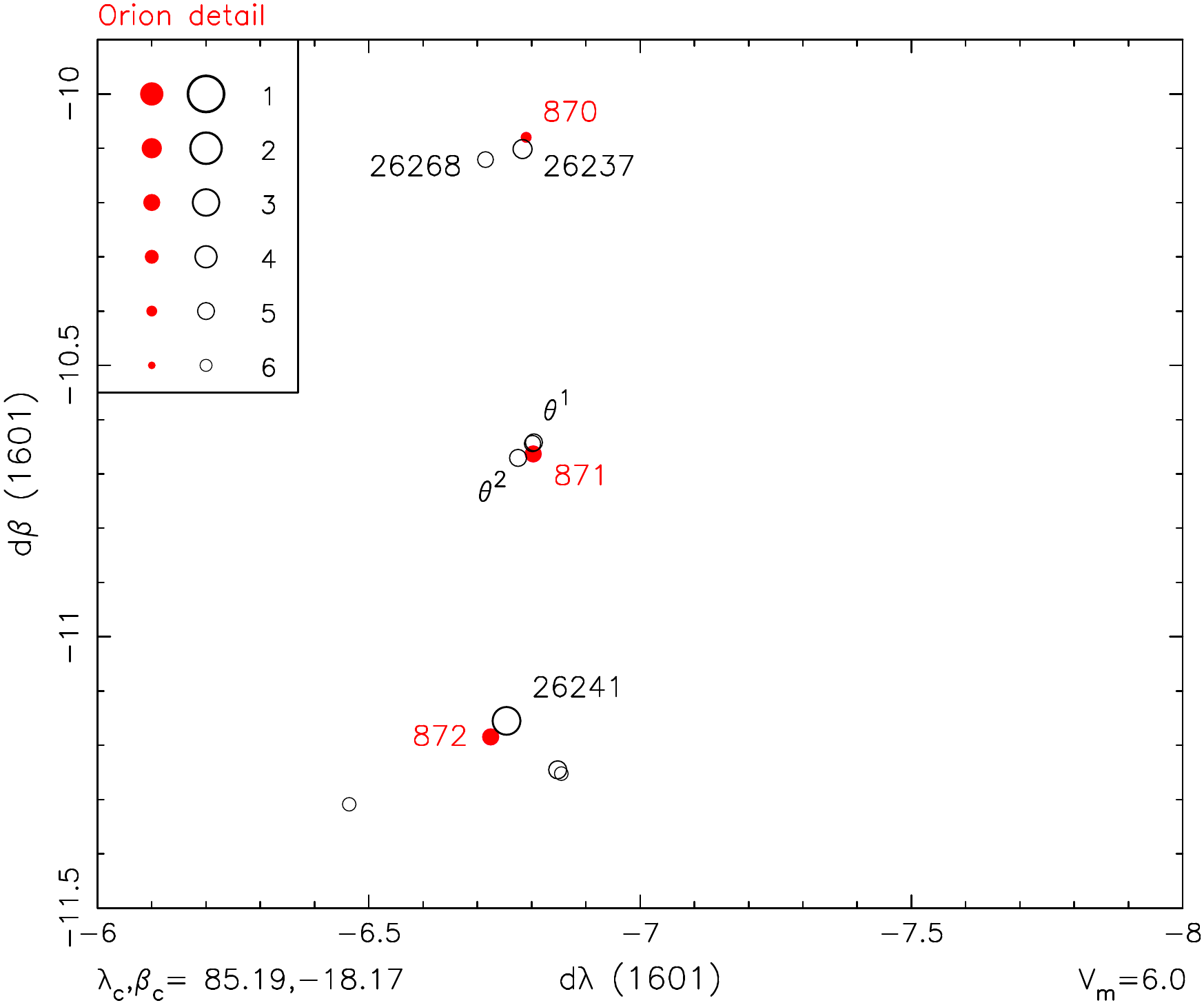}
\caption{The sword of Orion \label{f:oridetail}}
\end{figure}

\noindent K\,870 (Fig.\,\ref{f:oridetail}) is identified with
HIP\,26268 by Dreyer ($V$=5.2, $\Delta$=5\farcm1).  Our identification
HIP\,26237 is brighter and closer; note that the angle between
HIP\,26268 and HIP\,26237 is just 4\farcm2.

\noindent K\,871 is identified by Dreyer with HIP\,26235 (=
$\theta^2$\,Ori, $V$=5.0, $\Delta$=1\farcm7). Our identification
$\theta^1$\,Ori is equally bright and marginally closer
(Fig.\,\ref{f:oridetail}). Note that $\theta^1$\,Ori corresponds to
HIP\,26220 and HIP\,26221 (separation 13\arcsec). HIP\,26224 (at 21\arcsec,
$V$=6.7) is also part of $\theta^1$, but much fainter.

\begin{figure}
\includegraphics[angle=270,width=\columnwidth]{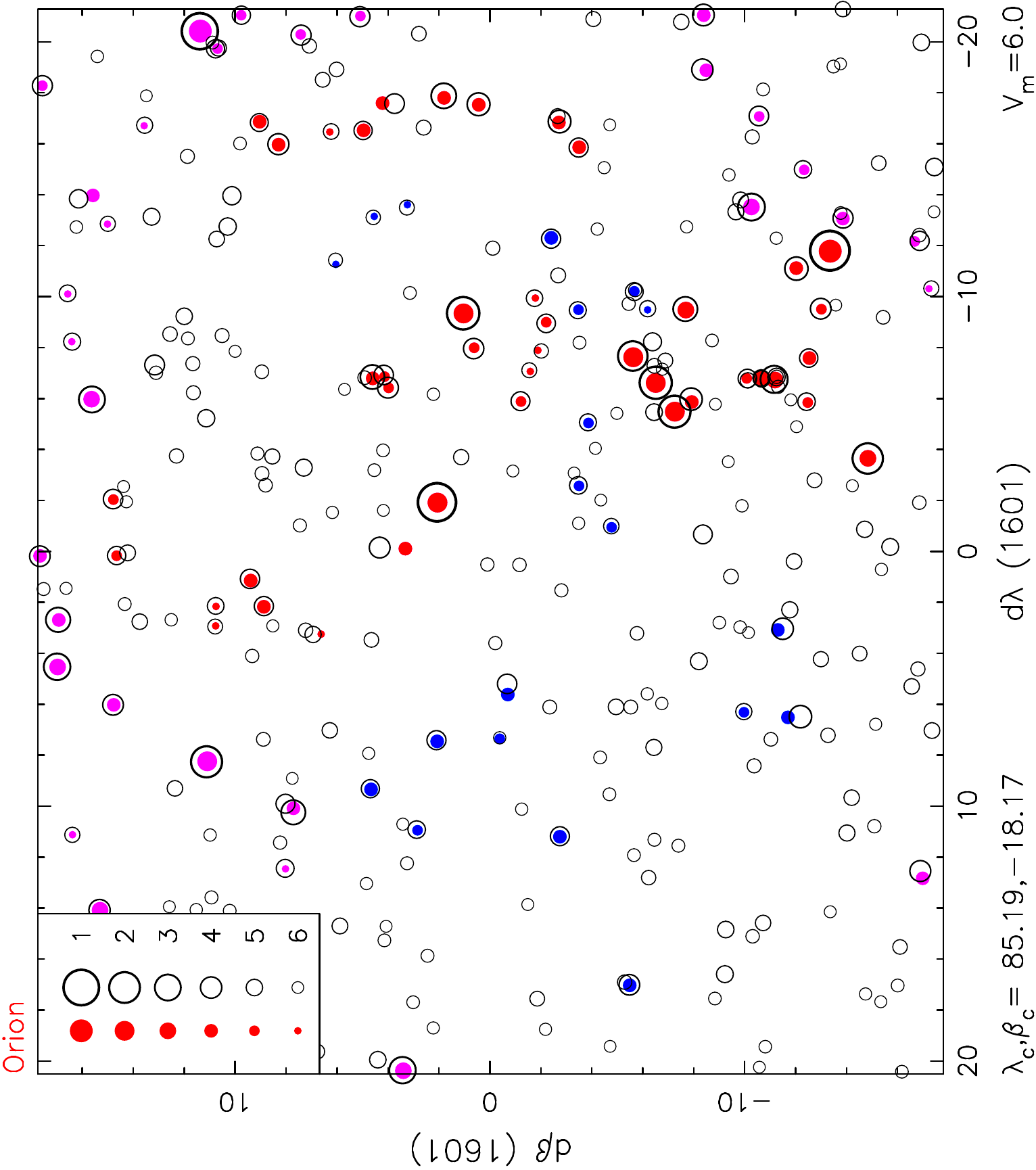}
\caption{Orion \label{f:orion}}
\end{figure}

\noindent K\,893, near +6,$-$10 in Fig.\,\ref{f:orion}, is identified
by Dreyer with HIP\,30700 ($V$=6.5, $\Delta$=29\farcm5). Our identification
is much closer and brighter.

\begin{figure}
\includegraphics[angle=270,width=\columnwidth]{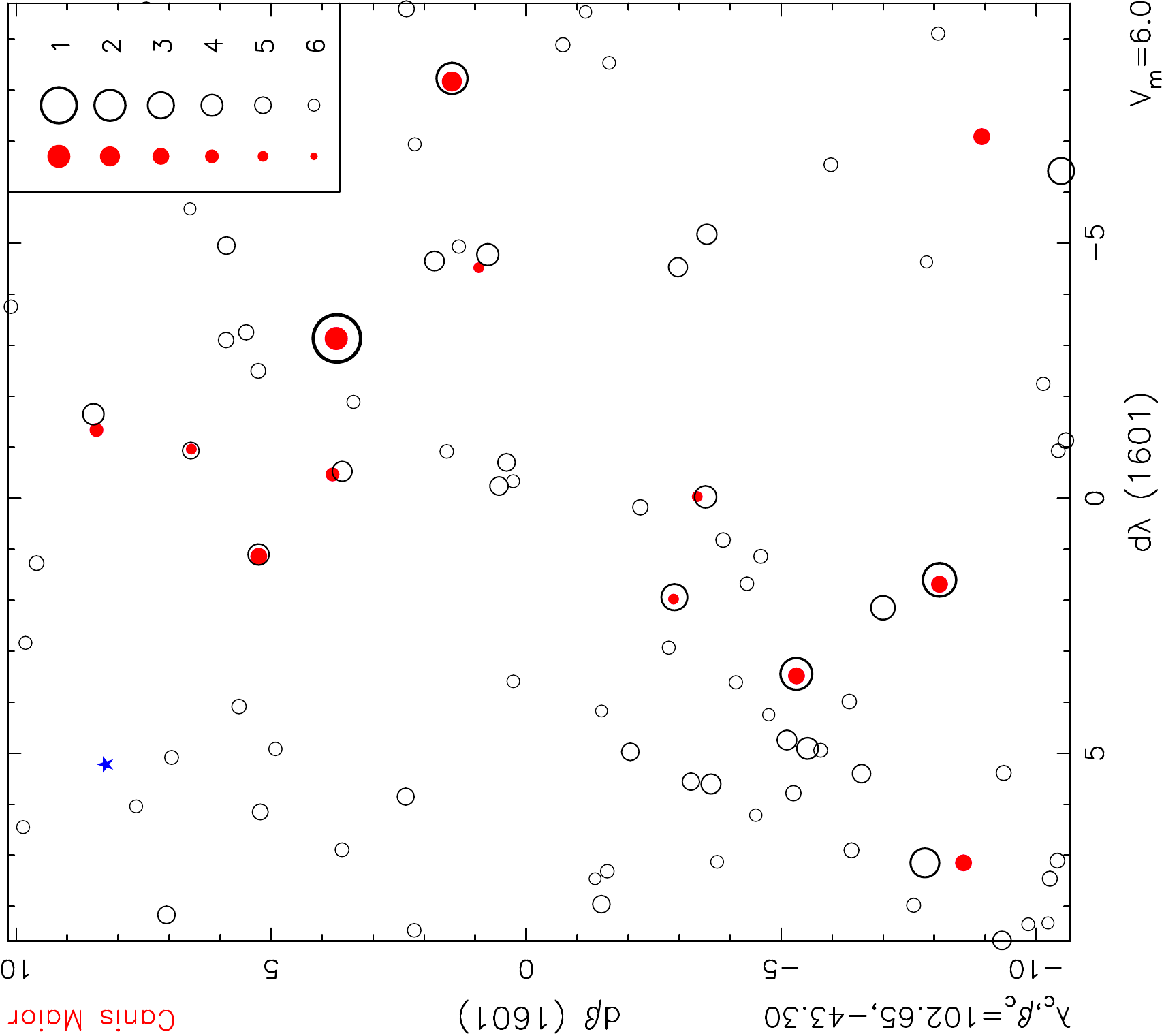}
\caption{Canis Maior. The blue $\star$ indicates the postition of K\,934
in \kepler. \label{f:canismaior}}
\end{figure}

\noindent K\,938, near $-$4.5,+1 in Fig.\,\ref{f:canismaior}) 
is identified by Dreyer with HIP\,31700 (= $\nu^3$\,CMa;
$V$=4.4, $\Delta$=52\farcm7) almost a degree north of it.
Our identification $\nu^2$\,CMa is brighter and much nearer.

\begin{figure}
\includegraphics[angle=270,width=\columnwidth]{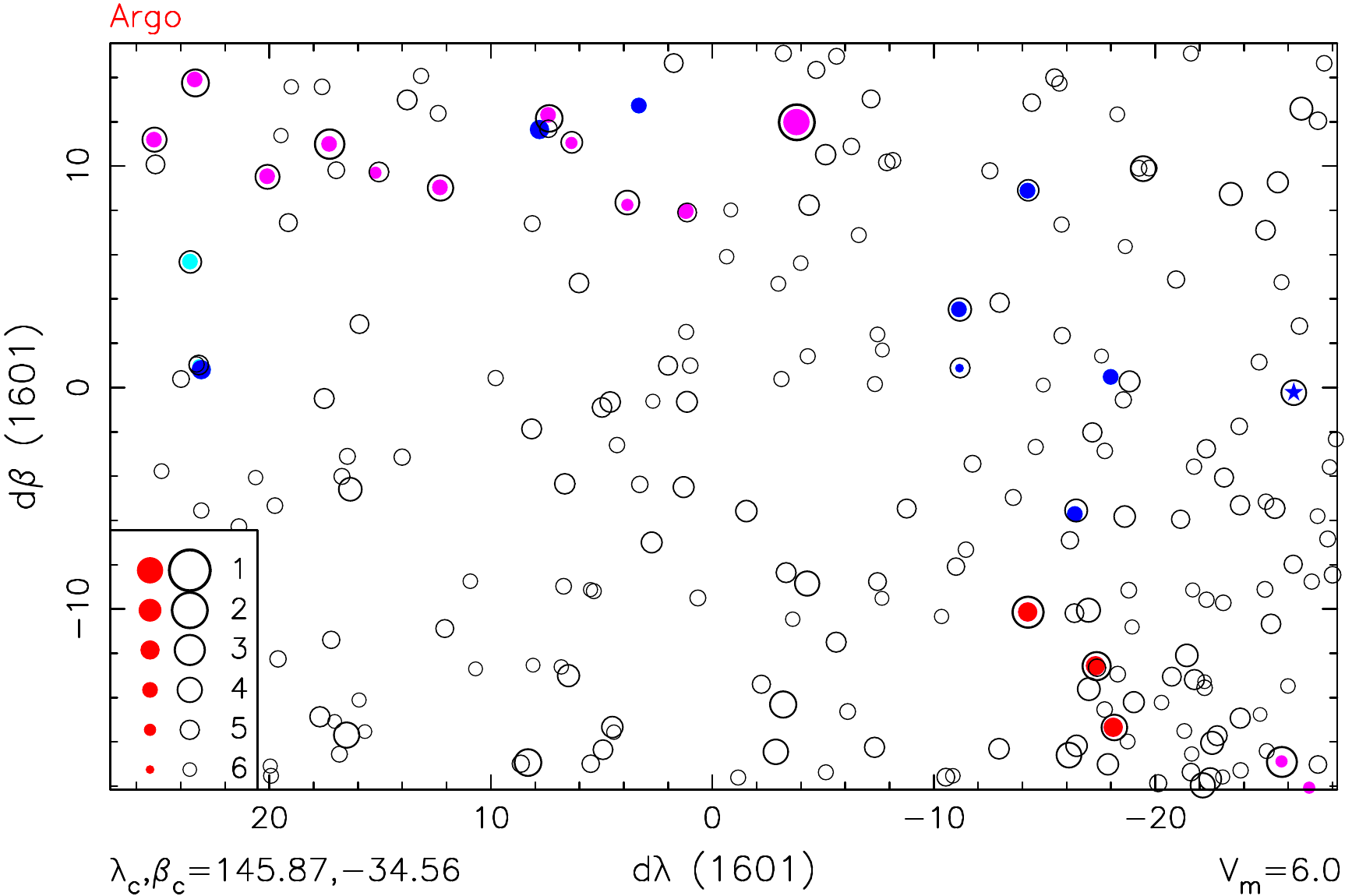}
\caption{Argo. K\,961 is near 23,1 (see also Fig.\,\ref{f:argbdetail}), its position 
as emended by Hevelius is indicated with the blue $\star$.  \label{f:argo}}
\end{figure}

\begin{figure}
\includegraphics[angle=270,width=\columnwidth]{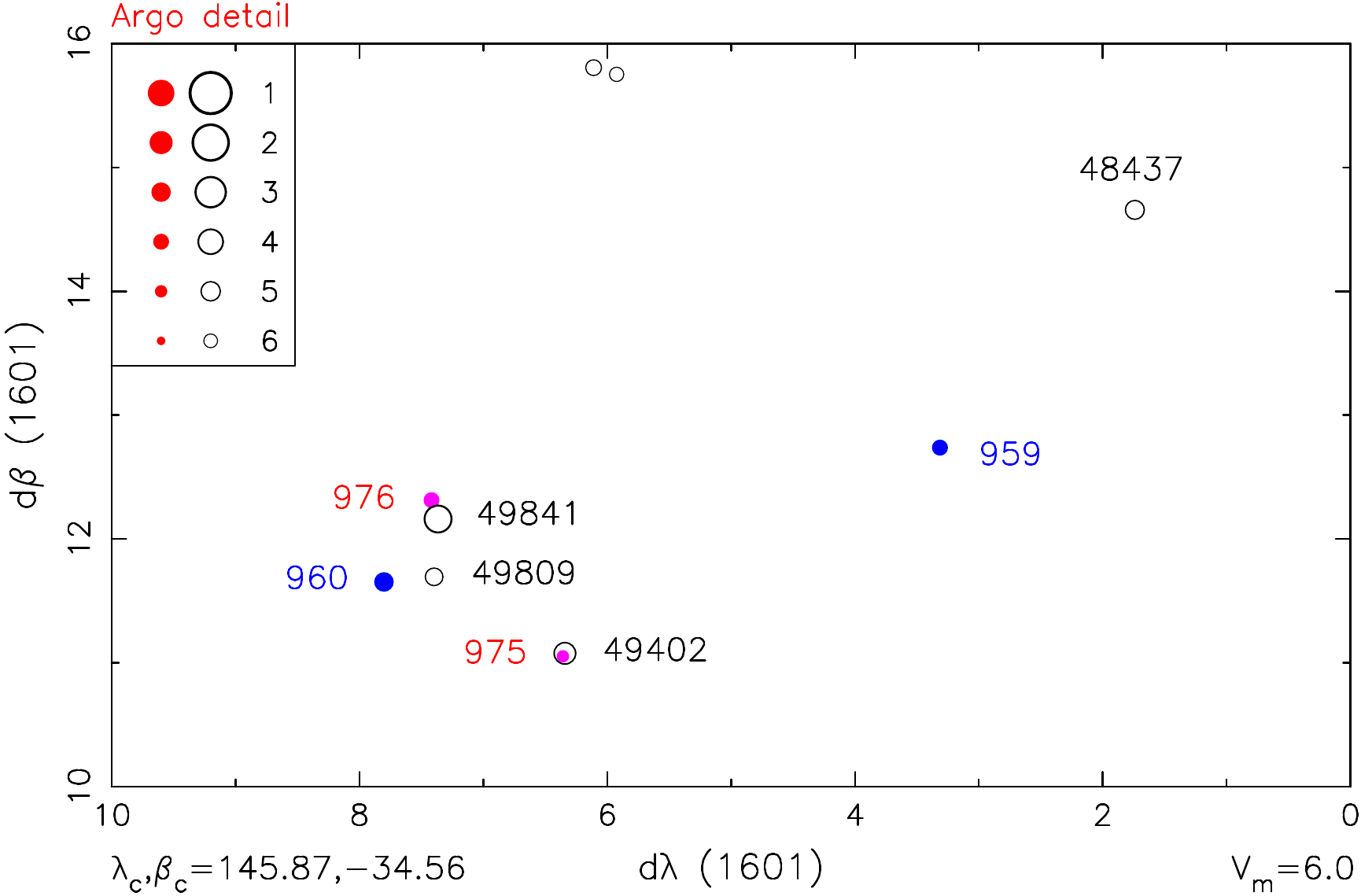}
\caption{Argo detail \label{f:argodetail}}
\end{figure}

\begin{figure}
\includegraphics[angle=270,width=\columnwidth]{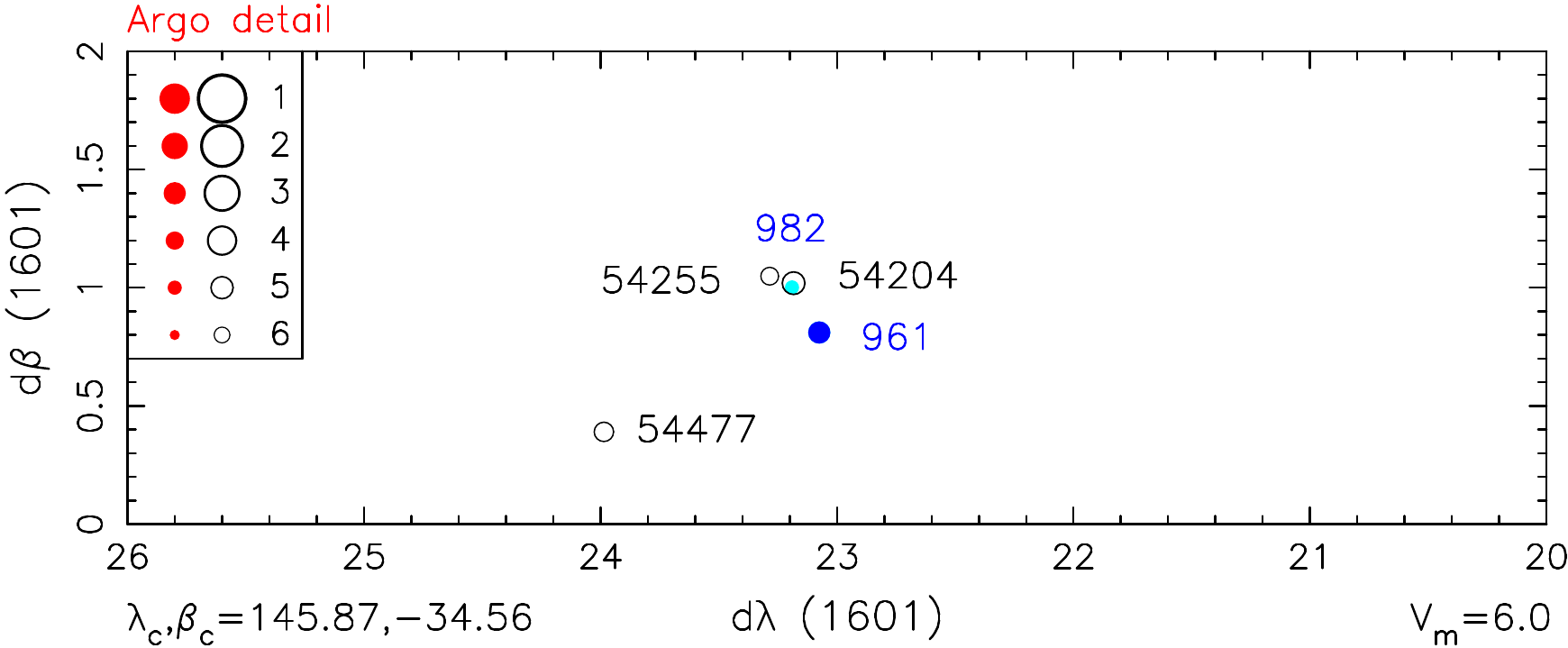}
\caption{Argo detail \label{f:argbdetail}}
\end{figure}

\noindent K\,959-961. \kepler\ annotates {\em Has tres trajecit
  Gr\"unbergerus ad finem Hydrae} (Gr\"unberger moves these three to
the end of Hydra). Figure\,\ref{f:argodetail} shows the area near
K\,959 and K\,960, which includes K\,975 and K\,976 (in Hydra). Our
preferred identification for each of these is with the nearest
HIPPARCOS star in the Figure. One problem with this is that we couple
the fainter of the pair K\,960/K\,976 to the brighter of the pair
HIP\,49809/HIP\,49841. Another problem is the description of K\,960 as
{\em Sequens earundem} (the following of the same two [sc.\ K\,959 and
  K\,960]) which appears inapt for a faint star between two brighter
ones.  Dreyer identifies both K\,960 and K\,976 with HIP\,49841, but
this has the problem that these two stars have significantly different
coordinates and magnitudes in the catalogue. The same problem arises
with the identification by Dreyer of both K\,961 and K\,982 with
9\,Crt = HIP\,54204 (see Fig.\,\ref{f:argbdetail}).  We prefer to
identify K\,961 with HIP\,54204, and the fainter K\,982 with the
fainter HIP\,54255. Hevelius and Rawlins correct the longitude of
K\,961, to $\lambda(1601)=113^\circ44\arcmin$ (by replacing $Z$=6 with
$Z$=4) which gives a close (0\farcm5) match to HIP\,37447 (=
$\alpha$\,Mon; see Fig.\,\ref{f:argo}).

K\,975, K,976: see K\,959, K\,960.

K\,982: see K\,961.

\begin{figure}
\includegraphics[angle=270,width=0.9\columnwidth]{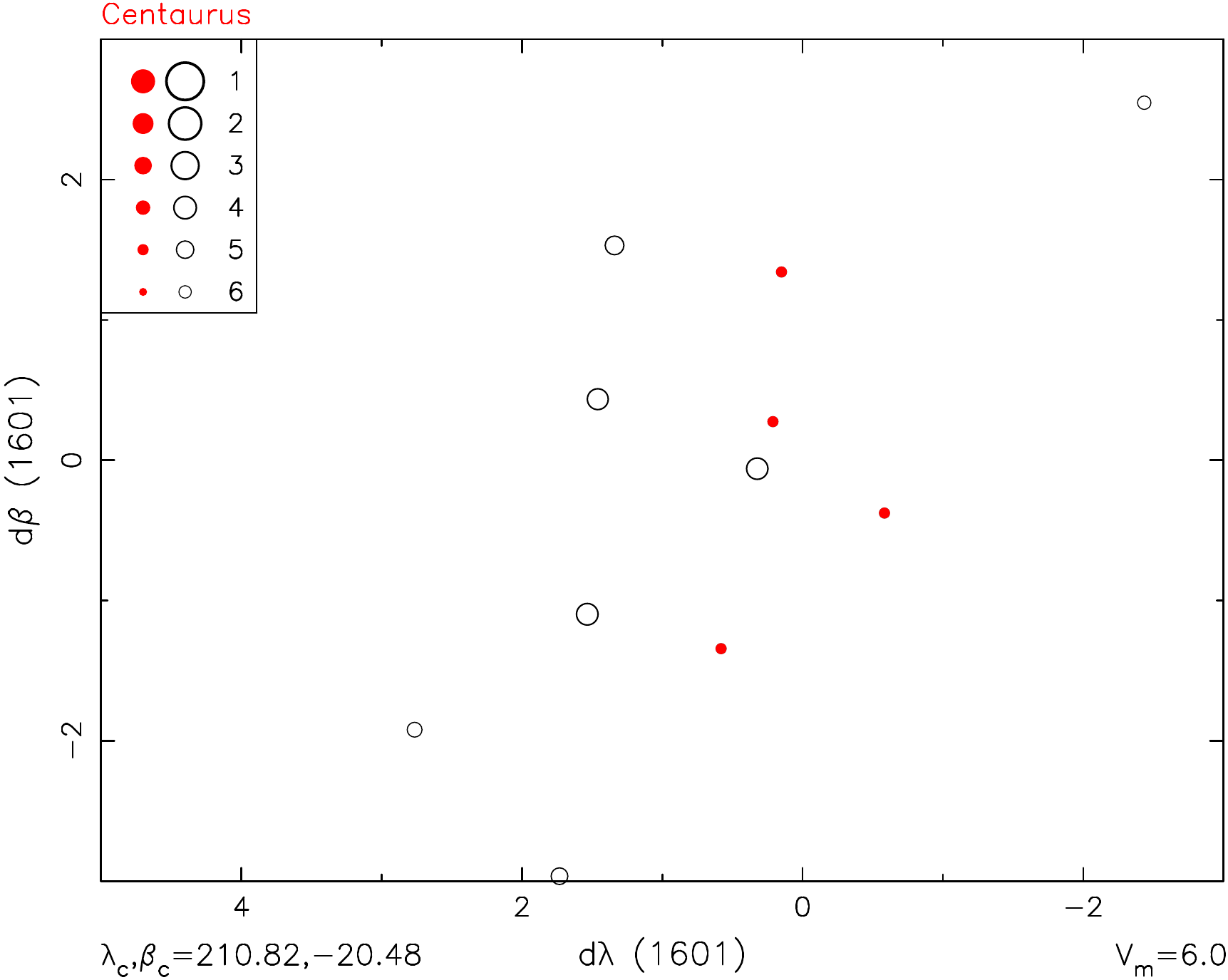}
\caption{Centaurus. The catalogue longitudes appear shifted by about a degree.
 \label{f:centaurus}}
\end{figure}

K\,1003, K\,1004 both are closest to HIP\,67153, near 0,0 in
Fig.\,\ref{f:centaurus}, but from the pattern of the whole
constellation it is seen that K\,1004 is HIP\,67669.

%% ++++++++++++++++++++++++++++++++++++++++++++++
\noindent M\,94 is almost identical to M\,102 = K\,584

\noindent M\,703 is an exact repeat of M\,701 = K\,339

\noindent M\,915 is an almost identical repeat of M\,908 = K\,908
\end{appendix}

\begin{appendix}
\section{Figures \label{s:figures}}

To illustrate and clarify our identifications we provide figures for
each constellation.  {\em It should be noted that the following
equations are used only for the Figures, i.e.\ for illustrative
purposes: to compute the angles between positions, e.g.\ to find the
nearest counterpart, we always use Eqs.\,\ref{e:angle} and
\ref{e:apprangle}.}  

In these figures the stars listed with the constellation in
\progym\ are shown red, those added in \kepler\ in blue, and other
stars listed in \progym\ and \kepler\ are shown light-red and
light-blue, respectively.  For all stars we use positions from
\keplere. In yellow we indicate stars from {\em Secunda Classis}
(which are discussed in Verbunt \&\ Van Gent, Paper\,III, in preparation).

To minimize deformation of the
constellations, we determine the center of the constellation
$\lambda_c,\beta_c$ from the extremes in $\lambda$ and $\beta$,
compute the rotation matrix
\begin{equation}
 {\cal{R}} \equiv \left( \begin{array}{ccc}
 \cos\lambda_c\cos\beta_c &  \sin\lambda_c\cos\beta_c & \sin\beta_c \\
-\sin\lambda_c            & \cos\lambda_c             & 0           \\
-\cos\lambda_c\sin\beta_c & -\sin\lambda_c\sin\beta_c & \cos\beta_c
                \end{array} \right)
\end{equation}
which moves this center to $(x,y,z)=(1,0,0)$, and then apply this
rotation to each of the stellar positions $\lambda_i,\beta_i$:
\begin{equation}
\left( \begin{array}{c} x \\ y \\ z \end{array} \right)
 = {\cal{R}} 
\left( \begin{array}{c}
  \cos\lambda_i\cos\beta_i \\ \sin\lambda_i\cos\beta_i \\ \sin\beta_i
                \end{array} \right)  
\end{equation}
The resulting $y,z$ values correspond roughly to differences in
longitude and latitude, exact at the center $\lambda_c,\beta_c$ and
increasingly deformed away from the center. We plot the rotated
positions of the stars in \keplere\ as $d\lambda\equiv y$ and
$d\beta\equiv z$ with filled circles.  The same rotation matrix
${\cal R}$ is applied to all stars down to a magnitude limit
$V_\mathrm{m}$ (usually $V_m=6.0$) from the {\em Hipparcos Catalogue}
and those in the field of view are plotted as open circles.  The
symbol sizes are determined from the magnitudes as indicated in the
legenda. The values used for $\lambda_\mathrm{c}$, $\beta_\mathrm{c}$
and $V_\mathrm{m}$ are indicated with each Figure.

Where necessary we show enlarged detail Figures; for easy comparison with
the Figures showing the whole constellation, these detail Figures use the
same rotation center (and thus rotation matrix).

\clearpage

\begin{figure}
\includegraphics[angle=270,width=\columnwidth]{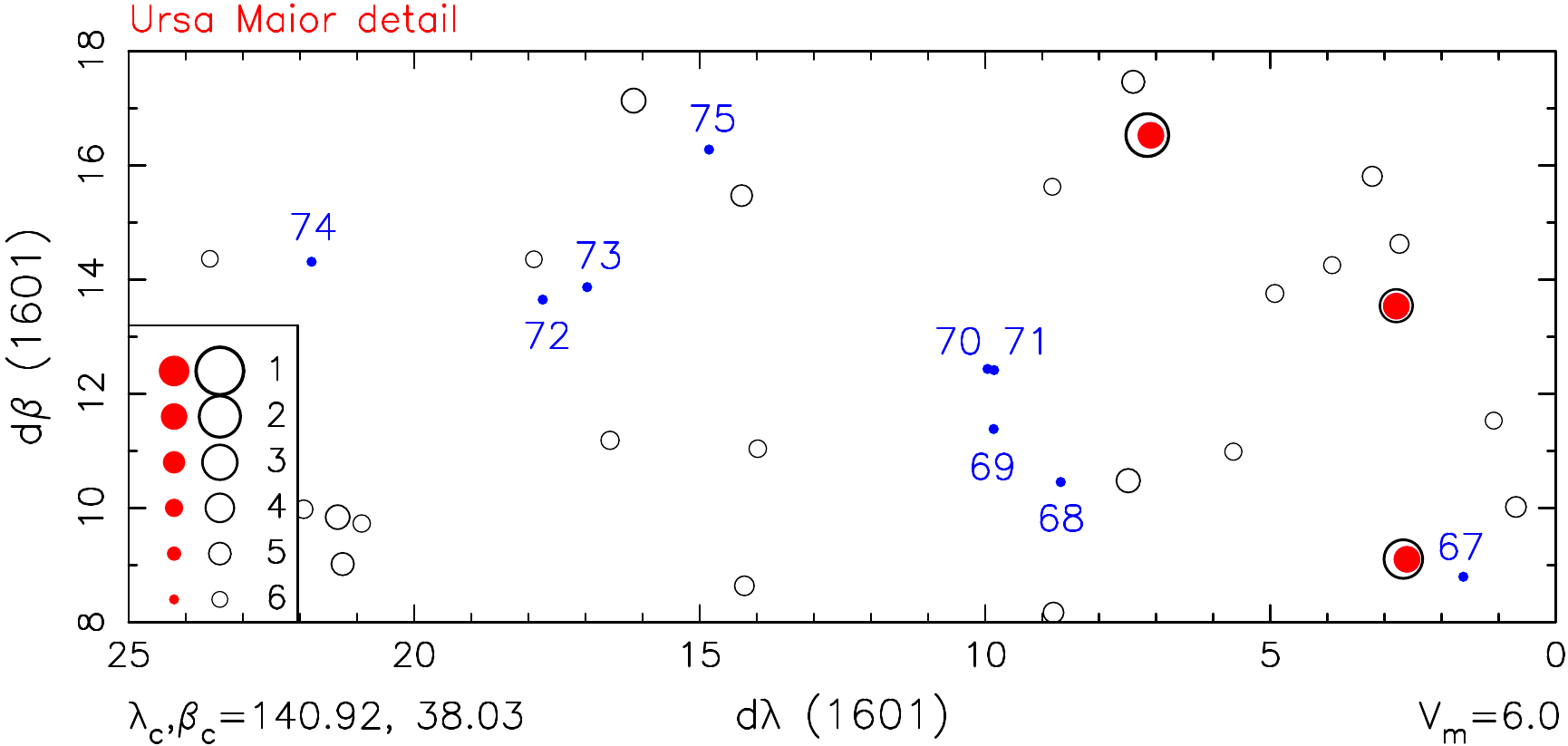}
\caption{Ursa Maior detail. No good counterparts are found for
K\,67-75 even if the search is extended down to $V$=6.5. Rawlins (1993) notes
that Brahe's positions in this area are in `extreme confusion'. Compare with
Fig.\,\ref{f:ursamaior}. \label{f:umadetail}}
\end{figure}

\begin{figure}
\centerline{\includegraphics[angle=270,width=0.9\columnwidth]{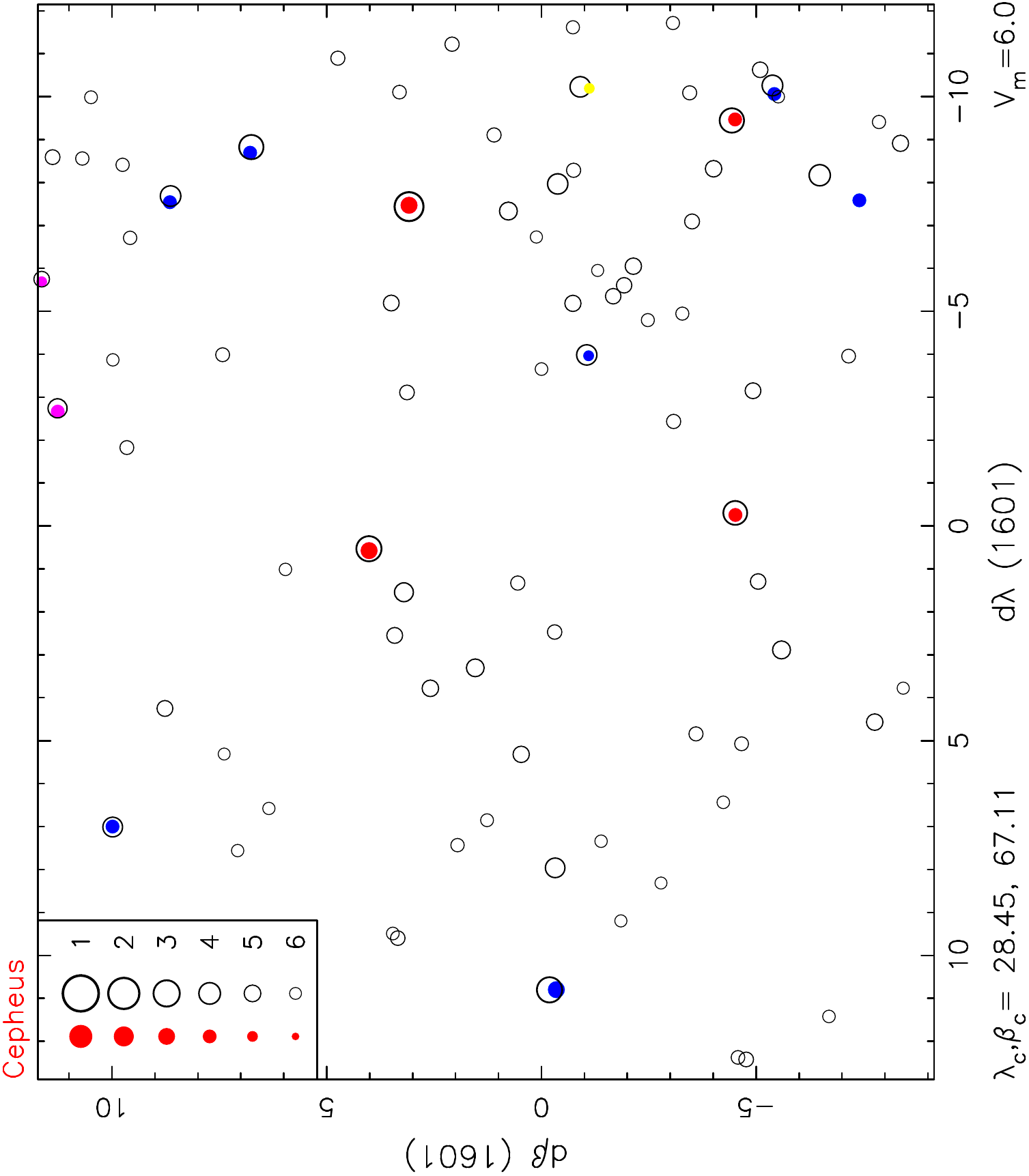}}
\caption{Cepheus.
 \label{f:cepheus}}
\end{figure}

\begin{figure}
\centerline{\includegraphics[angle=270,width=0.9\columnwidth]{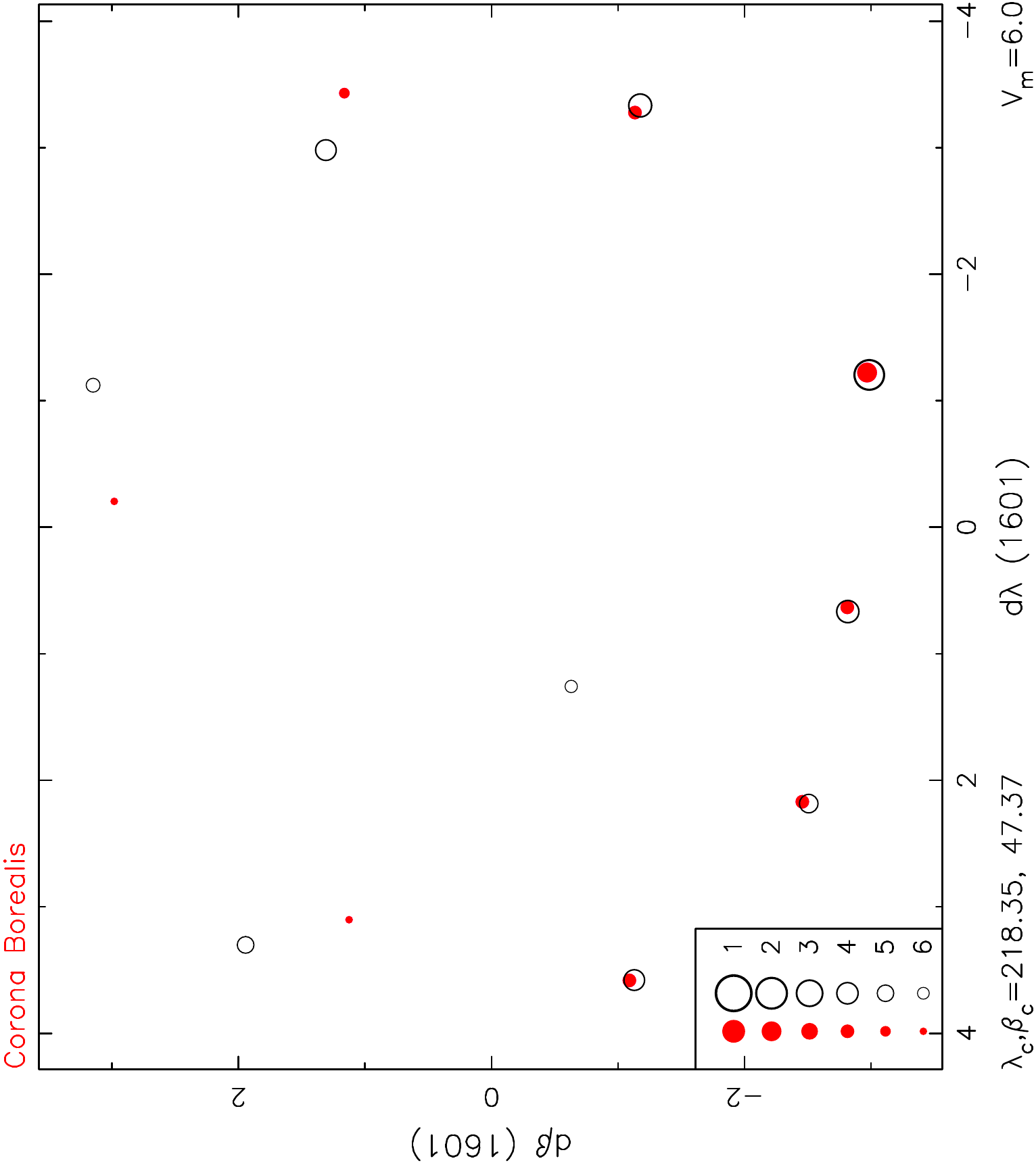}}
\caption{
Corona Borealis. \label{f:corbor}}
\end{figure}

\clearpage

\clearpage

\begin{figure}
\includegraphics[angle=270,width=\columnwidth]{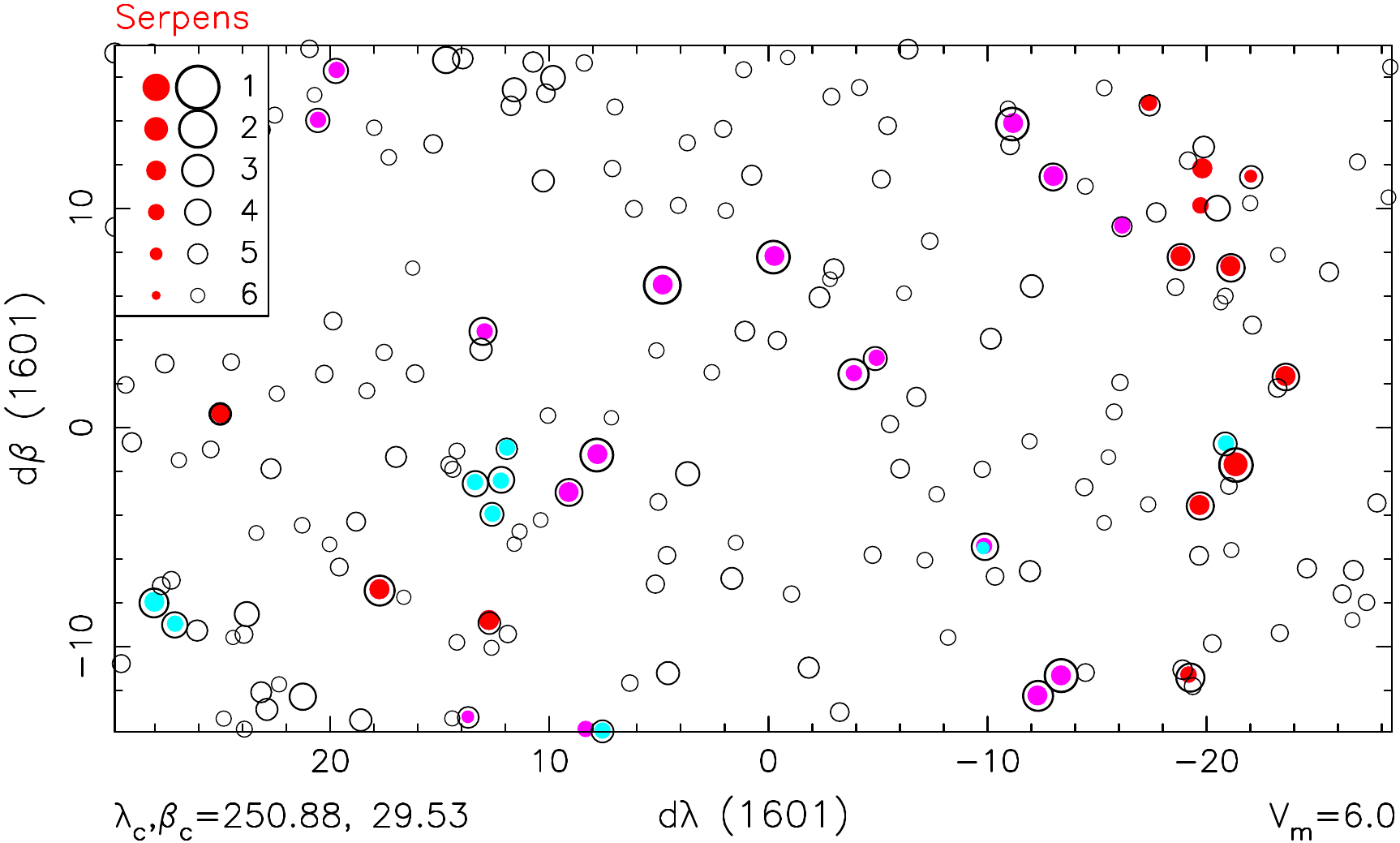}
\caption{Serpens. \label{f:serpens}}
\end{figure}

\begin{figure}
\centerline{\includegraphics[angle=270,width=0.9\columnwidth]{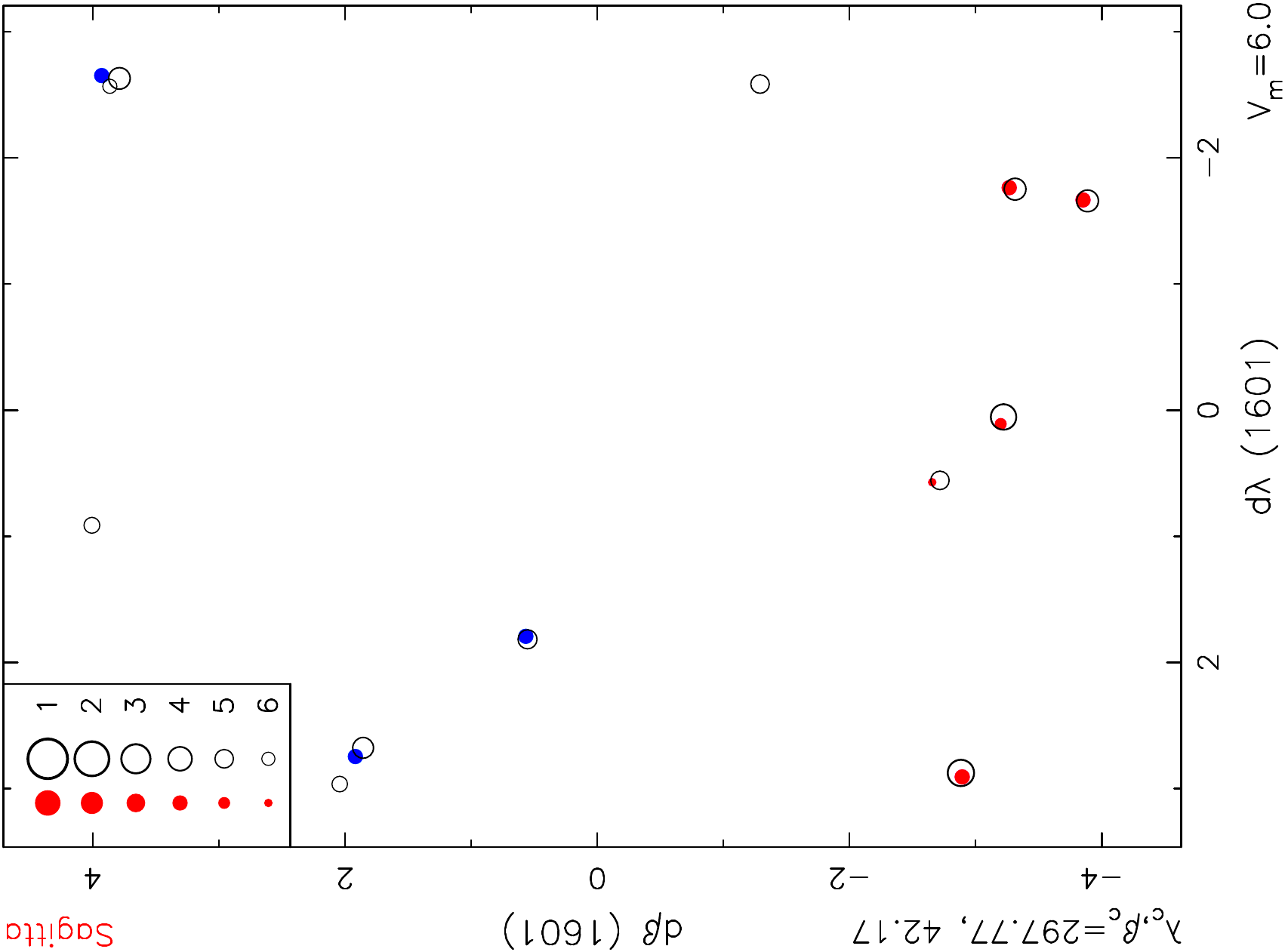}}
\caption{Sagitta.
\label{f:sagitta}}
\end{figure}

\clearpage

\begin{figure}
\centerline{\includegraphics[angle=270,width=0.9\columnwidth]{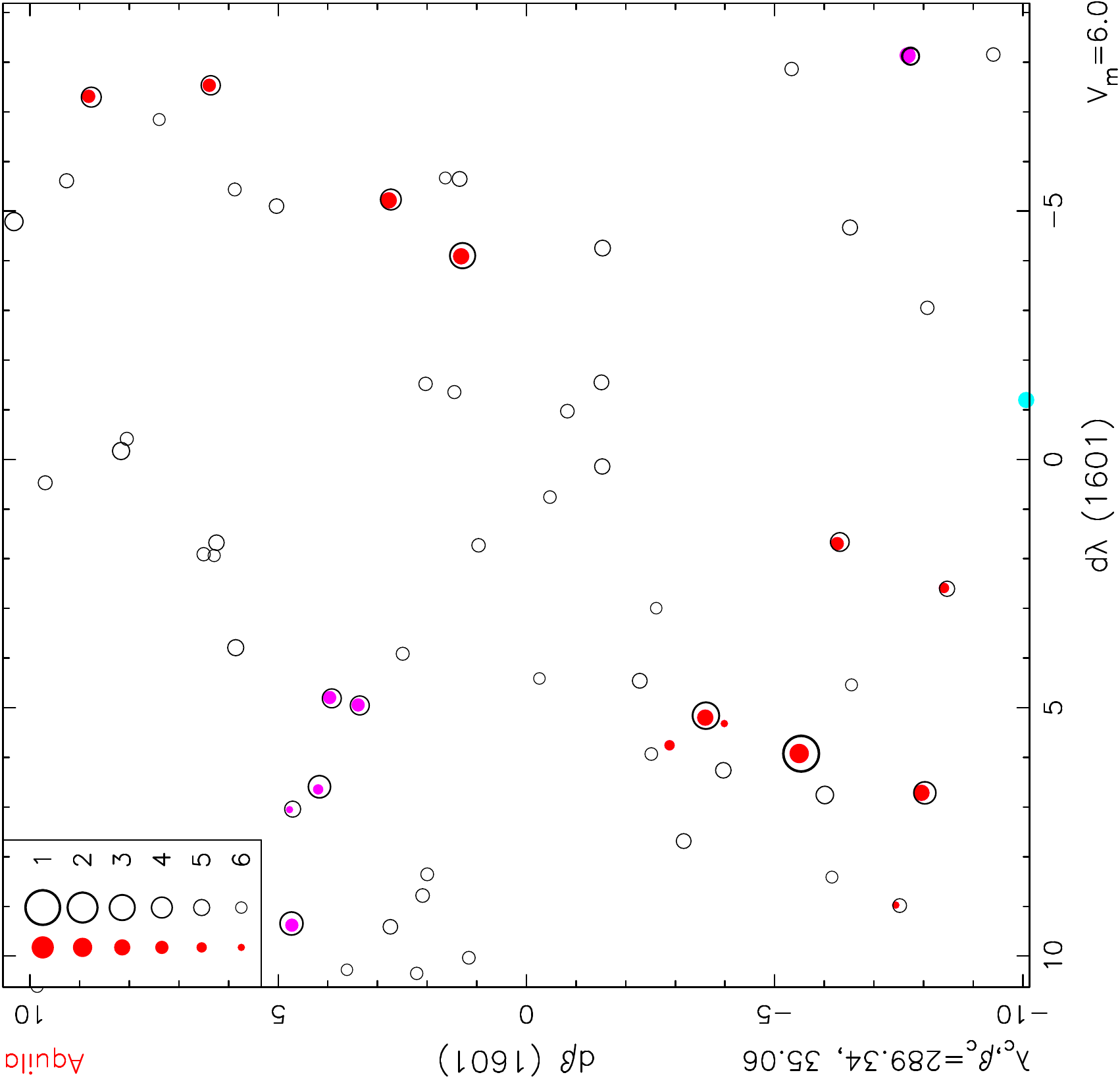}}
\caption{Aquila.
\label{f:aquila}}
\end{figure}

\begin{figure}
\centerline{\includegraphics[angle=270,width=0.9\columnwidth]{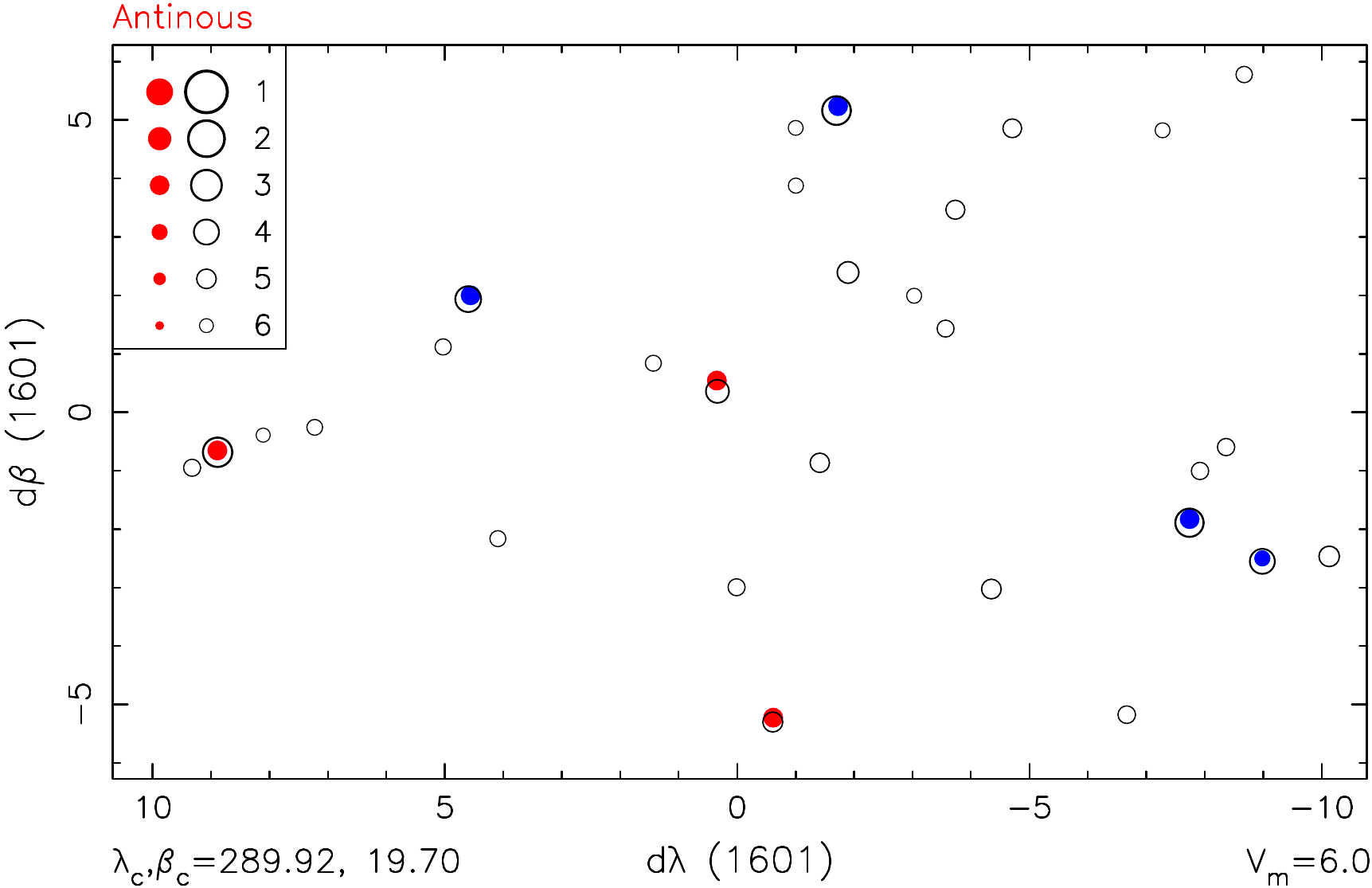}}
\caption{Antinous.
\label{f:antinous}}
\end{figure}

\begin{figure}
\centerline{\includegraphics[angle=270,width=0.95\columnwidth]{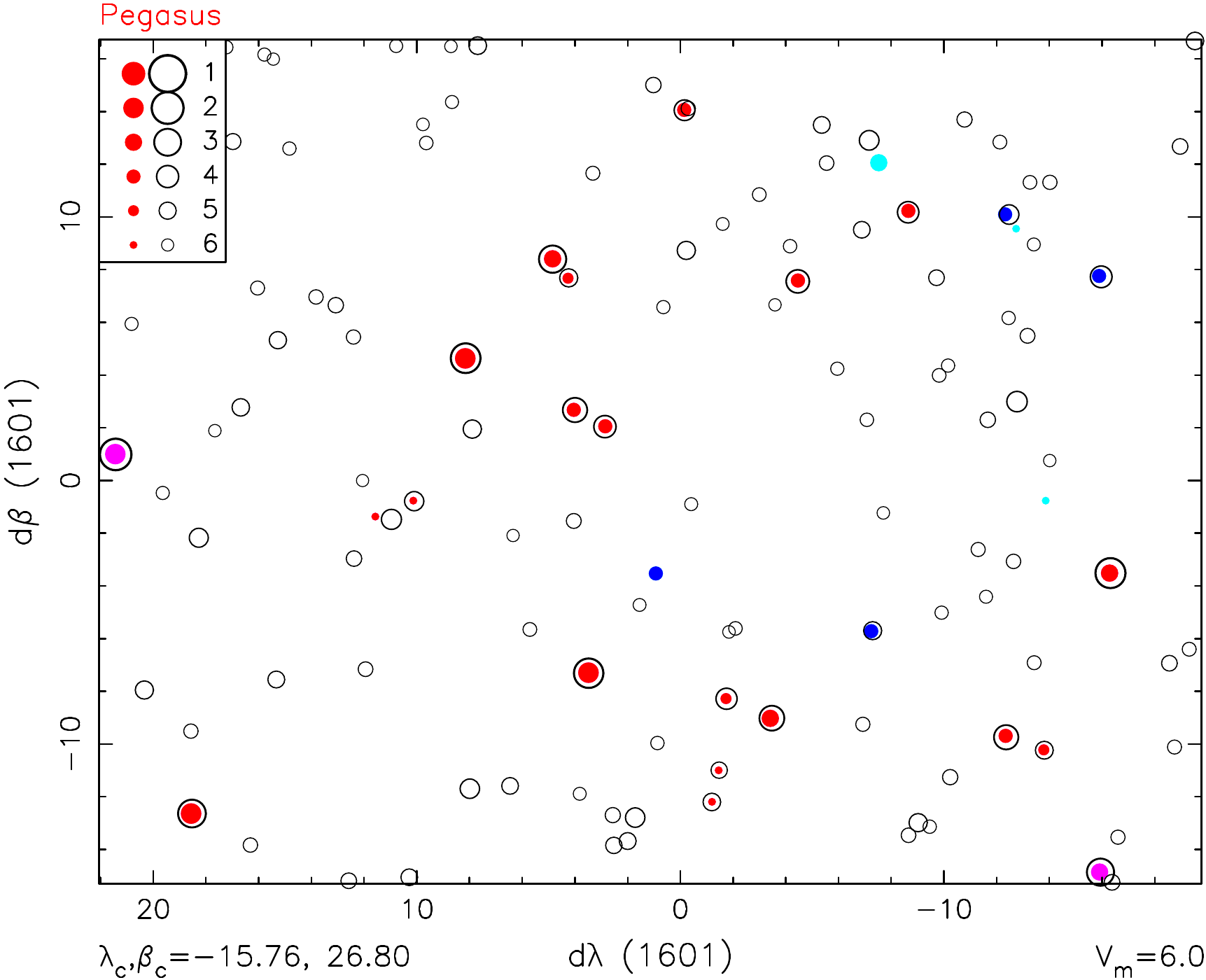}}
\caption{Pegasus.
\label{f:pegasus}}
\end{figure}

\clearpage 

\begin{figure}
\includegraphics[angle=270,width=0.9\columnwidth]{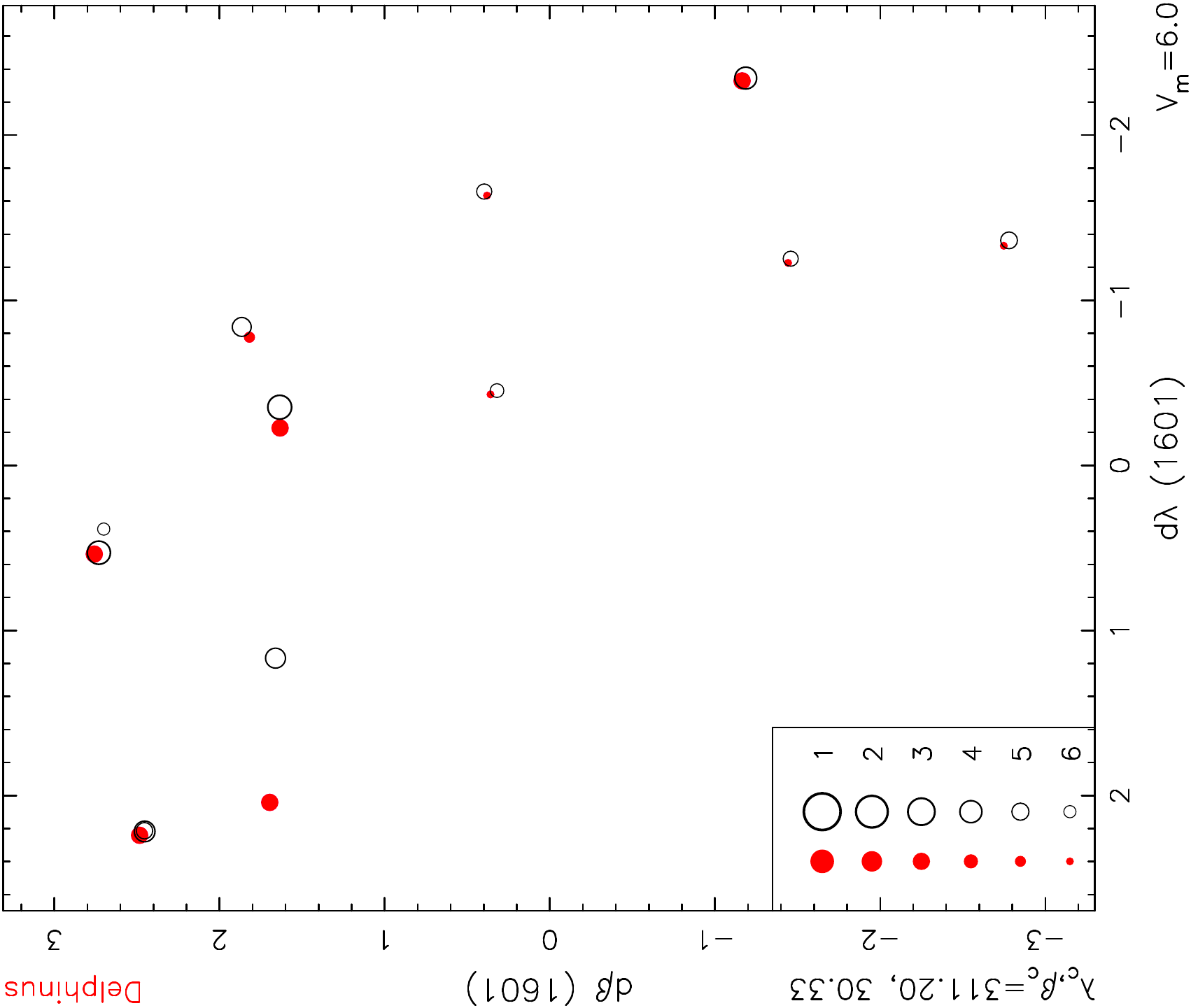}
\caption{Delphinus. 
\label{f:delphinus}}
\end{figure}

\begin{figure}
\centerline{\includegraphics[angle=270,width=0.6\columnwidth]{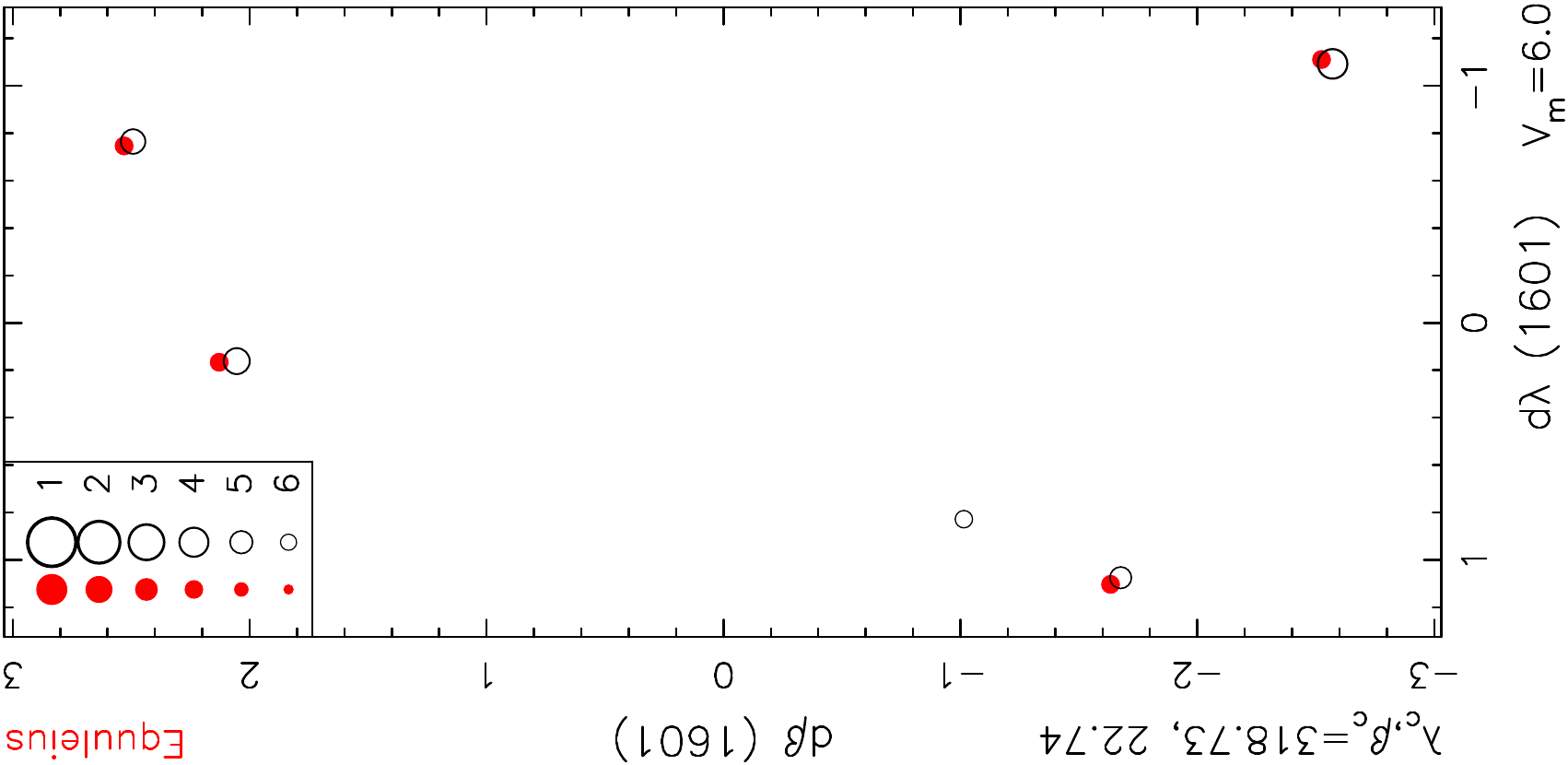}}
\caption{Equuleus. 
\label{f:equuleus}}
\end{figure}

\begin{figure}
\includegraphics[angle=270,width=\columnwidth]{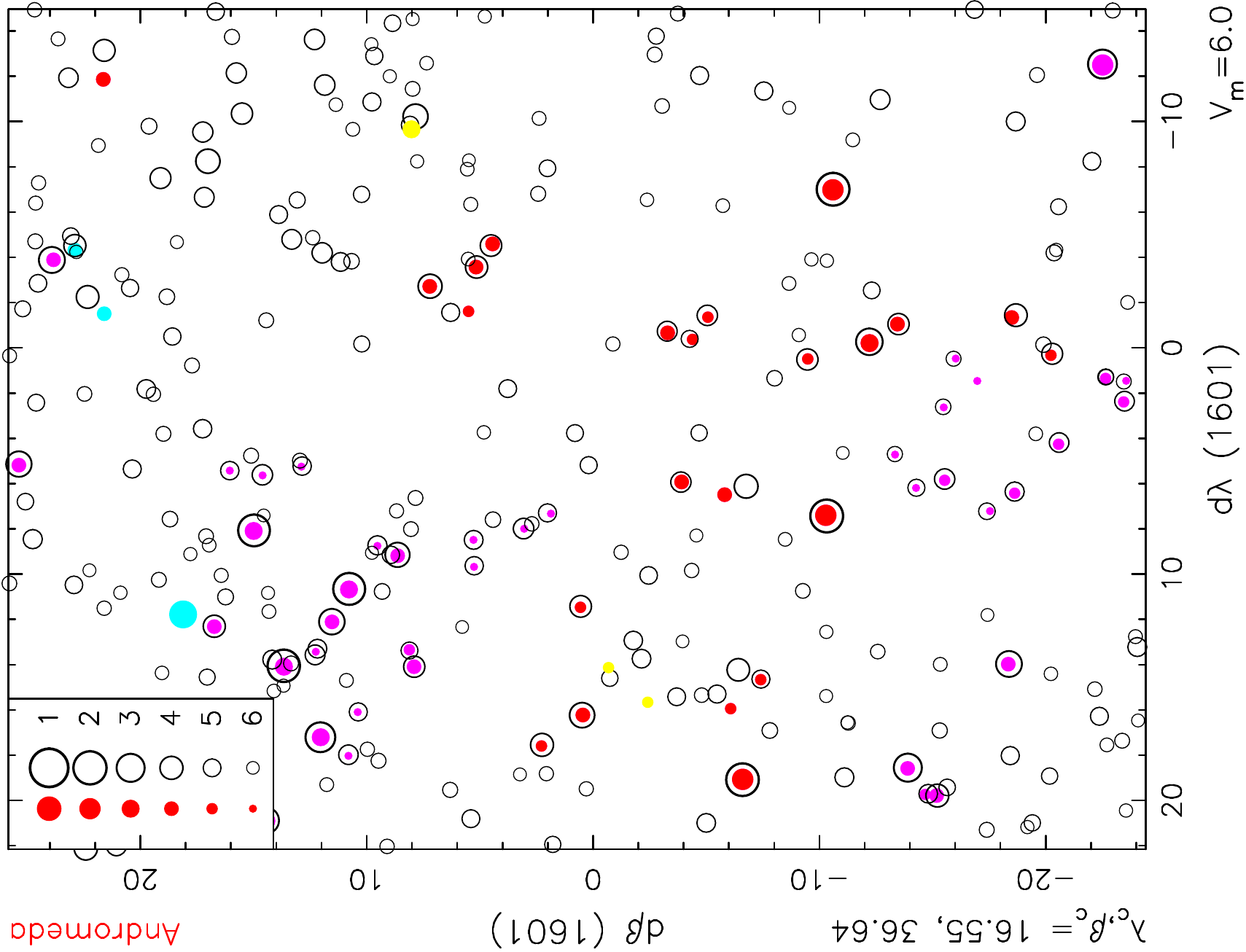}
\caption{Andromeda.
\label{f:andromeda}}
\end{figure}

\clearpage 

\begin{figure}
\includegraphics[angle=270,width=\columnwidth]{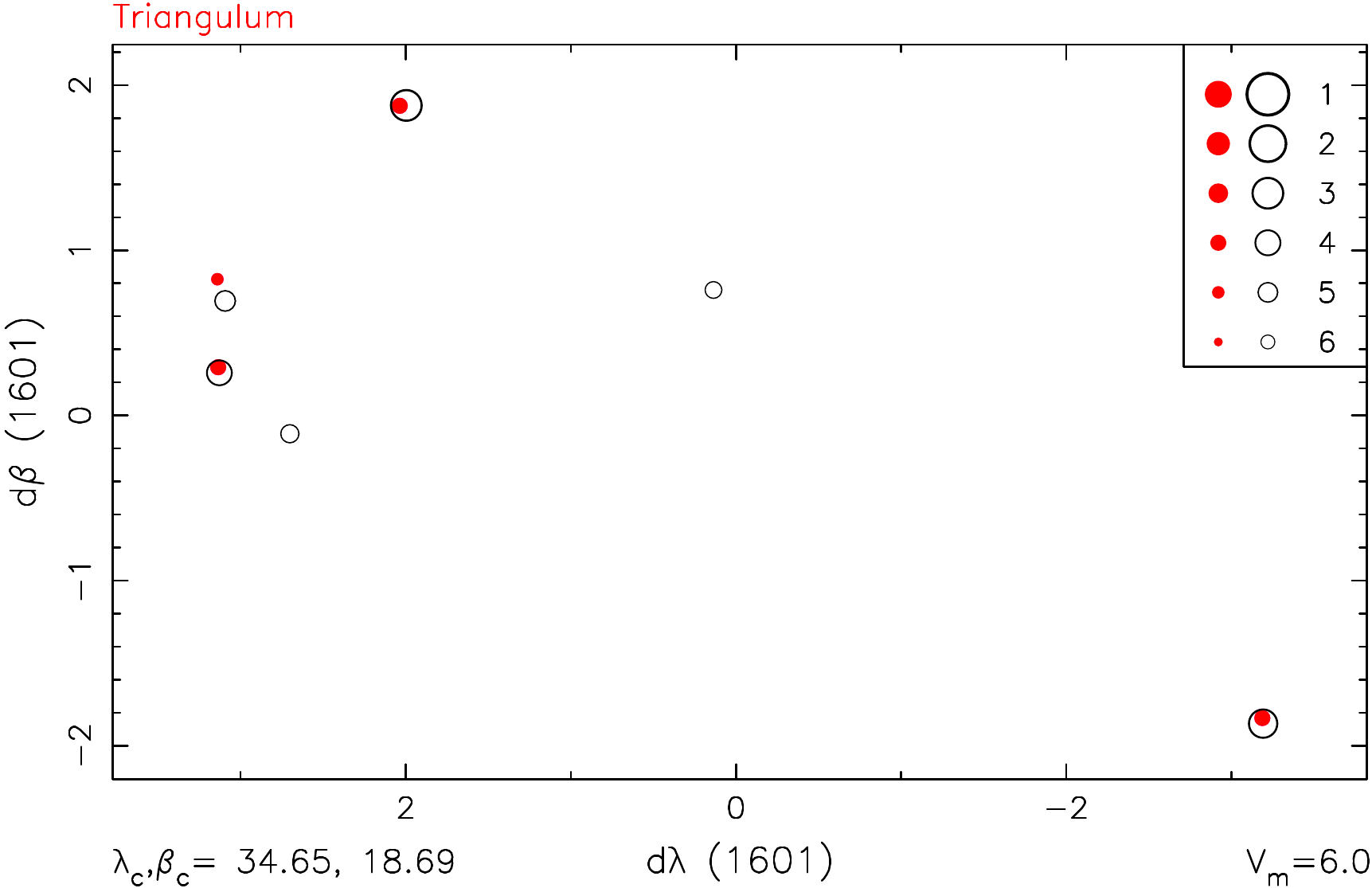}
\caption{Triangulum
\label{f:triangulum}}
\end{figure}

\begin{figure}
\includegraphics[angle=270,width=\columnwidth]{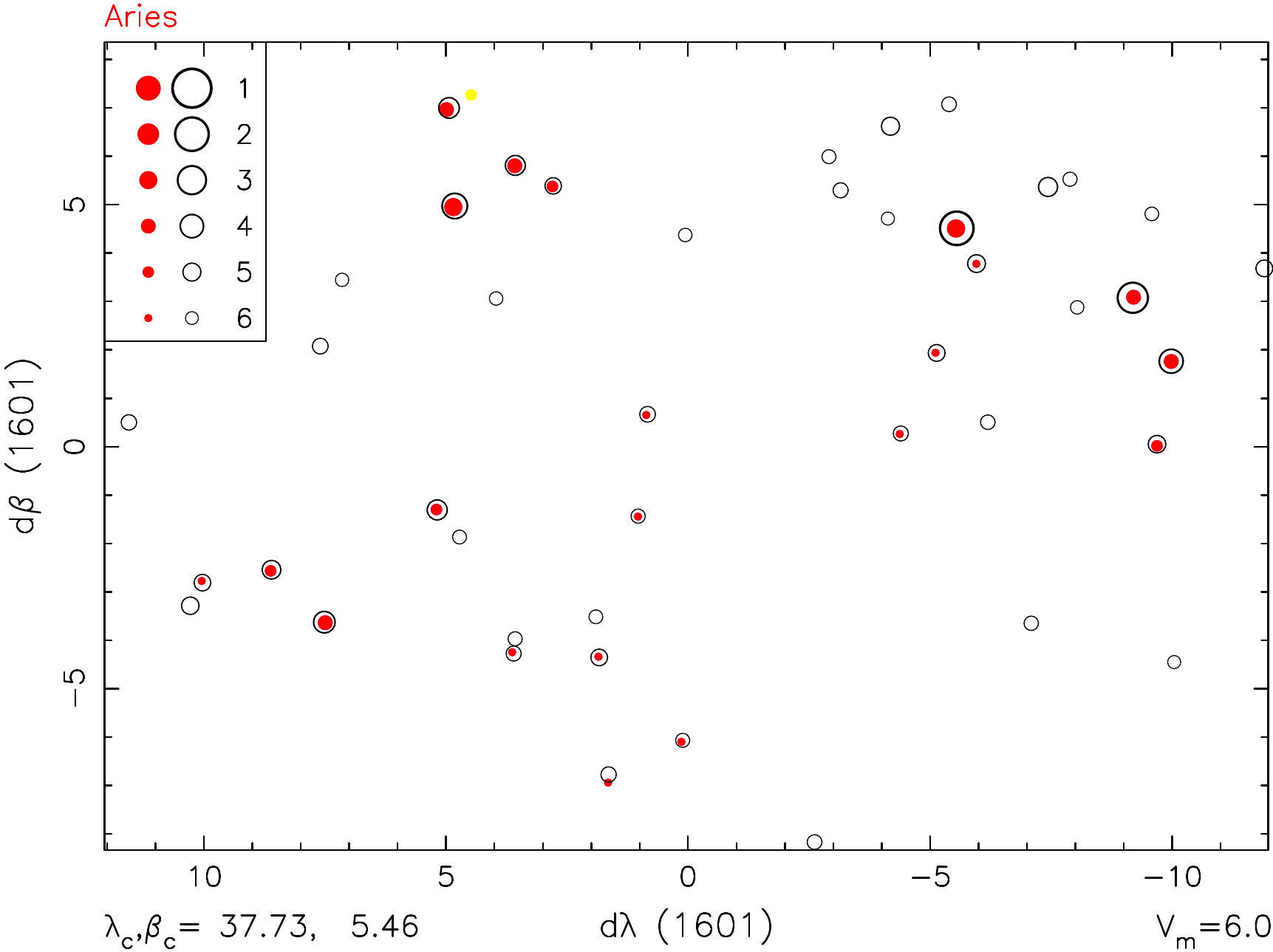}
\caption{Aries. \label{f:aries}}
\end{figure}

\clearpage

\begin{figure}
\includegraphics[angle=270,width=\columnwidth]{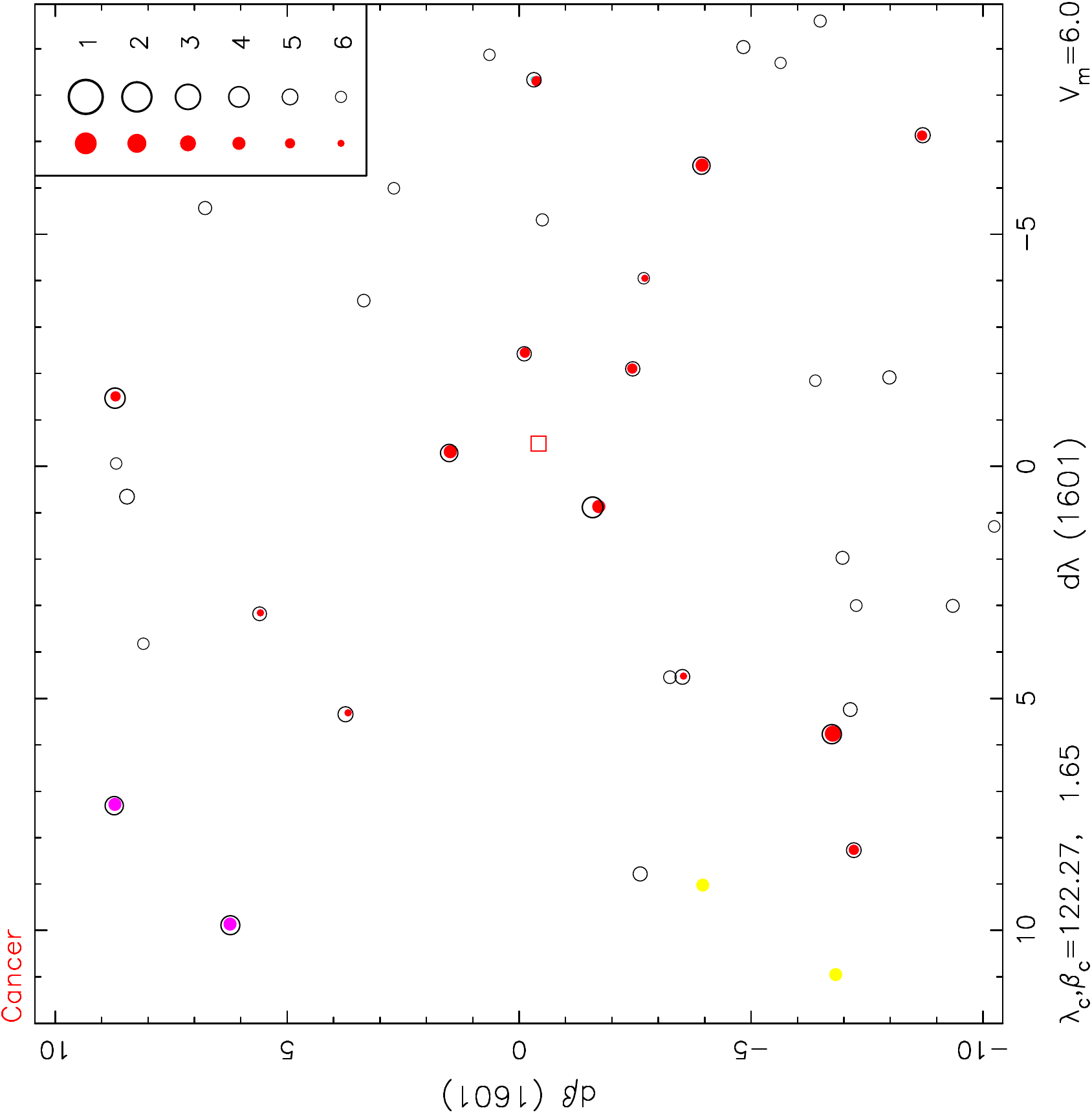}
\caption{Cancer. The red square indicates K\,577, Praesepe. \label{f:cancer}}
\end{figure}

\begin{figure}
\centerline{\includegraphics[angle=270,width=0.8\columnwidth]{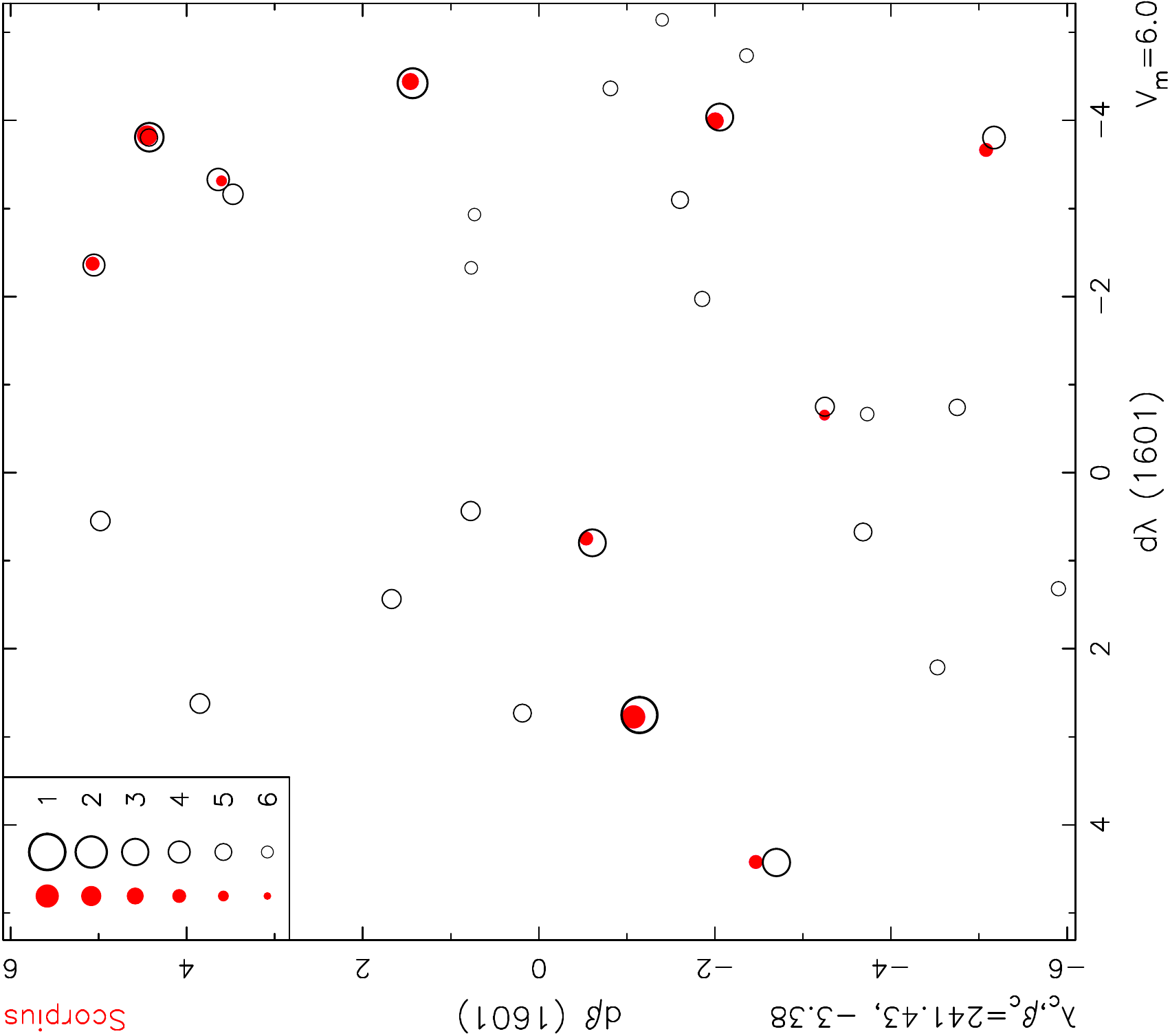}}
\caption{Scorpius. Only the northern stars are in Brahe's catalogue.
\label{f:scorpius}}
\end{figure}

\begin{figure}
\includegraphics[angle=270,width=\columnwidth]{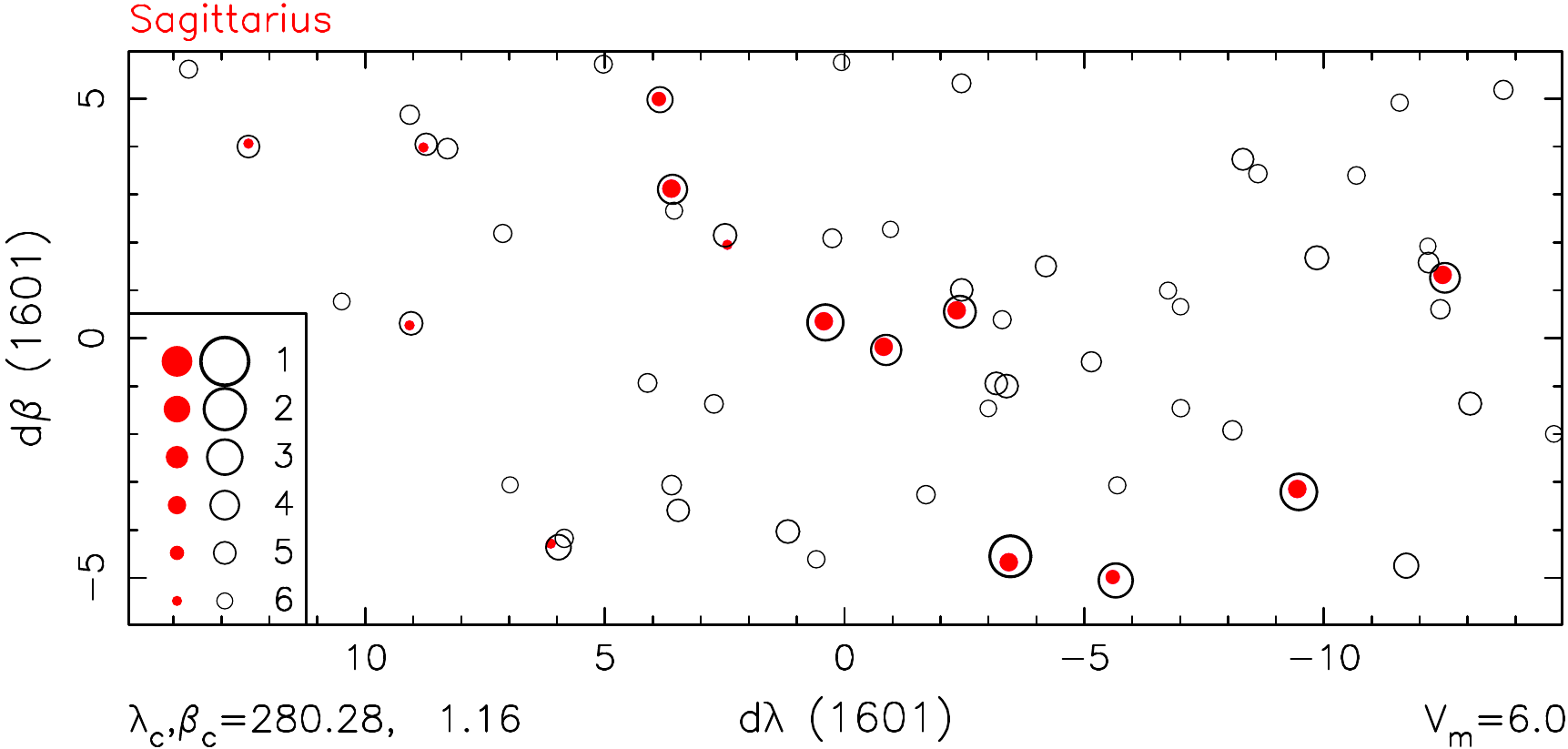}
\caption{Sagittarius. \label{f:sagittarius}}
\end{figure}

\clearpage

\begin{figure}
\includegraphics[angle=270,width=\columnwidth]{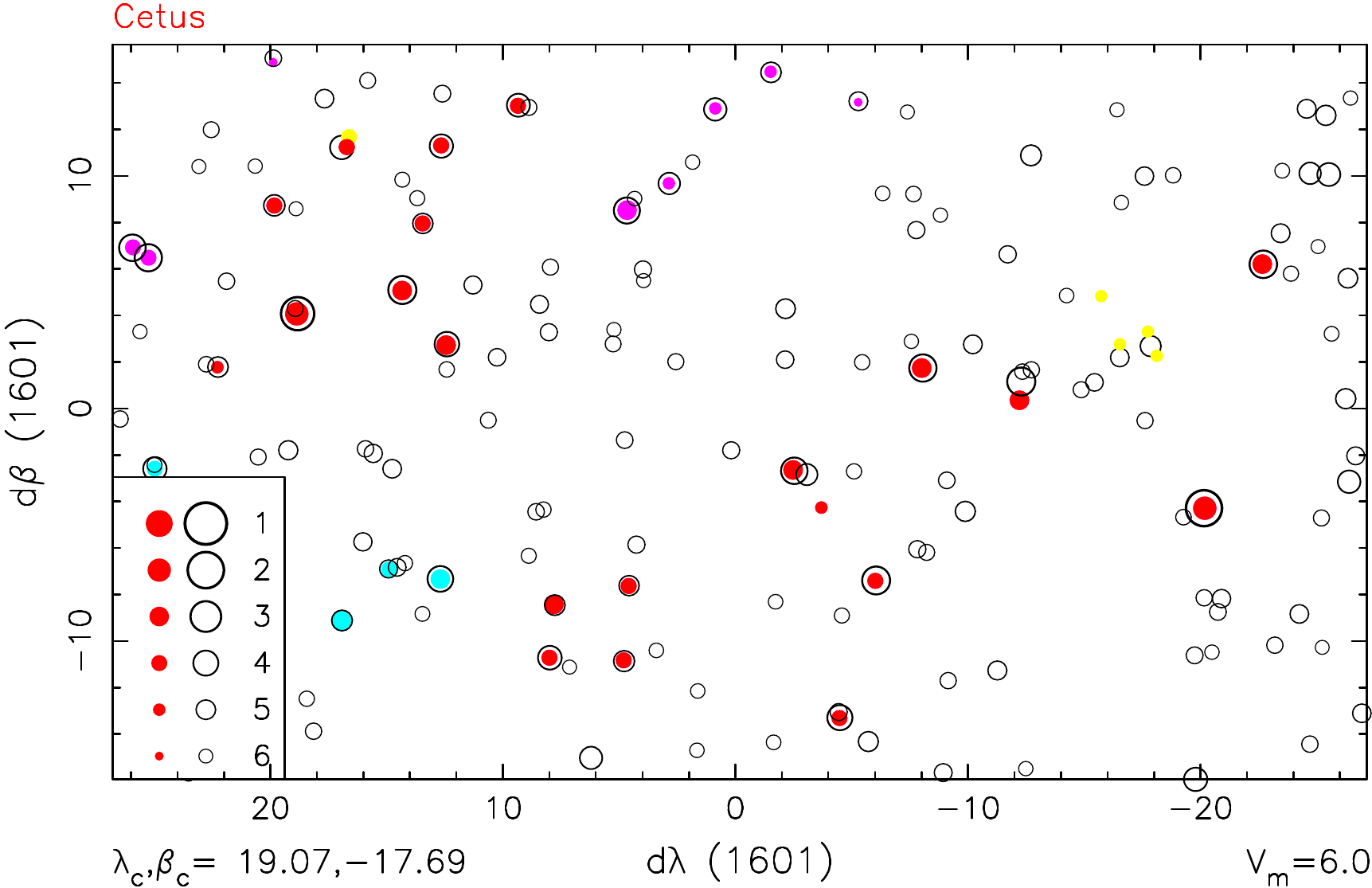}
\caption{Cetus \label{f:cetus}}
\end{figure}

\begin{figure}
\includegraphics[angle=270,width=\columnwidth]{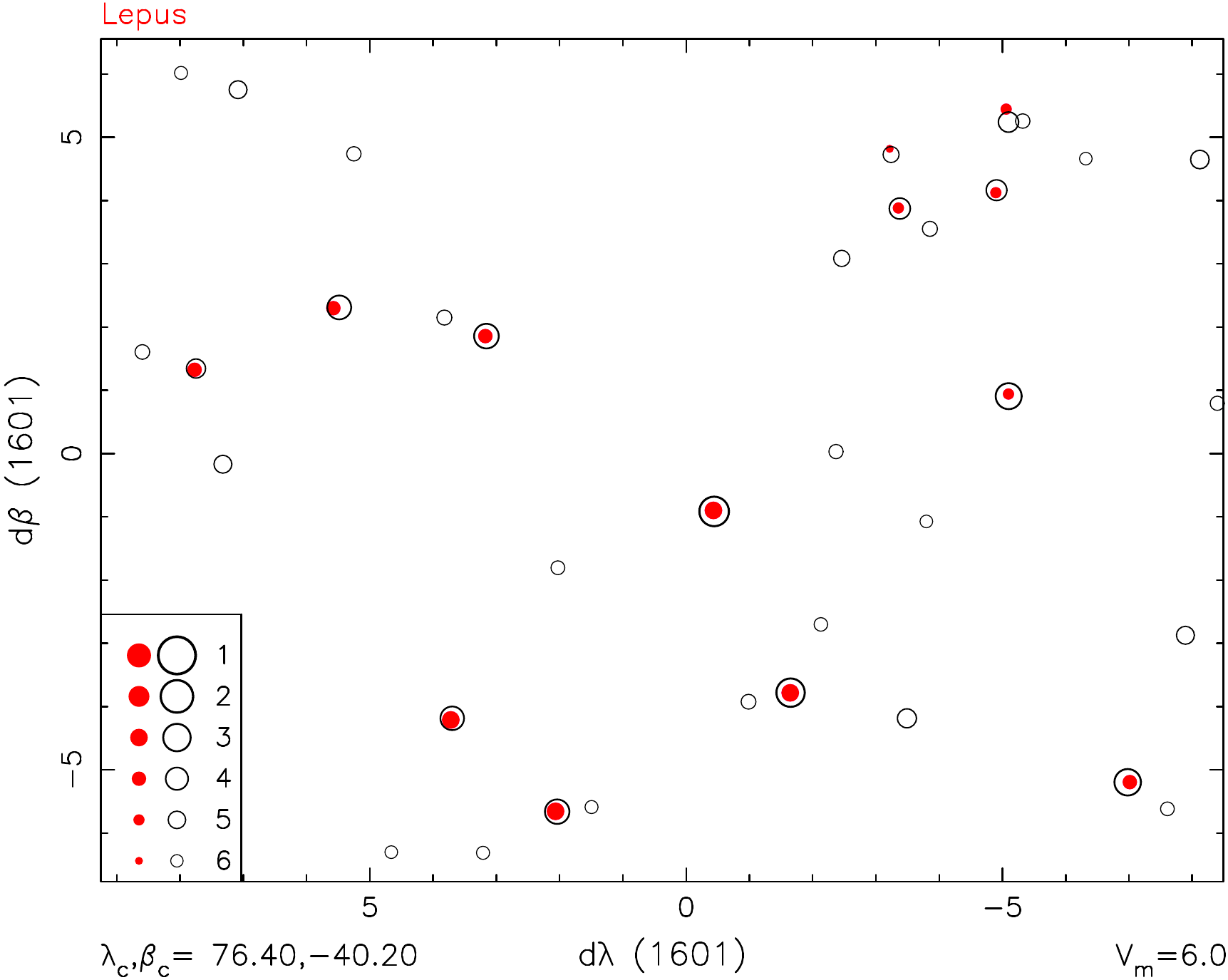}
\caption{Lepus \label{f:lepus}}
\end{figure}

\clearpage

\begin{figure}
\includegraphics[angle=270,width=\columnwidth]{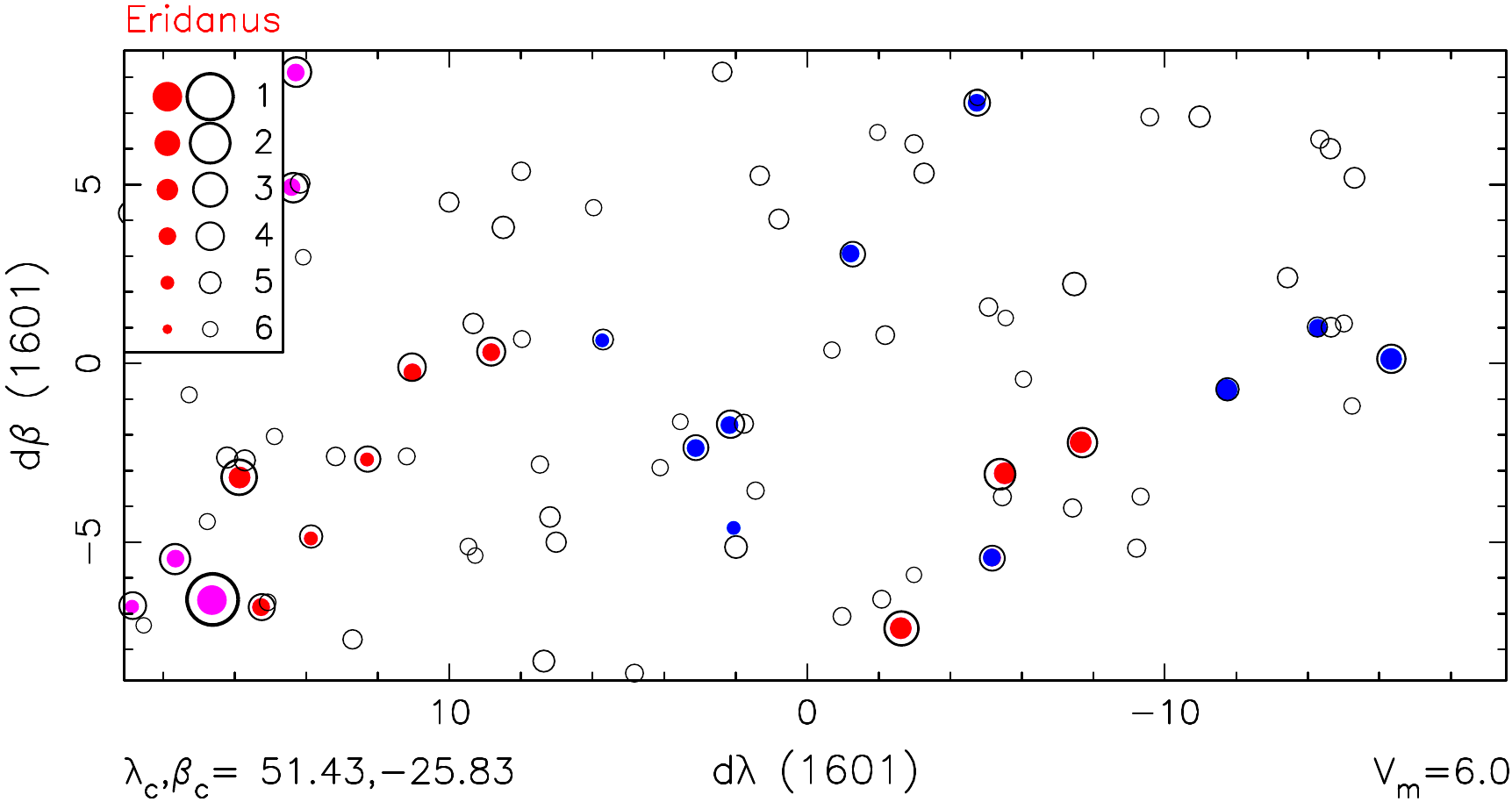}
\caption{Eridanus \label{f:eridanus}}
\end{figure}

\begin{figure}
\centerline{\includegraphics[angle=270,width=0.9\columnwidth]{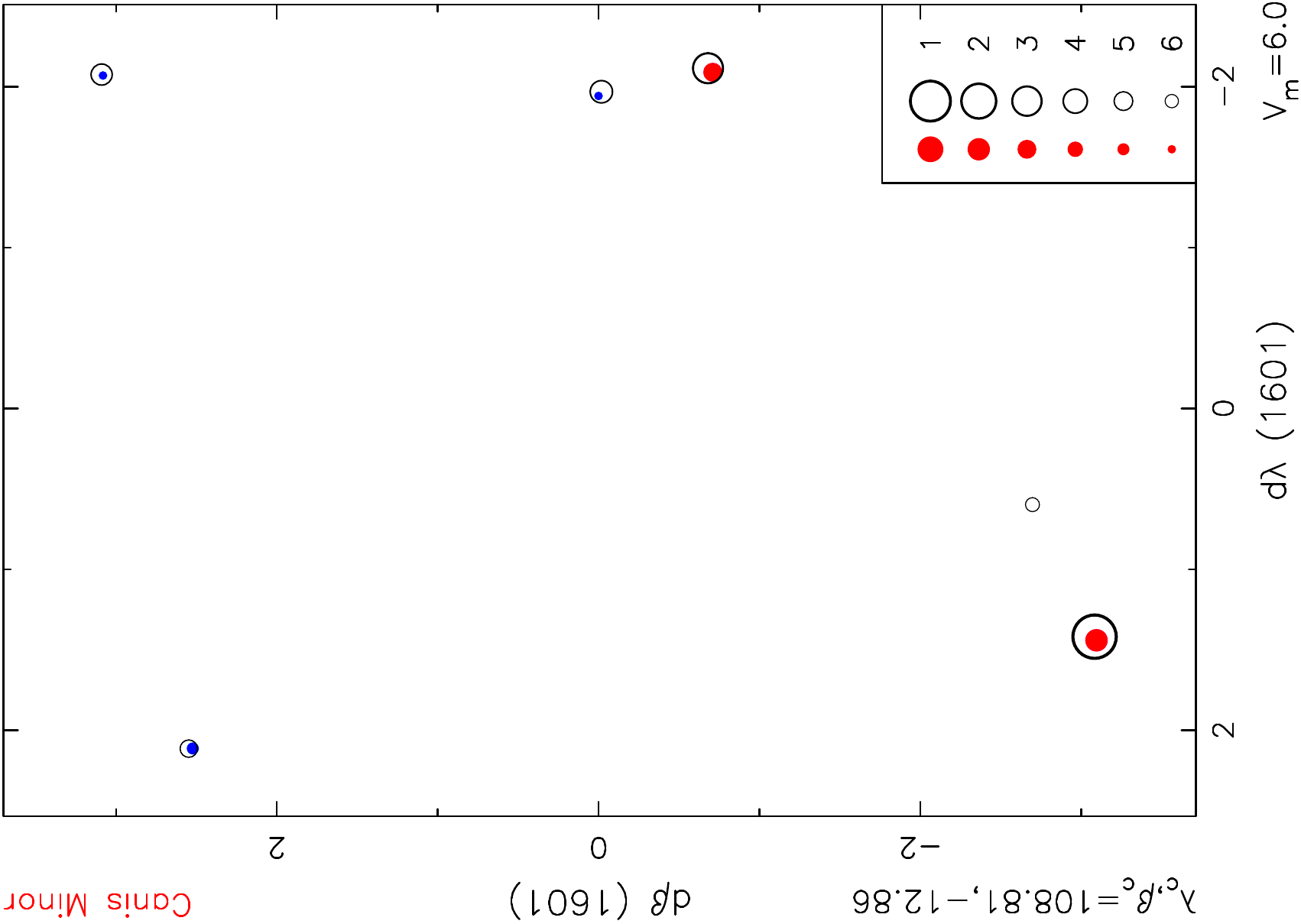}}
\caption{Canis Minor. \label{f:canisminor}}
\end{figure}

\clearpage

\begin{figure}
\includegraphics[angle=270,width=\columnwidth]{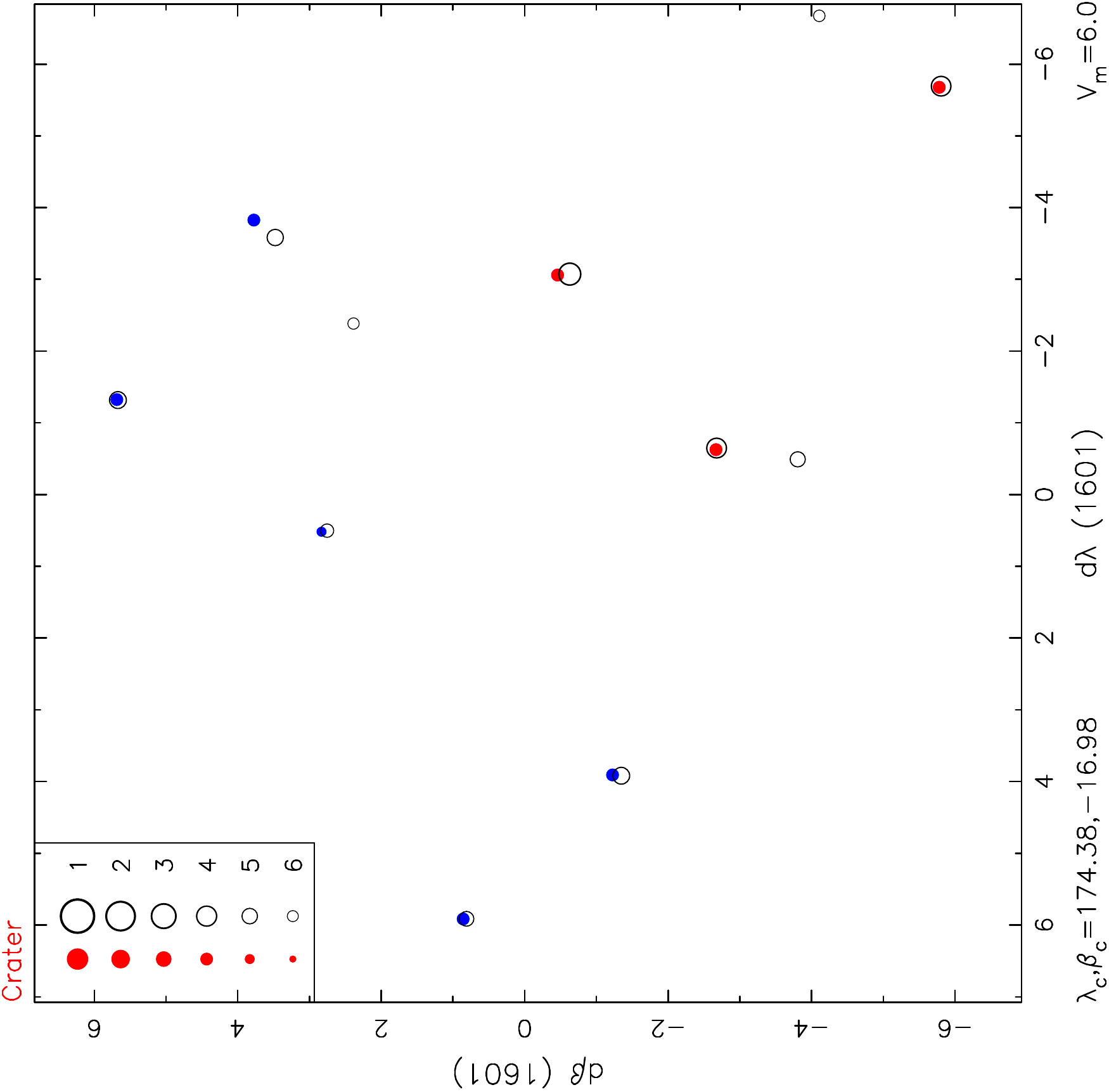}
\caption{Crater \label{f:crater}}
\end{figure}

\begin{figure}
\includegraphics[angle=270,width=\columnwidth]{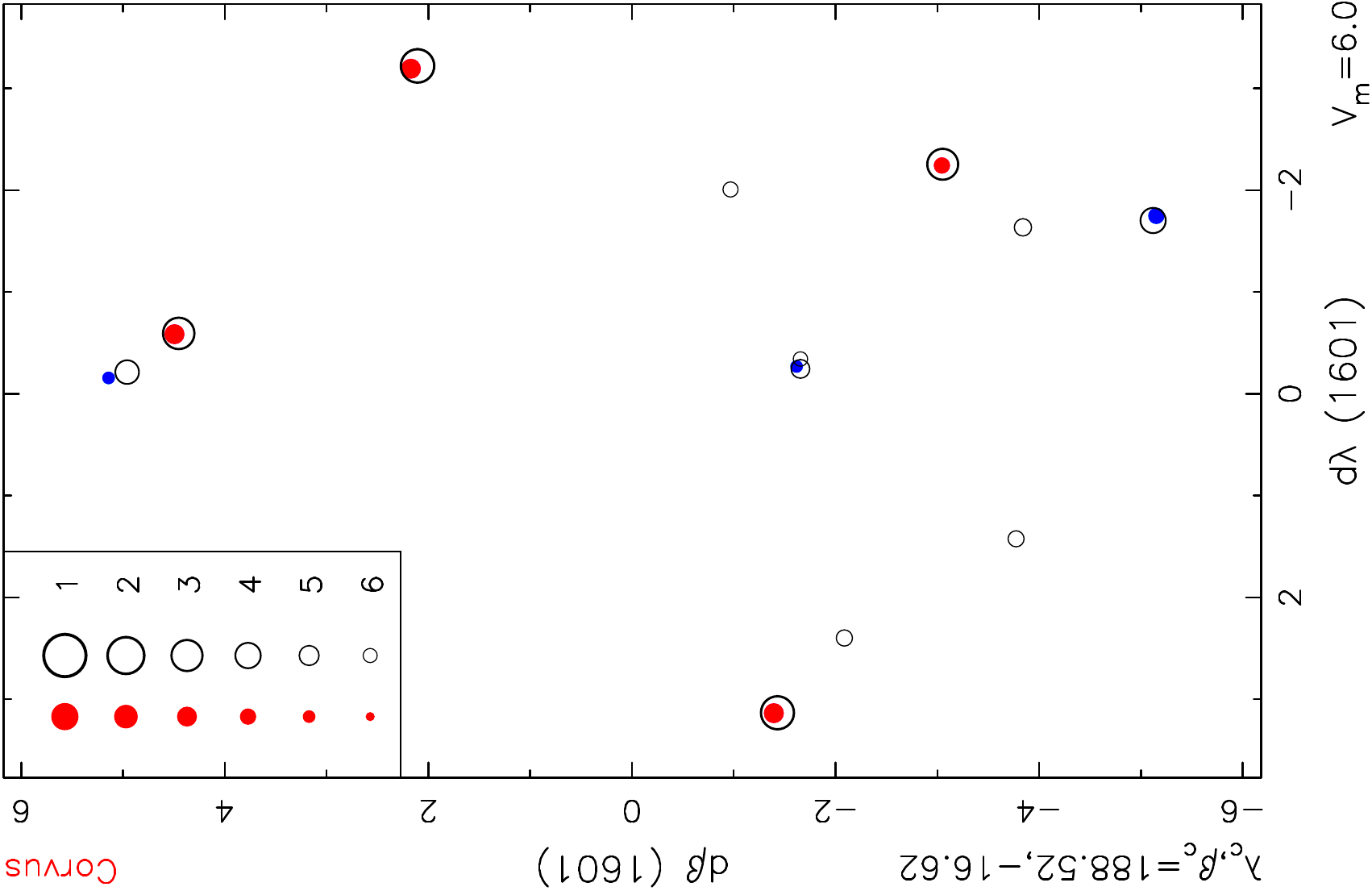}
\caption{Corvus \label{f:corvus}}
\end{figure}

\begin{figure*}
\includegraphics[angle=270,width=16.6cm]{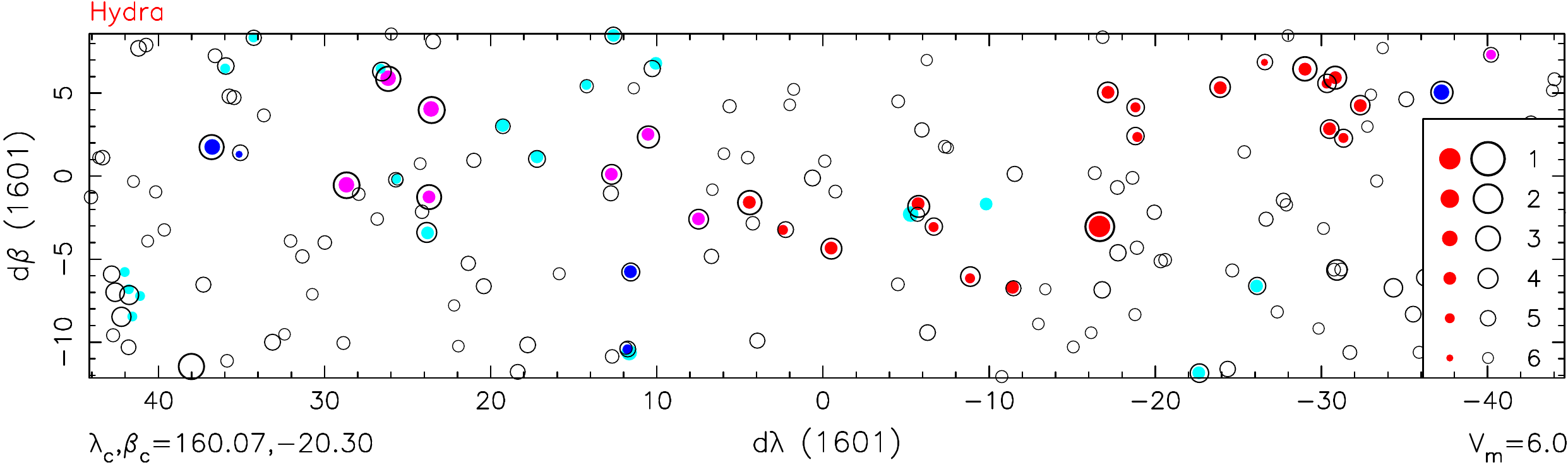}
\caption{Hydra \label{f:hydra}}
\end{figure*}

\end{appendix}

% -----------------------------------------
\end{document}